\documentclass[prb,aps,nofootinbib, twocolumn, amssymb,superscriptaddress,10pt]{revtex4-2}
\usepackage{amsmath}
\usepackage{amssymb}
\usepackage{amsthm}
\usepackage{amsfonts}%
\usepackage{enumerate}
\usepackage{latexsym}
\usepackage{color}
\usepackage{setspace} 
\usepackage{blindtext}
\usepackage{dsfont}
\usepackage{mathrsfs}
\usepackage[normalem]{ulem}

\usepackage{tabularx}
\newcolumntype{Y}{>{\centering\arraybackslash}X}

\usepackage{stackengine}

\usepackage{bm}
\usepackage{graphicx}
\usepackage{subfigure}

\usepackage{hyperref}
\hypersetup{
pdfnewwindow=true, colorlinks=true,
linkcolor=blue, anchorcolor=blue,
citecolor=blue, filecolor=blue,
menucolor=blue, urlcolor=blue} 

\usepackage{pifont}
\newcommand{\cmark}{\ding{51}}%
\newcommand{\xmark}{\ding{55}}%

\newcommand{\vk}{\mathbf{k}}

\newcommand{\vq}{\mathbf{q}}
\newcommand{\mk}{\mathbf{k}}

\begin{document}

\title{Phase diagram of twisted bilayer graphene at filling factor \texorpdfstring{$\nu=\pm3$}{nu=pm3}}
\author{Fang Xie}
\affiliation{Department of Physics, Princeton University, Princeton, New Jersey 08544, USA}
\affiliation{Department of Physics \& Astronomy, Rice University, Houston, Texas 77005, USA}
\author{Jian Kang}
\affiliation{School of Physical Science and Technology, ShanghaiTech University, Shanghai 201210, China}
\author{B. Andrei Bernevig}
\affiliation{Department of Physics, Princeton University, Princeton, New Jersey 08544, USA}
\affiliation{Donostia International Physics Center, P. Manuel de Lardizabal 4, 20018 Donostia-San Sebastian, Spain}
\affiliation{IKERBASQUE, Basque Foundation for Science, Bilbao, Spain}
\author{Oskar Vafek}
\affiliation{National High Magnetic Field Laboratory, Tallahassee, Florida, 32310, USA}
\affiliation{Department of Physics, Florida State University, Tallahassee, Florida 32306, USA}
\author{Nicolas Regnault}
\affiliation{Laboratoire de Physique de l'Ecole normale sup\'{e}rieure, ENS, Universit\'{e} PSL, CNRS, Sorbonne Universit\'{e}, Universit\'{e} Paris-Diderot, Sorbonne Paris Cit\'{e}, 75005 Paris, France}
\affiliation{Department of Physics, Princeton University, Princeton, New Jersey 08544, USA}

\date{\today}

\begin{abstract}
We study the correlated insulating phases of twisted bilayer graphene (TBG) in the absence of lattice strain at integer filling $\nu=\pm3$. 
Using the self-consistent Hartree-Fock method on a particle-hole symmetric model and allowing translation symmetry breaking terms, we obtain the phase diagram with respect to the ratio of $AA$ interlayer hopping $(w_0)$ and $AB$ interlayer hopping $(w_1)$. 
When the interlayer hopping ratio is close to the chiral limit ($w_0/w_1 \lesssim 0.5$), a quantum anomalous Hall state with Chern number $\nu_c = \pm 1$ can be observed consistent with previous studies. 
Around the realistic value $w_0/w_1 \approx 0.8$, we find a spin and valley polarized, translation symmetry breaking, state with $C_{2z}T$ symmetry, a charge gap and a doubling of the moir\'e unit cell, dubbed the $C_{2z}T$ stripe phase.
The real space total charge distribution of this $C_{2z}T$ stripe phase in the flat band limit does not have modulation between different moir\'e unit cells, although the charge density in each layer is modulated, and the translation symmetry is strongly broken.
Other symmetries, including $C_{2z}$, $C_{2x}$ and particle-hole symmetry $P$, and the topology of the $C_{2z}T$ stripe phase are also discussed in detail.
We observed braiding and annihilation of the Dirac nodes by continuously turning on the order parameter to its fully self-consistent value, and provide a detailed explanation of the mechanism for the charge gap opening despite preserving $C_{2z}T$ and valley $U(1)$ symmetries.
In the transition region between the quantum anomalous Hall phase and the $C_{2z}T$ stripe phase, we find an additional competing state with comparable energy corresponding to a phase with a tripling of the moir\'e unit cell. 

\end{abstract}

\maketitle

\section{Introduction}

Twisted bilayer graphene (TBG) at magic angle hosts a wealth of correlated insulating and superconducting phases, and as such is one of the most significant experimental discoveries in the recent years \cite{cao_correlated_2018, cao_unconventional_2018,xie2019spectroscopic, jiang_charge_2019, choi_imaging_2019, zondiner_cascade_2020,wong_cascade_2020,nuckolls_chern_2020,choi2020tracing,lu2019superconductors, yankowitz2019tuning, sharpe_emergent_2019, saito_independent_2020, stepanov_interplay_2020, arora_2020, serlin_QAH_2019, cao_strange_2020, polshyn_linear_2019, saito2020,das2020symmetry,saito2020isospin, wu_chern_2020,park2020flavour,Cao2021nematicity, das2022observation}. 
It also triggered a number of theoretical studies, in particular for the emerging strongly interacting insulating phases at integer fillings \cite{ochi_possible_2018,po_origin_2018,kang_strong_2019,xie_HF_2020, bultinck_ground_2020,liu2020theories,hejazi2020hybrid,cea_band_2020,zhang_HF_2020,liu2020nematic,xux2018,daliao_VBO_2019,daliao2020correlation,classen2019competing,kang_nonabelian_2020, soejima2020efficient,repellin_EDDMRG_2020, christos2020superconductivity, khalaf_charged_2020,Potasz_exact_2021,ourpaper3,ourpaper4,ourpaper5,ourpaper6,KWAN2021, kang_cascade_2021,  pan_dynamical_2022, Hofmann_fermionic_2022,zhang2021correlated}. 
Among them, a particularly interesting case corresponds to filling one of the eight active flat bands of TBG, namely the filling $\nu=-3$ (or its analogue for holes, namely $\nu = +3$).
There is a rich variety of candidate states for $\nu=\pm3$ depending on factors such as the strength of interlayer hoppings, strain of the lattice, or external fields.
The experimental results at this filling factor also depend on the specific setup: the quantum anomalous Hall (QAH) effect was observed when the sample is aligned to the hexagonal boron nitride (hBN) substrate~\cite{serlin_QAH_2019} or subject to an external magnetic field~\cite{choi2020tracing, saito2020, park2020flavour, das2022observation}, but not without the hBN alignment and at zero magnetic field~\cite{lu2019superconductors,yankowitz2019tuning}.

The nature of the insulating phase at $\nu=\pm3$ has been studied by various theoretical and numerical methods, including strong coupling expansion \cite{ourpaper4,kang_strong_2019}, mean field approximation \cite{zhang_HF_2020,kang_nonabelian_2020, xie_HF_2020, hejazi2020hybrid, KWAN2021, zhang2021correlated}, DMRG \cite{kang_nonabelian_2020,soejima2020efficient} and exact diagonalization \cite{ourpaper6,Potasz_exact_2021}, in which multiple types of candidate states are proposed. These theoretical studies have shown that the insulating states are close to Slater determinant wavefunctions with Chern number $\nu_C = \pm1$~\cite{kang_nonabelian_2020, Potasz_exact_2021, hejazi2020hybrid, ourpaper4, ourpaper6,soejima2020efficient,zhang_HF_2020} when the interlayer hoppings satisfy $w_0/w_1 \lesssim 0.5$, in which $w_0$ and $w_1$ are related to the values of $AA$ and $AB$ interlayer hoppings. 
However, the nature of the ground state at a larger --and perhaps more realistic-- value of $w_0/w_1 \approx 0.8$~\cite{Uchida_corrugation,Wijk_corrugation,jain_corrugation,koshino_maximally_2018} at $\nu=-3$, where the Chern insulator with $\nu_C=\pm 1$ disappears, is still a matter of debate~\cite{kang_nonabelian_2020, kang_strong_2019, KWAN2021, soejima2020efficient, hejazi2020hybrid, ourpaper5, ourpaper6}. For instance, the charge neutral excitations found in Ref.~\cite{ourpaper5} by perturbing the Chern insulator wavefunction with such larger values of $w_0/w_1$ contain states with negative energy at non-zero momentum, in agreement with exact diagonalization results~\cite{ourpaper6}. Condensation of such charge neutral modes at finite momentum would lead to states with broken translation symmetry. 
In Ref.~\cite{KWAN2021}, mean-field study also suggested Kekul\'e spiral state with broken translation symmetry in the presence of lattice heterostrain, although --in contrast to this work-- no translation symmetry breaking insulating state with zero Chern number was found without strain. 
In addition, semimetal states with broken rotation symmetry were also found to be highly energetically competitive in Refs.~\cite{kang_nonabelian_2020,soejima2020efficient}. 

To settle this question, we calculate the phase diagram of interacting TBG at integer filling $\nu=-3$ by varying the ratio $w_0/w_1$ in the absence of strain by using the self-consistent Hartree-Fock method. The self-consistent mean field order parameter allows hybridization between states with different momenta, which in turn allows translation symmetry breaking with enlarged unit cells. When the value of $w_0/w_1$ is small, we observe a QAH state, consistent with the results discussed in Refs.~\cite{kang_nonabelian_2020, Potasz_exact_2021, ourpaper4, ourpaper6,soejima2020efficient,zhang_HF_2020}. 
Near a realistic value of $w_0/w_1$, we find an energetically preferred phase with broken translation symmetry and a large charge gap, which, although similar, differs in detail from the previous proposals by Refs.~\cite{kang_strong_2019,kang_nonabelian_2020,zhang2021correlated}. 
This phase strongly hybridizes states whose momenta differ by $(\pi, 0)$ ($M$ point of moir\'e Brillouin zone). Therefore, it has a stripe shape in real space, and the new unit cell contains two moir\'e unit cells. 
Similar to Refs.~\cite{kang_strong_2019,kang_nonabelian_2020}, the total charge density distribution is \emph{identical} in every moir\'e unit cell when the dispersion of non-interacting flat bands is neglected, despite translation symmetry being strongly broken.
While the charge density distribution in one layer has modulation between different moir\'e unit cells, it is exactly compensated by the charge in the other layer, canceling the modulation of the total density.
Unlike the QAH phase, this stripe phase does not break the $C_{2z}T$ symmetry and therefore it cannot lead to an anomalous Hall effect. We verify this explicitly by computing the Wilson loops of the mean field bands, finding that the stripe phase does not carry a Chern number. 
We also study the process of the gap opening without breaking $C_{2z}T$ symmetry by gradually turning the interaction induced self-energy and moving away from the gapless non-interacting state. Depending on the path toward the fully interacting case, we can observe the Dirac points braiding and annihilation, which was first conjectured in Refs.~\cite{ahn_failure_2019,kang_nonabelian_2020, wu2019nonabelian}, and elaborate on the mechanism of the gap opening and the topology of the resulting $C_{2z}T$ stripe state. 
Between the $C_{2z}T$ stripe phase and the QAH phase, we also find a range of values of $w_0/w_1$ with multiple candidate states with comparable energies, including another translation symmetry breaking phase tripling the unit cell.

This article is organized as follows. In Sec.~\ref{sec:model} we briefly review the projected interacting Hamiltonian of TBG. We also discuss the folded moir\'e Brillouin zones which correspond to translation symmetry breaking considered in this article. Sec.~\ref{sec:hf_main_text} introduces the notations and concepts which are required to depict the Hartree-Fock mean field solutions. Then in Sec.~\ref{sec:phase_diagram}, we present the broken symmetries, band structures and topology of the various phases emerging at different values of $w_0/w_1$. We provide a detailed study of the $C_{2z}T$ stripe phase in Sec.~\ref{sec:c2t_stripe}. Finally, we summarize and discuss the results in Sec.~\ref{sec:conclusion}.

\section{Model}\label{sec:model}
In this section, we briefly introduce the notations of the interacting Hamiltonian of TBG projected into the flat bands. We also present the folded moir\'e Brillouin zones corresponding to enlarged unit cells that will be considered in this article.

\subsection{Non-interacting Hamiltonian}

We start with a short review of the non-interacting Hamiltonian of TBG \cite{bistritzer_moire_2011}. We will use the same notations as Ref. \cite{ourpaper3,ourpaper4, ourpaper6}:  $c^\dagger_{\vk, \alpha, s, \ell}$ denotes the electron creation operator, in which $\vk$ is the electron momentum measured from the single layer graphene $\Gamma$ point, $\alpha = A, B$ is the graphene sublattice, $s = \uparrow, \downarrow$ is the electron spin and $\ell=\pm 1$ refers to the graphene layer. 
The low energy behavior of electrons in single layer graphene is well-captured by the states around the two Dirac points $K$ and $K'$. Thus, it is reasonable to use the basis of the Bistritzer-MacDonald model. By focusing on one valley $K$, we define vectors $\vq_{j} = C_{3z}^{j-1}(\mathbf{K}_- - \mathbf{K}_+)$, which represent the difference between Dirac points in top and bottom layers due to the twisting. The vector $\mathbf{K}_\ell$ is the momentum of the Dirac point $K$ in layer $\ell$, and $|\mathbf{K}_\ell| = 1.703\,\textrm{\AA}^{-1}$. The reciprocal vectors of the moir\'e lattice, denoted by $\mathcal{Q}_0$, are spanned by basis vectors $\tilde{\mathbf{b}}_{1} = \vq_{2} - \vq_{3}$ and $\tilde{\mathbf{b}}_{2} = \vq_{2} - \vq_{1}$. 
The momenta lattices $\mathcal{Q}_\pm = \mathcal{Q}_0 \pm \vq_1$ form a hexagonal lattice in momentum space, which stand for the copies of Dirac points from the top and bottom layers in repeated moir\'e Brillouin zone, respectively.

Parameterizing the electron operators as follows:
\begin{equation}
    c^\dagger_{\vk, \mathbf{Q}, \eta, \alpha, s} = 
    c^\dagger_{\eta\mathbf{K}_{\eta\cdot \ell} + \vk - \mathbf{Q}, \alpha, s, \eta\cdot \ell}~~~\text{if}~\mathbf{Q}\in\mathcal{Q}_{\ell}\,,
\end{equation}
in which $\eta = \pm$ stands for the valley index, the second quantized non-interacting Hamiltonian of TBG can be written as
\begin{equation}
    \hat{H}_0 = \sum_{\substack{\vk\in{\rm MBZ}\\\mathbf{Q},\mathbf{Q}'\in\mathcal{Q}_\pm\\\eta s \alpha \beta}}\Big[h^{(\eta)}_{\mathbf{Q} \mathbf{Q}'}(\vk)\Big]_{\alpha\beta}c^\dagger_{\vk,\mathbf{Q},\eta, \alpha, s}c_{\vk,\mathbf{Q}', \eta, \beta, s}\,,
\end{equation}
where MBZ stands for moir\'e Brillouin zone. The ``first quantized'' single-body Hamiltonians of TBG $h^{(\eta)}(\vk)$, which is also known as Bistritzer-MacDonald (BM) Hamiltonian \cite{bistritzer_moire_2011}, is given by the following equations:
{\small\begin{align}
    h^{(+)}_{\mathbf{Q} \mathbf{Q}'}(\vk) &= v_F \bm{\sigma}\cdot(\vk - \mathbf{Q})\delta_{\mathbf{Q}, \mathbf{Q}'} + \sum_{j=1}^3T_j \delta_{\mathbf{Q} - \mathbf{Q}',\pm\vq_j}\,,\\
    h^{(-)}_{\mathbf{Q} \mathbf{Q}'}(\vk) &= -v_F \bm{\sigma}^*\cdot(\vk - \mathbf{Q})\delta_{\mathbf{Q}, \mathbf{Q}'} + \sum_{j=1}^3\sigma_xT_j\sigma_x \delta_{\mathbf{Q} - \mathbf{Q}',\pm\vq_j}\,,
\end{align}
}in which $\bm{\sigma} = (\sigma_x, \sigma_y)$ and $\bm{\sigma}^* = (\sigma_x, -\sigma_y)$, and Fermi velocity $v_F = 6104.5\rm meV\cdot \textrm{\AA}$. The matrices $T_j$, which describe the strength of the interlayer hoppings, are given by the following equation:
{\small
\begin{equation}
    T_j = w_0 \sigma_0 + w_1 \left[\cos\frac{2\pi (j - 1)}{3}\sigma_x + \sin\frac{2\pi (j - 1)}{3}\sigma_y\right]\,.
\end{equation}
}Here $w_0$ and $w_1$ are proportional to the interlayer tunneling amplitudes in the $AA$ and $AB$ stacking regions in moir\'e unit cell, respectively. 
In this paper, we fix the value of $AB$ hopping $w_1 = 110\,\rm meV$ and twist angle $\theta = 1.07^\circ$, and we use $w_0/w_1 \in [0, 1]$ as a tunable parameter of our non-interacting Hamiltonian $H_0$. A realistic value of $w_0/w_1$ is expected to be around $0.7\sim 0.8$ due to the lattice corrugation \cite{Uchida_corrugation,Wijk_corrugation,jain_corrugation,koshino_maximally_2018}.

By diagonalizing the single-body Hamiltonian, we can obtain the band structure $\epsilon_{\vk, m, \eta}$ and single-body wavefunctions $u_{\mathbf{Q}\alpha, m\eta}(\vk)$ of TBG:
\begin{equation}\label{eqn:diag_bm_ham}
    \sum_{\mathbf{Q}'\beta} h^{(\eta)}_{\mathbf{Q}\alpha,\mathbf{Q}'\beta}(\vk) u_{\mathbf{Q}' \beta, m\eta}(\vk) = \epsilon_{\vk, m, \eta}u_{\mathbf{Q} \alpha, m \eta}(\vk)\,,
\end{equation}
where $m$ is the energy band index. The non-interacting Hamiltonian can be written in the eigenstate basis:
\begin{equation}\label{eqn:h0}
    \hat{H}_0 = \sum_{\vk\in{\rm MBZ}}\sum_{\eta, s} \sum_{m\neq 0} \epsilon_{\vk,m,\eta}  c^\dagger_{\vk, m, \eta, s}c_{\vk, m, \eta, s}\,.
\end{equation}
These electron operators in the energy band basis $c^\dagger_{\vk, m, \eta, s}$ are given by:
\begin{align}
    c^\dagger_{\vk, m, \eta, s} &= \sum_{\mathbf{Q}\alpha}u_{\mathbf{Q} \alpha, m \eta}(\vk) c^\dagger_{\vk, \mathbf{Q}, \eta, \alpha, s}\,,\label{ed:eqn:band_to_planewave}\\
    c^\dagger_{\vk, \mathbf{Q}, \eta, \alpha, s} &= \sum_{m} u^*_{\mathbf{Q}\alpha, m\eta}(\vk) c^\dagger_{\vk, m, \eta, s} \,.\label{ed:eqn:planewave_to_band}
\end{align} 
We fix the gauge choice of the single body wavefunctions $u_{\mathbf{Q}\alpha, m\eta}(\vk)$ as described in Ref. \cite{ourpaper3,ourpaper4,ourpaper5,ourpaper6}, such that the sewing matrix of $C_{2z}T$ symmetry is given by identity matrix. Thus, the operators $c^\dagger_{\vk,m,\eta,s}$ will not change under $C_{2z}T$ transformation:
\begin{equation}\label{eqn:c2t_operator}
    (C_{2z}T)c^\dagger_{\vk m \eta s}(C_{2z}T)^{-1} = c^\dagger_{\vk m \eta s}\,.
\end{equation}
Except for $C_{2z}T$, the single valley non-interacting Hamiltonian also has $C_{3z}$, $C_{2x}$ and a particle hole symmetry $P$. These symmetries are discussed in detail in App.~\ref{sec:app_ham_sym}.

As discussed in Ref. \cite{bistritzer_moire_2011}, there are two flat bands around the first magic angle at charge neutrality per spin and valley, separated by a gap from other remote bands. Therefore, we can project the non-interacting Hamiltonian Eq.~(\ref{eqn:h0}) into these eight total flat bands:
\begin{equation}\label{eqn:h0_proj}
    H_0 = \sum_{\vk\in{\rm MBZ},\eta s}\sum_{m=\pm1}\epsilon_{\vk, m, \eta}c^\dagger_{\vk, m, \eta, s}c_{\vk, m, \eta, s}\,.
\end{equation}

\subsection{Interacting Hamiltonian}
We consider the density-density interaction projected into the TBG flat bands. The projected interacting Hamiltonian reads \cite{ourpaper3}:
\begin{equation}\label{eqn:interacting_ham}
    H_I = \frac{1}{2\Omega_{\rm tot}}\sum_{\mathbf{G} \in \mathcal{Q}_0,\vq \in {\rm MBZ}}V(\vq + \mathbf{G})\overline{\delta\rho}_{\vq + \mathbf{G}}\overline{\delta\rho}_{-\vq - \mathbf{G}},
\end{equation}
in which $V(\vq)$ is the Fourier transform of screened Coulomb potential. In this article, we consider double gated TBG, leading to a Fourier transform interaction given by $V(\vq) = \pi \xi^2U_\xi \tanh(\xi q/2)/(\xi q/2)$, where $\xi = 10\,\rm nm$ is the distance between the two gates and $U_\xi = 24\,\rm meV$. The operator $\overline{\delta\rho}_{\vq + \mathbf{G}}$ represents the relative electron density measured from charge neutrality, after being projected into the TBG flat bands:
\begin{align}
    \overline{\delta\rho}_{\vq + \mathbf{G}} =& \sum_{\vk\in{\rm MBZ}}\sum_{mn\eta s}M^{(\eta)}_{mn}(\vk, \vq + \mathbf{G})\nonumber\\
    &\times \left(c^\dagger_{\vk + \vq, m, \eta, s} c_{\vk, n, \eta, s}-\frac12 \delta_{\mathbf{q}, 0}\delta_{m,n}\right)\,,
\end{align}
\begin{equation}
    M^{(\eta)}_{mn}(\vk, \vq + \mathbf{G}) = \sum_{\mathbf{Q}\alpha} u^*_{\mathbf{Q}\alpha, m\eta}(\vk + \vq + \mathbf{G}) u_{\mathbf{Q}\alpha, n\eta}(\vk)\,,\label{eqn:def_form_factor}
\end{equation}
where $u_{\mathbf{Q}\alpha, m\eta}(\vk)$ is the wavefunction of the BM Hamiltonian defined in Eq.~(\ref{eqn:diag_bm_ham}). Since $\overline{\delta\rho}_{\mathbf{q}+\mathbf{G}}$ is defined as the relative density measured from the charge neutrality, the interacting Hamiltonian in Eq.~(\ref{eqn:interacting_ham}) has a many-body particle-hole symmetry, which leads to identical phases at $\nu$ and $-\nu$. As such, our results at $\nu=-3$ will also be valid for $\nu=3$. These wavefunctions depend on the interlayer hopping parameter $w_0/w_1$. Thus, the interacting Hamiltonian $H_I$ will also depend on $w_0/w_1$.

By adding the non-interacting term in Eq.~(\ref{eqn:h0_proj}), we obtain the total Hamiltonian:
\begin{equation}
    H = t H_0 + H_I\,.
\end{equation}
Here we introduce a parameter $t\in[0, 1]$ to control the relative strength of the flat band kinetic energy. In this article, we will mostly focus on the flat band limit, {\it i.e.}, $t=0$, unless otherwise stated. 

\subsection{Folded moir\'e Brillouin zone}

As suggested by the presence of Fermi pockets of charge $\pm 1$ excitations, and negative excitations in the charge neutral spectra for a range of values of $w_0/w_1$ away from the chiral limit \cite{ourpaper5,ourpaper6}, it is reasonable to expect that the system will host translation symmetry breaking ground states. Therefore, we account for translation symmetry breaking orders by considering enlarged unit cells, or folded moir\'e Brillouin zones. Each type of translation symmetry breaking order is associated with a specific pair of momenta $\mathbf{Q}_{1,2}$. Any two momenta that differ by an integer multiple of these vectors should be identified as the same point in the folded Brillouin zone:
\begin{equation}
    \vk_1 - \vk_2 = l_1 \mathbf{Q}_1 + l_2 \mathbf{Q}_2\,~~l_1, l_2 \in \mathbb{Z}\,.
\end{equation}
The vectors $\mathbf{Q}_{1,2}$ are the basis vectors of folded moir\'e Brillouin zone. We define the following quantity:
\begin{equation}
    N_F = |\tilde{\mathbf{b}}_1 \times \tilde{\mathbf{b}}_2|\big{/}|\mathbf{Q}_1\times \mathbf{Q}_2|\,,
\end{equation}
as the number of times the moir\'e Brillouin zone is folded, with $\tilde{\mathbf{b}}_{1,2}$ the reciprocal vectors of the original moir\'e lattice. Therefore, every momentum $\mathbf{k} \in {\rm MBZ}$ can always be represented by a momentum value $\bm{\kappa}$ in the folded (small) moir\'e Brillouin zone (FMBZ) together with an integer $b$ (dubbed {\it subband} index):
\begin{equation}\label{eqn:fmbz_momentum}
    \mathbf{k} = \bm{\kappa} + \mathbf{Q}_b\,,~~  \bm{\kappa}\in{\rm FMBZ},~b = 1,2,\cdots, N_F\,,
\end{equation}
in which $\mathbf{Q}_b = l_1 \mathbf{Q}_1 + l_2 \mathbf{Q}_2$ stand for all the $N_F$ reciprocal vectors of the FMBZ in the 1st MBZ. We focus on the eight simplest ({\it i.e.}, the smallest $N_F$ values, up to $N_F = 4$) types of Brillouin zone folding vectors $\mathbf{Q}_{1,2}$, and their notations and factor of Brillouin zone folding $N_F$ are shown in Table \ref{tab:unit_cell_choices}.

\begin{center}
    \begin{table}[t]
        \begin{tabularx}{\linewidth}{Y|Y|Y|Y}
        \hline
        notations & $\mathbf{Q}_1$ & $\mathbf{Q}_2$ & $N_F$\\
        \hline\hline
        $(2\times 1)$ & $\frac12 \tilde{\mathbf{b}}_1$ & $\tilde{\mathbf{b}}_2$ & 2 \\
        \hline
        $(1\times 2)$ & $\tilde{\mathbf{b}}_1$ & $\frac12\tilde{\mathbf{b}}_2$ & 2 \\
        \hline
        $(3\times 1)$ & $\frac13 \tilde{\mathbf{b}}_1$ & $\tilde{\mathbf{b}}_2$ & 3 \\
        \hline
        $(1\times 3)$ & $\tilde{\mathbf{b}}_1$ & $\frac13 \tilde{\mathbf{b}}_2$ & 3 \\
        \hline
        $(2\times 2)$ & $\frac12\tilde{\mathbf{b}}_1$ & $\frac12 \tilde{\mathbf{b}}_2$ & 4\\
        \hline
        $(4\times 1)$ & $\frac14\tilde{\mathbf{b}}_1$ & $\tilde{\mathbf{b}}_2$ & 4\\
        \hline
        $(1\times 4)$ & $\tilde{\mathbf{b}}_1$ & $\frac14\tilde{\mathbf{b}}_2$ & 4\\
        \hline
        $(\sqrt3\times \sqrt3)$ & $\frac13(\tilde{\mathbf{b}}_1 + \tilde{\mathbf{b}}_2)$ & $\frac13(\tilde{\mathbf{b}}_1 - \tilde{\mathbf{b}}_2)$ & 3\\
        \hline
        \end{tabularx}
    \caption{The enlarged unit cell choices. The first column shows the notation we use for each type of enlarged unit cells. The second and third columns provide the basis vectors of the folded moir\'e Brillouin zones. The fourth column gives the factor of folding $N_F$, which represents the amount of moir\'e unit cells in each enlarged unit cell.}
    \label{tab:unit_cell_choices}
    \end{table}
\end{center}

\section{Hartree-Fock}\label{sec:hf_main_text}
In this section, we provide an overview of the concepts and notations that will be required to describe the Hartree-Fock results in Sec.~\ref{sec:phase_diagram}. We perform the numerical Hartree-Fock mean field calculation on a $C_{3z}$ rotation symmetric discrete $N_L \times N_L$ momentum lattice in the \emph{unfolded} MBZ. For convenience, we define the total amount of momentum points in MBZ as $N_M = N_L^2$. Hence, the momentum values in MBZ are given by the following set:
{\small\begin{equation}
    {\rm MBZ} = \left\{\vk\Big{|}\vk = \frac{k_1}{N_L} \tilde{\mathbf{b}}_1 + \frac{k_2}{N_L}\tilde{\mathbf{b}}_2; k_1, k_2 = 0, 1, \cdots, N_L-1\right\}\,.
\end{equation}}
Thus, there will be $N_M$ states in each energy band. In this article, we mostly focus on the integer filling factor $\nu=-3$. At this filling factor, the total number of electrons in the narrow bands is $N = N_M$.

As shown in Eq.~(\ref{eqn:fmbz_momentum}), for a given choice of enlarged unit cell, the FMBZ is a subset of MBZ, and each momentum $\vk \in {\rm MBZ}$ can be represented by a momentum $\bm{\kappa} \in {\rm FMBZ}$ and a subband index $b$. Thus, a single body state can be represented by five quantum numbers: momentum $\bm{\kappa}\in{\rm FMBZ}$, subband index $b = 1, 2, \cdots, N_F$, energy band index $m=\pm 1$, valley $\eta = \pm$ and spin $s=\uparrow\downarrow$.

The Hartree-Fock order parameter with broken translation symmetry has the following form:
\begin{align}
    \Delta_{bm\eta s; b'n \eta' s'}(\bm{\kappa}) =& \langle c^\dagger_{\bm{\kappa} + \mathbf{Q}_b,m\eta s}c_{\bm{\kappa} + \mathbf{Q}_{b'}, n\eta's'} \rangle\nonumber\\
    & - \frac{1}{2}\delta_{bb'}\delta_{mn}\delta_{\eta\eta'}\delta_{ss'} \,,~~\bm{\kappa} \in {\rm FMBZ}\,.\label{eqn:def_order_parameter}
\end{align}
For each momentum $\bm{\kappa}$, the order parameter $\Delta(\bm{\kappa})$ is a $8N_F\times 8N_F$ matrix. The Hartree-Fock Hamiltonians $\mathcal{H}^{(HF)}_{bm\eta s; b'n\eta's'}(\bm{\kappa})$, which are also $8N_F\times 8N_F$ matrices, can be written as functions of momentum $\bm{\kappa}$ and the order paramter $\Delta(\bm{\kappa})$. The explicit expression of the Hartree-Fock Hamiltonians and the iterative self-consistent method are discussed in detail in App.~\ref{sec:hf_app}. By diagonalizing the Hartree-Fock Hamiltonian, we obtain the Hartree-Fock band dispersion $E_i(\bm{\kappa})$ and its corresponding HF wavefunction $\phi_{bm\eta s, i}(\bm{\kappa})$:
\begin{equation}
    E_i(\bm{\kappa}) \phi_{bm\eta s, i}(\bm{\kappa}) = \sum_{b'n\eta's'}\mathcal{H}^{(HF)}_{bm\eta s; b'n\eta's'}(\bm{\kappa})\phi_{b'n\eta's', i}(\bm{\kappa})\,.
\end{equation}

To characterize a given Hartree-Fock mean field solution, we define several quantities. The first quantity is the translation symmetry breaking strength $\mathcal{T}$ which is defined as the norm of the off-diagonal elements of the order parameter in the subband indices. It can be written as:
\begin{equation}\label{eqn:def_trans_breaking}
    \mathcal{T} = \frac{1}{N_M} \sum_{\bm{\kappa}\in{\rm FMBZ}}\sum_{b\neq b'}\sum_{mn,\eta\eta',ss'} |\Delta_{bm\eta s; b'n\eta's'}(\bm{\kappa})|^2\,.
\end{equation}
For a translation symmetric solution, the off-diagonal elements in $b,b'$ vanish and $\mathcal{T}=0$. When $\mathcal{T} \neq 0$, the solution breaks the translation symmetry by one moir\'e unit cell. 

We can also define a quantity to measure the strength of $C_{2z}T$ symmetry breaking. The projected interacting Hamiltonian is written by fermion operators with fixed $C_{2z}T$ sewing matrices. Thus, the creation/annihilation operators are invariant under the $C_{2z}T$ transformation as shown in Eq.~(\ref{eqn:c2t_operator}). It is also an anti-unitary transformation. Hence, a mean-field state is invariant under $C_{2z}T$ only when its order parameter has no imaginary part.
However, the $C_{2z}T$ symmetry is defined from the non-interacting TBG Hamiltonian for single spin and valley, it is actually a spinless operation. Due to the spin and valley $U(4)$ symmetry at the flat band limit \cite{bultinck_ground_2020, ourpaper3, kang_strong_2019}, imaginary parts can be introduced into the spin and valley components of the order parameter under certain $U(4)$ rotation without breaking $C_{2z}T$. Therefore, we first do a partial trace over the spin, valley and subband indices of the order parameter, and then we use the norm of the imaginary part of this {\it reduced order parameter} to measure the strength of $C_{2z}T$ symmetry breaking. It can be defined as the following equation:
\begin{equation}
    \mathcal{C} = \frac{1}{N_MN_F}\sum_{\bm{\kappa}\in{\rm FMBZ}}\sum_{mm'}\Bigg{(}\mathrm{Im}\sum_{b\eta s}\Delta_{bm\eta s;b m'\eta s}(\bm{\kappa})\Bigg{)}^2\,.
\end{equation}
If the solution does not break the $C_{2z}T$ symmetry, then the reduced order parameter will be real, and thus we have $\mathcal{C} = 0$.

Another quantity we use to describe the mean field solution is the charge gap $E_G$. For integer filling $\nu=-3$, once the moir\'e Brillouin zone is folded by $N_F$ times, there will be $N_F$ bands occupied in the folded Brillouin zone. For these symmetry breaking solutions, we define the charge gap as the difference between the bottom of the lowest conduction band ($(N_F + 1)$-th band from bottom) and the top of the highest valence band ($N_F$-th band from bottom).

\section{Phase diagram}\label{sec:phase_diagram}
In Sec.~\ref{sec:gs_bands}, we discuss the ground states appearing with different values of $w_0/w_1$, their broken symmetry and the band structures. We also study the $C_{2z}T$ symmetry and the topology of these states in Sec.~\ref{sec:c2t_topo_phase_diagram}.

\begin{figure*}[t]
    \centering
    \includegraphics[width=\linewidth]{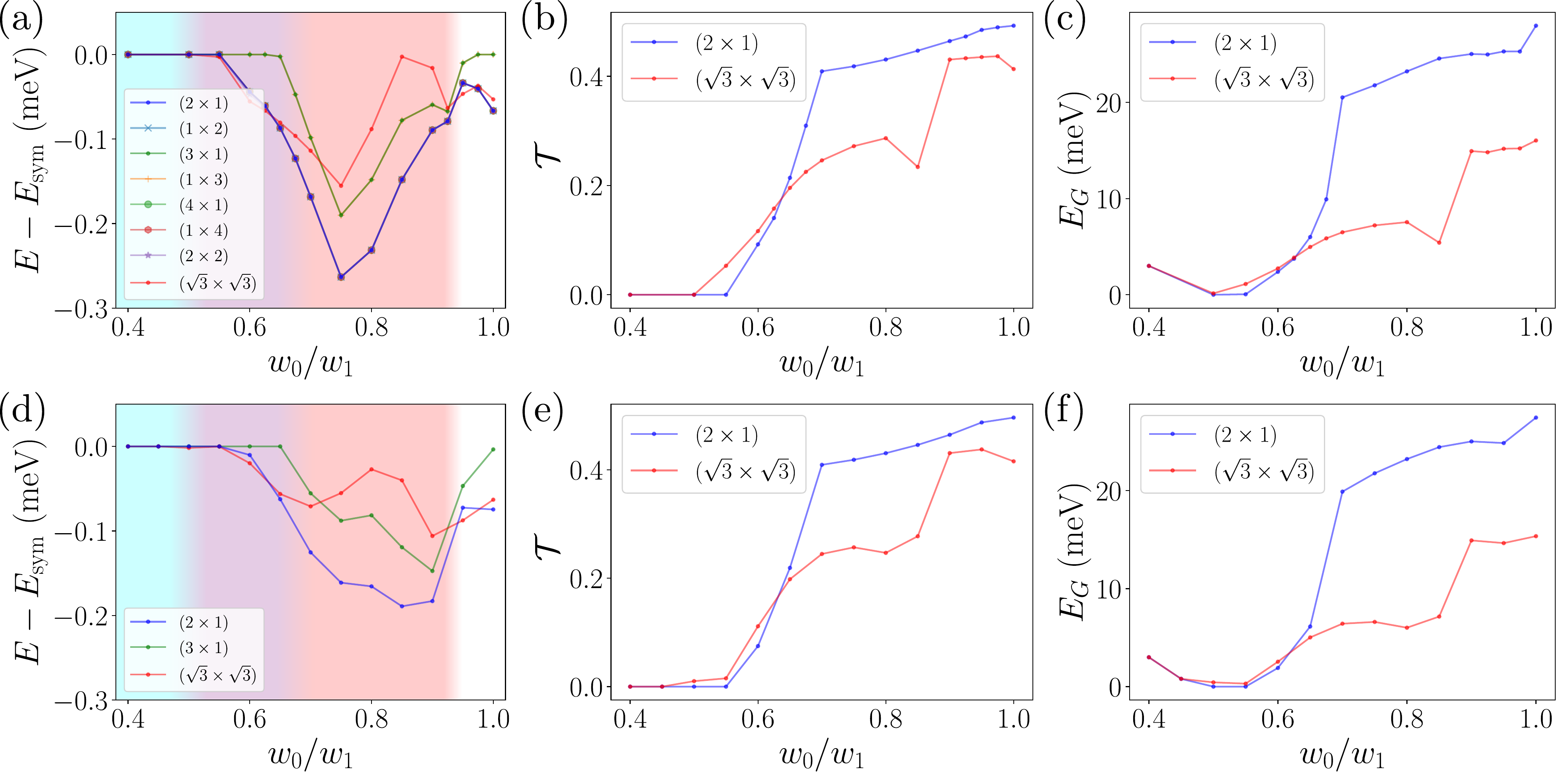}
    \caption{(a) We provide the energy difference per moir\'e unit cell $(E - E_{\rm sym})$ of density wave states with different possible enlarged unit cell choices as a function of $w_0/w_1$, calculated on $12\times 12$ momentum lattice, in which $E_{\rm sym}$ is the energy of translation symmetric solution. The shaded colors in the background represent different phases when the value of $w_0/w_1$ changes. In the region labeled by light blue, the mean field solution does not break the translation symmetry. In the purple region, the state we obtained with the lowest energy has enlarged unit cell $(\sqrt3 \times \sqrt3)$. However, the energy with $(2\times 1)$ unit cell is only slightly higher, which means the purple region has competing states with different enlarged unit cells. In the red region, the ground states we obtained has enlarged unit cell $(2\times1)$, which is a stripe phase in real space. (b) The strength of the translation symmetry breaking $\mathcal{T}$ of the two types of enlarged unit cells as a function of $w_0/w_1$. (c) The Hartree-Fock band gap $E_G$ of the two types of enlarged unit cells ($(\sqrt3\times \sqrt3)$ and $(2\times 1)$) as a function of $w_0/w_1$. (d) The energy difference per moir\'e unit cell of density wave states with enlarged unit cell choices $(2\times 1)$, $(3\times 1)$ and $(\sqrt{3}\times \sqrt{3})$ as a function of $w_0/w_1$ on a $18\times 18$ momentum lattice. In subfigures (e-f), we also show the Hartree-Fock band gap $E_G$ and the symmetry breaking strength $\mathcal{T}$ and $\mathcal{C}$ as functions of $w_0/w_1$ on the $18\times 18$ momentum lattice with two enlarged unit cell choices $(2\times1)$ and $(\sqrt{3} \times \sqrt{3})$.}
    \label{fig:mean_field_phase_diag}
\end{figure*}

\begin{figure*}[t]
    \centering
    \includegraphics[width=\linewidth]{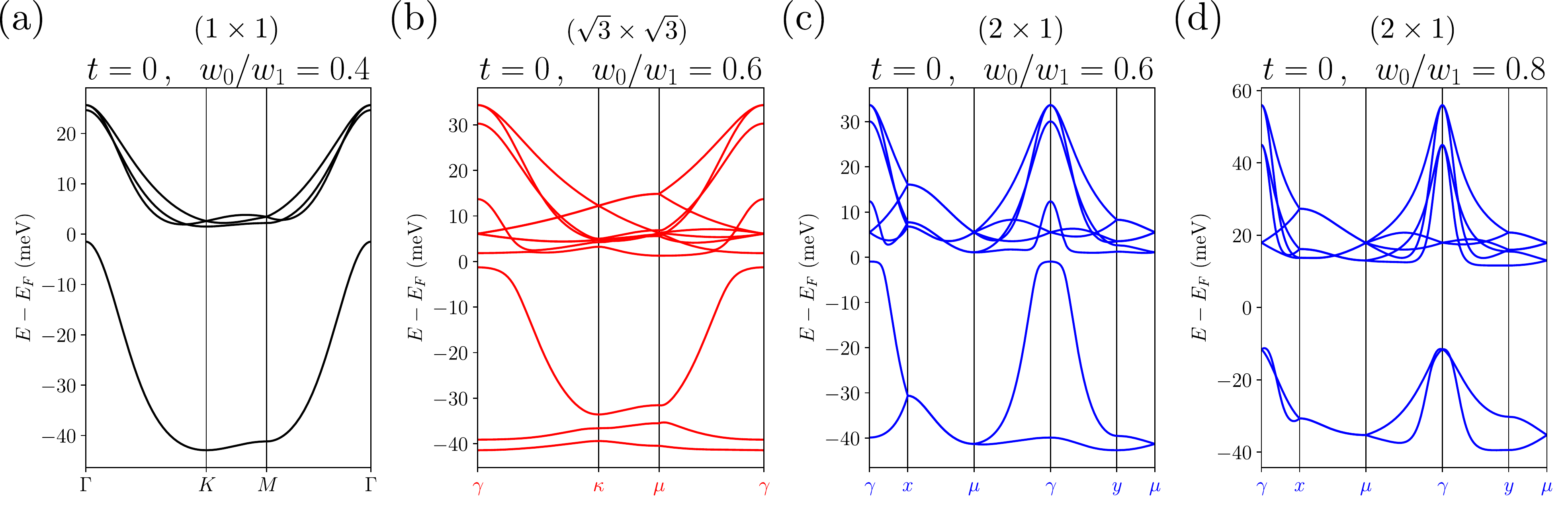}
    \caption{The Hartree-Fock band structures obtained on $18 \times 18$ momentum lattice at flat band limit. (a) The HF band structure without translation symmetry breaking at $w_0/w_1 = 0.4$. (b) The HF band structure with enlarged unit cell $(\sqrt3 \times \sqrt3)$ at $w_0/w_1 = 0.6$. (c) The HF band structure with enlarged unit cell $(2 \times 1)$ at $w_0/w_1 = 0.6$. (d) The HF band structure with enlarged unit cell $(2\times 1)$ at $w_0/w_1 = 0.8$. The definitions of high symmetry points of these folded moir\'e Brillouin zones are shown in Fig.~\ref{fig:folded_brillouin_zone}.}
    \label{fig:mean_field_bands_18x18}
\end{figure*}

\begin{figure}[t]
    \centering
    \includegraphics[width=\linewidth]{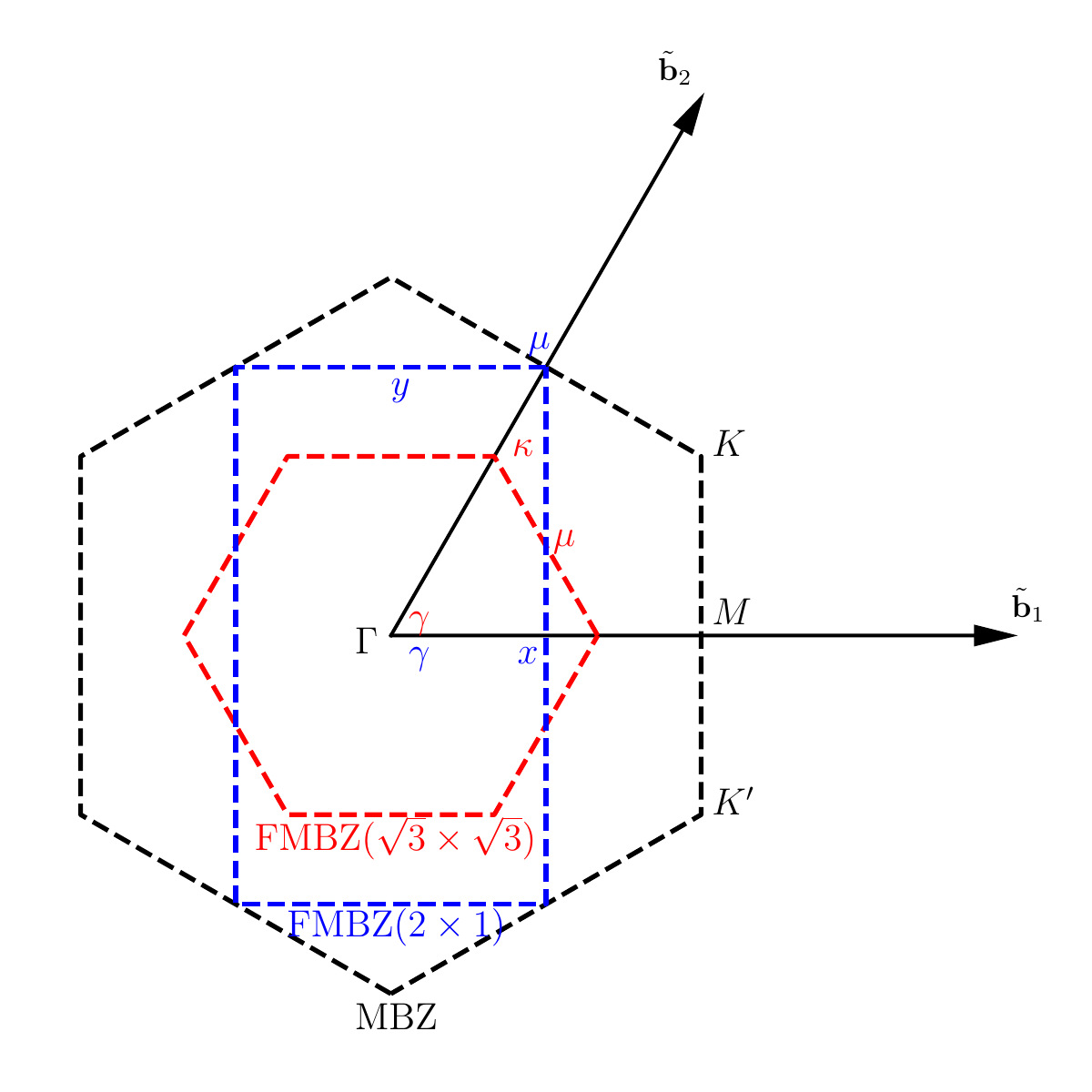}
    \caption{The moir\'e Brillouin zone (black), the folded Brillouin zone with $(2\times 1)$ unit cell choice (blue) and the folded Brillouin zone with $(\sqrt{3}\times \sqrt{3})$ unit cell (red). The high symmetry points of these different Brillouin zones are also represented by different colors, namely $\Gamma$, $K$, $K'$ and $M$ for the MBZ, $\gamma$, $x$, $\mu$ and $y$ for the FMBZ of $(2\times 1)$ unit cell, and $\gamma$, $\kappa$, $\kappa'$ and $\mu$ for the FMBZ of $(\sqrt3 \times \sqrt3)$ unit cell. Vectors $\tilde{\mathbf{b}}_1$ and $\tilde{\mathbf{b}}_2$ are the reciprocal lattice basis.}
    \label{fig:folded_brillouin_zone}
\end{figure}

\subsection{Ground states and band structures}\label{sec:gs_bands}

By performing the mean field calculation using the Hartree-Fock Hamiltonians with different choices of $\mathbf{Q}_{1,2}$ vectors shown in Table \ref{tab:unit_cell_choices}, and comparing the energy of different solutions, we are able to obtain a phase diagram with a varying value of $w_0/w_1$. In the following paragraphs, we ignore the effect of the flat band dispersion ($t=0$) and assume our order parameter $\Delta(\bm{\kappa})$ is polarized in valley $\eta = +$, unless otherwise stated. More precisely, we assume that the order parameter satisfies the following condition:
\begin{equation}
    \langle c^\dagger_{\bm{\kappa} + \mathbf{Q}_bm\eta s}c_{\bm{\kappa}+\mathbf{Q}_{b'}m'\eta's'}\rangle = 0\,,~~{\rm if}~\eta=-{\rm ~or~}\eta'=-\,.
\end{equation}
For each enlarged unit cell choice and value of $w_0/w_1$, we choose 10 random initial conditions and perform self-consistent iterations to ensure the solutions are converging properly.

\subsubsection{Ground states}\label{sec:phases}

In Fig.~\ref{fig:mean_field_phase_diag} (a), we show the energy (compared to the solution without translation symmetry breaking) as a function of $w_0/w_1 \in [0.4, 1]$ for different choices of enlarged unit cells on a $12\times 12$ momentum lattice, which are represented by using different colors. We are able to identify three different regions, which are labeled by light blue, purple and red in Fig.~\ref{fig:mean_field_phase_diag} (a). When $w_0/w_1 \lesssim 0.5$ (represented by light blue), the ground state corresponds to the Chern insulator Slater determinant state, {\it i.e.}, with no translation symmetry breaking and Chern number $\nu_C = \pm 1$ (see Sec.~\ref{sec:c2t_topo_phase_diagram}), in agreement with Refs.~\cite{kang_nonabelian_2020, Potasz_exact_2021, ourpaper4, ourpaper6,soejima2020efficient,hejazi2020hybrid,zhang_HF_2020}. While it is not shown, this region actually extends to the chiral limit $w_0/w_1 = 0$. 
In the interval $0.5 \lesssim w_0/w_1\lesssim 0.65$ (represented by purple), the energies of translation symmetry breaking solutions with enlarged unit cells such as $(\sqrt3 \times \sqrt3)$ and $(2\times 1)$ become lower than the energy of the translation invariant solution. We also notice that the solutions with enlarged unit cell $(\sqrt{3}\times \sqrt{3})$ is usually energetically preferred: its energy is around $0.01\,\rm meV$ per moir\'e unit cell lower than the states with enlarged unit cell $(2\times1)$. Note that the Chern insulator solution without translation symmetry breaking still remains competitive in this intermediate region with an energy difference of only $0.05\,\rm meV$ per moir\'e unit cell. Therefore, competing states may coexist in the purple region of the phase diagram, and it is difficult to conclude what is the exact nature of this phase from Hartree-Fock, as already hinted by the exact diagonalization \cite{ourpaper6} and DMRG results~\cite{kang_nonabelian_2020}. 

If we further increase the value of $w_0/w_1$ to the interval $0.7\lesssim w_0/w_1 \lesssim 0.9$ (represented by red), the $(2\times 1)$ enlarged unit cell solution (or solution with $(1\times 2)$ which can be related by $C_{3z}$ rotation) clearly has the lowest ground state energy. The unit cell $(2\times 1)$ implies that it breaks the translation symmetry of the original moir\'e unit cell (see Sec.~\ref{sec:real_space}), and therefore we call the red region as $C_{2z}T$ \emph{stripe phase}, whose properties will be discussed in Sec.~\ref{sec:c2t_stripe}. Except for this $C_{2z}T$ stripe phase, another state with $(3\times 1)$ unit cell also has a lower energy than the state with $(\sqrt{3}\times\sqrt{3})$ enlarged unit cell. The energy difference between the state with $(3\times 1)$ enlarged unit cell and the $C_{2z}T$ stripe state is $\sim 0.08~\rm meV$ per moir\'e unit cell, which is clearly larger than the energy difference between the $(2\times 1)$ and $(\sqrt{3}\times \sqrt{3})$ enlarged unit cell states in the purple (intermediate) region. Therefore, the $C_{2z}T$ stripe phase in the red region is unambiguously preferred, as opposed to the situation in the intermediate (purple) region. When $w_0/w_1 \gtrsim 0.9$, the energies with different enlarged unit cells become comparable again, which leads to strong competition between the states with $(\sqrt3\times \sqrt3)$ unit cells and $(2\times1)$ unit cells. 

We also notice that the solutions using $(1\times 2)$, $(4\times 1)$ and $(1\times 4)$ unit cells always have the same ground state energy, implying that they are all equivalent solutions under certain $C_{3z}$ rotation or moir\'e unit cell translation. For the enlarged unit cell choice $(2\times 2)$, we obtained a solution whose energy per moir\'e unit cell is only $0.003 \rm\, meV\, (0.0013\%)$ lower than the $(2\times 1)$ solution at $w_0/w_1 = 0.6$, and the difference is barely visible in Fig.~\ref{fig:mean_field_phase_diag}(a). But for all the other values of $w_0/w_1$ that we have considered, the enlarged unit cell $(2\times 2)$ gives us the same solution as $(2\times 1)$, $(1\times 2)$, $(4\times 1)$ or $(1\times 4)$ unit cell choices.

Among the eight types of enlarged unit cells defined in Table \ref{tab:unit_cell_choices}, we found $(2\times 1)$ and $(\sqrt3\times \sqrt3)$ are energetically preferred in our phase diagram. Moreover, the state with $(3\times 1)$ enlarged unit cell is also a relevant candidate in the red region. 
For this reason, we solely focus on these three foldings to study the finite size effect, by solving the energies of self-consistent equations on a larger momentum lattice ($18\times 18$) in Fig.~\ref{fig:mean_field_phase_diag} (d). The phase diagram on the $18\times 18$ lattice is qualitatively similar to the results on the $12\times 12$ momentum lattice. The QAH state can still be observed in the light blue region ($w_0/w_1 \lesssim 0.5$), and multiple competing states in the purple region ($0.5 \lesssim w_0/w_1 \lesssim 0.65$). The $C_{2z}T$ stripe phase is still clearly preferred in the red region. The state with $(3\times 1)$ enlarged unit cell, although having a relatively low energy in the red region ($0.7\lesssim w_0/w_1 \lesssim 0.9$), is still around $\sim 0.1~\rm meV$ higher than the $C_{2z}T$ stripe phase. Thus, the $C_{2z}T$ stripe phase is indeed the best candidate ground state when $0.7 \lesssim w_0/w_1 \lesssim 0.9$.

In spite of the fact that the phase diagrams obtained on $12\times 12$ and $18\times 18$ lattices are qualitatively similar, the details of these phases are slightly different, especially in the purple region. For example, the state with $(\sqrt{3}\times \sqrt{3})$ enlarged unit cell has a lower energy than the translation invariant solution on the $18\times 18$ lattice, but no translation symmetry breaking is observed on the $12\times 12$ lattice.

\subsubsection{Translation symmetry breaking, charge gap and band structures}\label{sec:gap_band}

From now on, we will only consider the two favored foldings $(\sqrt3\times \sqrt3)$ and $(2\times 1)$. We calculate the values of translation symmetry breaking strength $\mathcal{T}$ using the solutions on $12\times 12$ and $18\times 18$ momentum lattices, which can be found in Figs. \ref{fig:mean_field_phase_diag} (b) and (e). In the intermediate regime and in the stripe phase (purple and red regions), the translation symmetry breaking $\mathcal{T}$ becomes non-zero and increases with increasing $w_0/w_1$. The values of the charge gap $E_G$ of the solutions on $12\times 12$ and $18\times 18$ momentum lattice can be found in Figs. \ref{fig:mean_field_phase_diag} (c) and (f). In the QAH phase (blue region), the charge gap descreases with the increasing $w_0/w_1$, while in the stripe phase (red region), the gap increases with increasing $w_0/w_1$. In the intermediate region (purple), these competing states all have small gaps.

In Fig.~\ref{fig:mean_field_bands_18x18}, we provide several Hartree-Fock bands in the folded Brillouin zones obtained from the simulation on $18\times 18$ momentum lattices to illustrate the typical HF band structure in the different regions of the phase diagram. The band structure of the quantum anomalous Hall state at $w_0/w_1 = 0.4$ is shown in Fig.~\ref{fig:mean_field_bands_18x18} (a), which agrees with the result obtained in Refs.~\cite{ourpaper5, kang_cascade_2021}. 
Indeed, the charge excitations shown in Fig.~11b of Ref.~\cite{ourpaper5} also has 3 particle bands. The QAH state does not break the translation symmetry, thus the HF bands are shown along the high symmetry lines in the moir\'e Brillouin zone. Figs. \ref{fig:mean_field_bands_18x18} (b) and (c) are the Hartree-Fock bands in the purple region both obtained at $w_0/w_1 = 0.6$ with enlarged unit cell choices $(\sqrt3 \times \sqrt3)$ and $(2\times 1)$, respectively. The corresponding high symmetry points are represented using red and blue greek letters, whose definitions can be found in Fig.~\ref{fig:folded_brillouin_zone}. We observe that these two competing states both have small gap, and they also have similar band widths. In Fig.~\ref{fig:mean_field_bands_18x18} (d), we show the HF band structure in the stripe phase (red region) at $w_0/w_1 = 0.8$ with folded moir\'e Brillouin zone of unit cell $(2\times 1)$. Clearly, the charge gap in the stripe phase is much larger than the intermediate competing region (purple). 

We also studied the spin texture of the occupied bands of the stripe phase -- and as discussed below, relaxed the assumption of valley polarization and exact flat bands ($t=0$) -- which shows that the stripe phase is \emph{fully spin and valley polarized}.
In other words, the order parameter satisfies the following conditions under a proper spin $SU(2)$ rotation:
\begin{align}
    &\langle c^\dagger_{\bm{\kappa} + \mathbf{Q}_b, m, \eta, s}c_{\bm{\kappa} + \mathbf{Q}_{b'}, n, \eta', s'} \rangle= 0\,,\nonumber\\
    &{\rm if}~(\eta,s)\neq (+,\uparrow){\rm ~or~}(\eta',s')\neq (+,\uparrow)\,.
\end{align}
We provide detailed numerical results of the spin distribution of several solutions in App.~\ref{sec:add_polarization}.

As mentioned previously, these calculations were performed assuming valley polarization and in the flat band limit. To test these hypotheses, we also performed Hartree-Fock calculation without these assumptions at the representative values of the phase diagram $w_0/w_1 = 0.4, 0.6$ and $0.8$, albeit on a smaller momentum lattice. We obtain identical phases at these $w_0/w_1$ values, ensuring that these assumptions are valid. A detailed study is also provided in App.~\ref{sec:add_polarization}.

\subsection{\texorpdfstring{$C_{2z}T$}{C2zT} symmetry and topology}\label{sec:c2t_topo_phase_diagram}

\begin{figure}[t]
    \centering
    \includegraphics[width=0.75\linewidth]{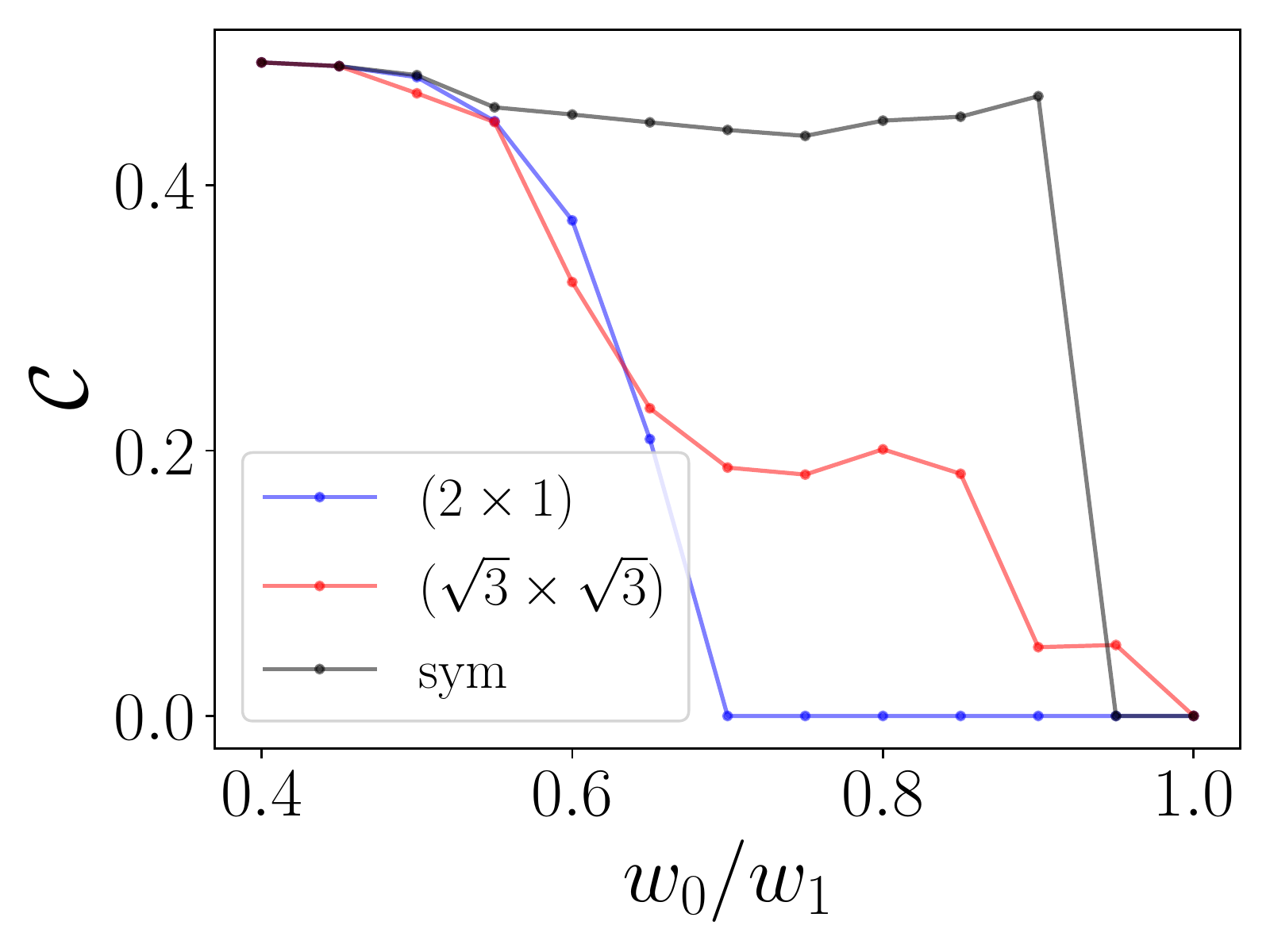}
    \caption{The strength of the $C_{2z}T$ symmetry breaking $\mathcal{C}$ of the two types of enlarged unit cells $(\sqrt3 \times \sqrt3)$ and $(2\times 1)$ as a function of $w_0/w_1$. We also show the value of $\mathcal{C}$ for translation symmetric solution in black. This figure is obtained on a $18\times 18$ momentum lattice at flat band limit.}
    \label{fig:c2t_breaking_phase_diagram}
\end{figure}

\begin{figure*}[t]
    \centering
    \includegraphics[width=\linewidth]{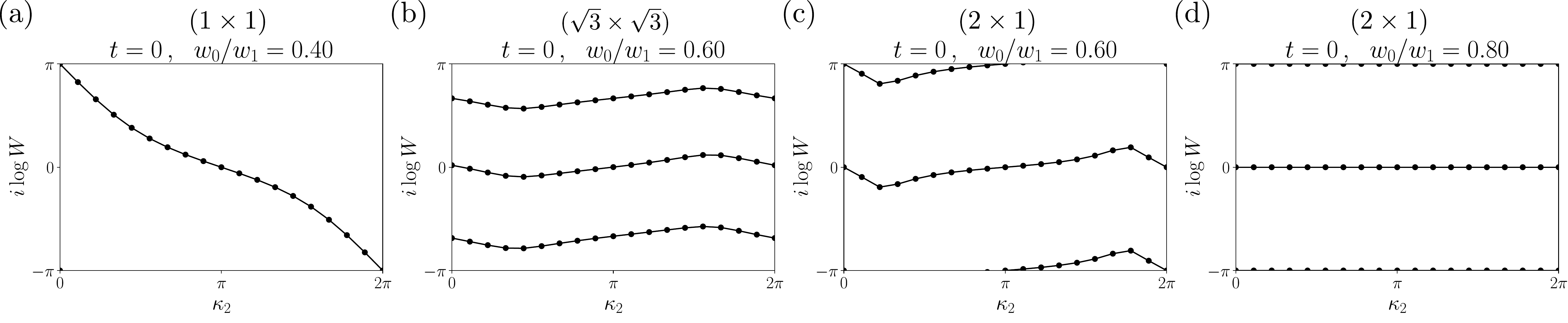}
    \caption{Wilson loop eigenvalue exponents of mean-field energy bands evaluated on $18\times 18$ momentum lattices. Top panel: (a) The Wilson loops of the lowest HF band at $w_0/w_1 = 0.4$. At this value of $w_0/w_1$, the translation symmetry is not broken, and the non-trivial winding number shown by the Wilson loop indicates that this state is a quantum anomalous Hall state. (b) The Wilson loop of the three lowest bands with enlarged unit cell $(\sqrt3\times\sqrt3)$ at $w_0/w_1 = 0.6$. (c) The Wilson loop of the two lowest bands with enlarged unit cell $(2\times 1)$ at $w_0/w_1 = 0.6$. (d) The Wilson loop eigenvalues of the two lowest bands with enlarged unit cell $(2\times 1)$ at $w_0/w_1 = 0.8$. The perfectly flat Wilson loop is an important property of the $C_{2z}T$ symmetry in the $C_{2z}T$ stripe phase.}
    \label{fig:wilson}
\end{figure*}

We also evaluated the value of $\mathcal{C}$ as a function of $w_0/w_1$ for the solutions obtained with enlarged unit cell choices $(2\times 1)$ and $\left(\sqrt3 \times \sqrt3\right)$ on a $18\times 18$ momentum lattice. The results can be found in Fig.~\ref{fig:c2t_breaking_phase_diagram}. In the light blue region with small $w_0/w_1 \lesssim 0.5$, the $C_{2z}T$ symmetry is strongly broken, which is an important property of Chern insulator states. When $w_0/w_1$ gets larger, the $C_{2z}T$ breaking of both $\left(\sqrt3 \times \sqrt3\right)$ and $(2\times 1)$ enlarged unit cell solutions become significantly smaller. More interestingly, for the solution with unit cell choice $(2\times 1)$, the $C_{2z}T$ breaking strength drops to zero in the stripe phase.

Restoration of $C_{2z}T$ symmetry implies that the Chern number must vanish in the stripe phase. In addition to checking the strength of $C_{2z}T$ symmetry breaking, we are also able to study the topological winding numbers directly from the mean field solutions. By using the Hartree-Fock eigenvectors $\phi_{bm\eta s, i}(\bm{\kappa})$ and single body wavefunctions of BM Hamiltonian $u_{\mathbf{Q}\alpha, m\eta}(\bm{\kappa} + \mathbf{Q}_b)$, we are able to rewrite the wavefunction of an eigenstate in Hartree-Fock band structure in the plane wave basis $\Phi_{\mathbf{Q}\alpha, b, \eta, s; i}(\vk)$. For enlarged unit cell choices $(2\times 1)$ and $(\sqrt3 \times \sqrt3)$, we use the following notation to parametrize the FMBZ: $\bm{\kappa} = \frac{\kappa_1}{2\pi}\mathbf{Q}_1 + \frac{\kappa_2}{2\pi}\tilde{\mathbf{b}}_2$. And we evaluate the Wilson loop along the direction of $\mathbf{Q}_1$ in the $N_F$ occupied HF bands, which we denote by $W(\kappa_2)$.
We also provide a detailed discussion of Wilson loops in App.~\ref{sec:wilson_app}.

The Wilson loop matrix $W(\kappa_2)$ is unitary and its eigenvalues are always given by $e^{-i\chi}, \chi \in [-\pi, \pi)$. We numerically calculate the Wilson loop eigenvalue exponents $\chi$ on $18\times 18$ momentum lattice. The Wilson loop eigenvalue exponents at $w_0/w_1 = 0.4, 0.6$ and $0.8$ can be found in Fig.~\ref{fig:wilson}. Fig.~\ref{fig:wilson} (a) shows the Wilson loop in the light blue region at $w_0/w_1 = 0.4$. The non-trivial winding number confirms that the light blue region is indeed a quantum anomalous Hall phase, which has already been widely studied previously \cite{zhang_HF_2020,soejima2020efficient,kang_nonabelian_2020,ourpaper6}. Figs. \ref{fig:wilson} (b) and (c) show the Wilson loops of the two low energy states at $w_0/w_1 = 0.6$ with enlarged unit cell choices $(\sqrt3\times \sqrt3)$ and $(2\times1)$, respectively. We found that the non-zero Chern number has already vanished in this competing region. Finally in Fig.~\ref{fig:wilson} (d), we present the Wilson loop for the $C_{2z}T$ stripe phase at $w_0/w_1 = 0.8$. The eigenvalues of Wilson loop spectrum in the $C_{2z}T$ stripe phase is completely flat, which is a consequence of the $C_{2z}T$ symmetry  \cite{ahn_failure_2019,song_all_2019,xie_superfluid_2020} and the translation symmetry breaking along $\tilde{\mathbf{a}}_1$, as discussed in App.~\ref{sec:app_wilson_loop}.

\begin{figure}[!htbp]
    \centering
    \includegraphics[width=\linewidth]{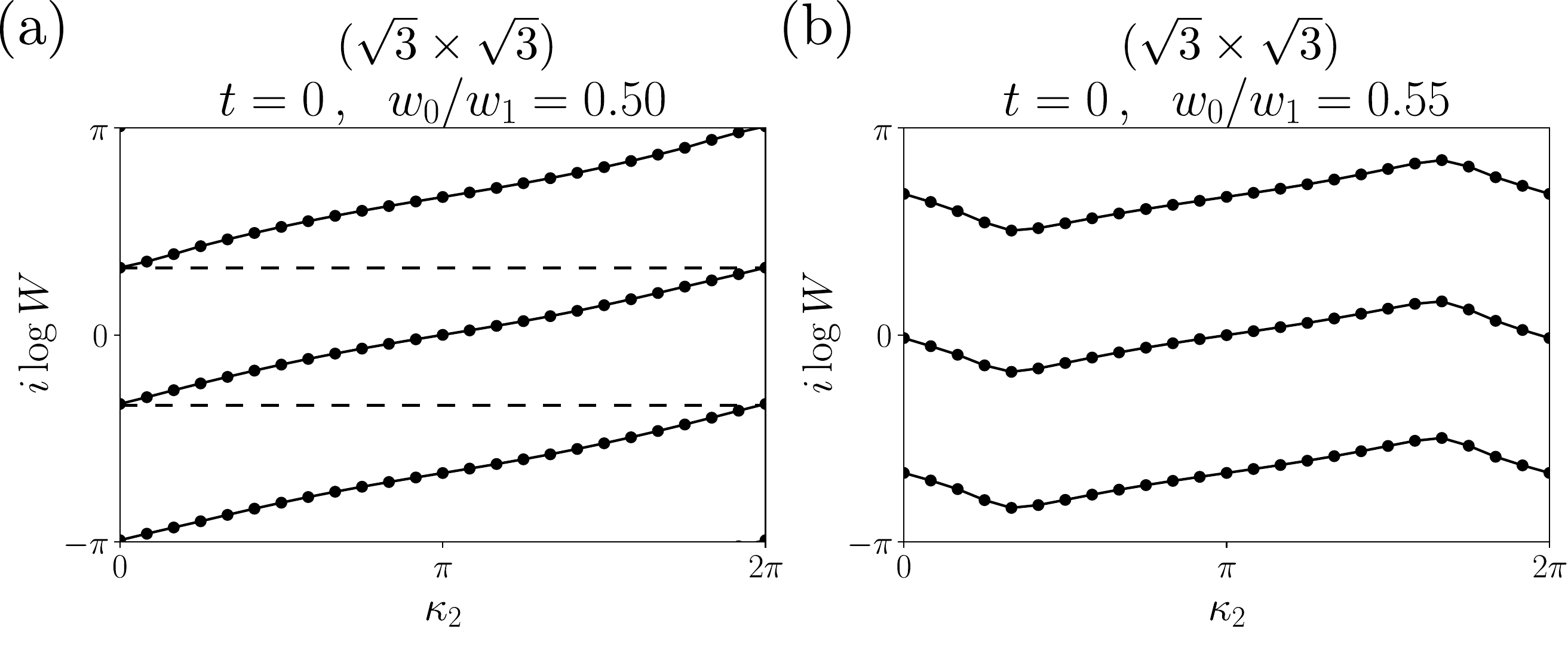}
    \caption{Wilson loop eigenvalue exponents of mean-field energy bands evaluated on $24\times 24$ momentum lattice with enlarged unit cell $(\sqrt3\times \sqrt3)$ at $w_0/w_1 = 0.5$ (a) and $w_0/w_1 = 0.55$ (b).}
    \label{fig:wilson_extra}
\end{figure}

As we mentioned in Sec.~\ref{sec:gs_bands}, the charge gap of the mean field solutions is small in the competing region between the QAH and $C_{2z}T$ stripe phases. Therefore the wavefunctions are varying fast around the $\gamma$ point in the FMBZ. Hence, we should use a denser momentum mesh for calculating the Wilson loops in the competing region. We evaluated the Wilson loops of the mean-field solutions with enlarged unit cell $(\sqrt3 \times \sqrt3)$ on $24\times 24$ momentum lattice at $w_0/w_1 = 0.5$ and $0.55$, which can be found in Fig.~\ref{fig:wilson_extra}. Both the solutions at these two values of $w_0/w_1$ have non-vanishing break the translation symmetry ($\mathcal{T} \neq 0$). We find that the Hartree-Fock bands still carry non-zero winding number at $w_0/w_1 = 0.5$, but the winding number vanishes at $w_0/w_1 = 0.55$. This observation implies that the disappearance of Chern number happens in the competing region of the phase diagram.

\section{\texorpdfstring{$C_{2z}T$}{C2T} stripe phase}\label{sec:c2t_stripe}
In this section, we discuss the $C_{2z}T$ symmetric stripe phase that we obtained for $w_0/w_1 \gtrsim 0.65$. As mentioned in Sec.~\ref{sec:gap_band} and discussed in App.~\ref{sec:add_polarization}, the $C_{2z}T$ stripe phase is spin and valley polarized regardless of whether the flat band kinetic energy is taken into account or neglected. Therefore, we are able to perform the mean field calculation on a even larger momentum lattice by assuming that the system is fully polarized in valley $\eta = +$ and spin $s = \uparrow$, and the following discussion is based on our numerical solution on a $36\times 36$ momentum lattice. We characterize this phase by studying its symmetries and real space charge distributions. Moreover, we propose a mechanism based on Dirac nodes motion to understand the development of charge gap in the $C_{2z}T$ stripe phase. 

\subsection{Symmetry}\label{sec:c2t_stripe_sym}
\begin{figure}[t]
    \centering
    \includegraphics[width=\linewidth]{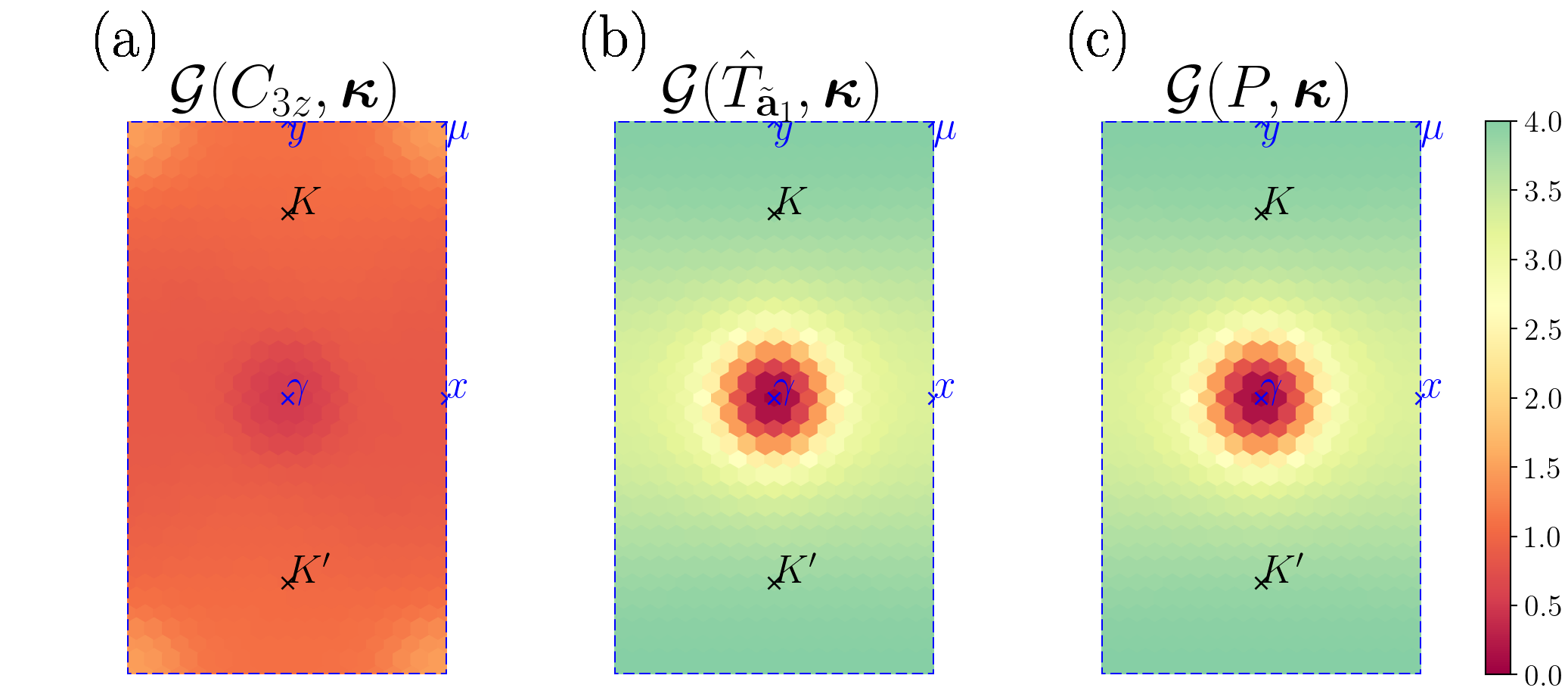}
    \caption{The symmetry breaking strength values $\mathcal{G}(C_{3z},\bm{\kappa})$, $\mathcal{G}(\hat{T}_{\tilde{\mathbf{a}}_1}, \bm{\kappa})$ and $\mathcal{G}(P, \bm{\kappa})$ calculated from the Hartree-Fock solution at $w_0/w_1 = 0.8$ at flat band limit on a $36\times 36$ lattice. We also numerically checked the values of $\mathcal{G}(C_{2z}T, \bm{\kappa})$, $\mathcal{G}(C_{2x}, \bm{\kappa})$ and $\mathcal{G}(\hat{T}_{\tilde{\mathbf{a}}_1}P, \bm{\kappa})$ are equal to zero up to machine precision $(<10^{-15})$ in the FMBZ. Although both $\hat{T}_{\tilde{\mathbf{a}}_1}$ and $P$ symmetries are broken, their product $\hat{T}_{\tilde{\mathbf{a}}_1}P$ is still conserved.}
    \label{fig:flatband_stripe_sym_break}
\end{figure}

\begin{center}
    \begin{table}[t]
        \resizebox{\linewidth}{!}{
        \begin{tabular}{c||c|c|c|c|c|c}
            \hline
            $g$ & $C_{2z}T$ & $C_{3z}$ & $C_{2x}$ & $\hat{T}_{\tilde{\mathbf{a}}_1}$ & $P$ & $\hat{T}_{\tilde{\mathbf{a}}_1} P$\\
            \hline\hline
            $[H_0, g]_\zeta$ & $+$ & $+$ & $+$ & $+$ & $-$ & $-$ \\
            $[H_I, g]_\zeta$ & $+$ & $+$ & $+$ & $+$ & $+$ & $+$\\
            coordinate $\mathbf{r}$ & $-\mathbf{r}$ & $C_{3z}\mathbf{r}$ & $C_{2x}\mathbf{r}$ & $\mathbf{r} + \tilde{\mathbf{a}}_1$ & $-\mathbf{r}$ & $-\mathbf{r} - \tilde{\mathbf{a}}_1$\\
            momentum $\vk$ & $\vk$ & $C_{3z}\vk$ & $C_{2x}\vk$ & $\vk$ & $-\vk$ & $-\vk$ \\
            sublattice $\alpha$ & $-\alpha$ & $\alpha$ & $-\alpha$ & $\alpha$ & $\alpha$ & $\alpha$ \\
            layer $\ell$ & $\ell$ & $\ell$ & $-\ell$ & $\ell$ & $-\ell$ & $-\ell$\\
            \hline\hline
            \stackanchor{stripe}{$\frac{w_0}{w_1}=0.8,~t=0$} & \cmark & \xmark & \cmark & \xmark & \xmark & \cmark \\
            \hline
            \stackanchor{stripe}{$\frac{w_0}{w_1}=0.8,~t=1$} & \cmark & \xmark & \cmark & \xmark & \xmark & \xmark\\
            \hline
            \stackanchor{QAH}{$\frac{w_0}{w_1}=0.4,~t=0$} & \xmark & \cmark & \xmark & \cmark & \cmark & \cmark\\
            \hline
            \stackanchor{QAH}{$\frac{w_0}{w_1}=0.4,~t=1$} & \xmark & \cmark & \xmark & \cmark & \xmark & \xmark\\
            \hline\hline
        \end{tabular}}
        \caption{Six types of lattice symmetries of the projected interacting Hamiltonian of TBG: $C_{2z}T$, $C_{2x}$, $C_{3z}$, $P$ and $\hat{T}_{\tilde{\mathbf{a}}_1}$ and $\hat{T}_{\tilde{\mathbf{a}}_1}P$ for the spin and valley and polarized mean-field solutions. The first and second rows indicate whether a given symmetry $g$ is commuting $(+, [H, g] = 0)$ or anti-commuting $(-, \{H, g\}=0)$ with the kinetic and interacting Hamiltonian. The third to sixth rows show how the real space coordinate $\mathbf{r}$, momentum $\vk$, sublattice $\alpha$ and graphene layer $\ell$ change under the given symmetries. The seventh to tenth rows show whether this symmetry is conserved in the mean field solutions for $C_{2z}T$ stripe and QAH phases without and with kinetic energy, respectively.}
        \label{tab:real_space_sym}
    \end{table}
\end{center}

First, we analyze the real space lattice symmetries of the self-consistent Hartree-Fock solution. Since the $C_{2z}T$ stripe phase at around $w_0/w_1 = 0.8$ is spin and valley polarized as observed in the numerical simulation, we only focus on the lattice symmetries for the single valley Hamiltonian: $C_{2z}T$, $C_{3z}$, $C_{2x}$ and $P$ (particle-hole symmetry). Notice that $C_{2z}T$, $C_{3z}$ and $C_{2x}$ commute with both the kinetic Hamiltonian $H_0$ and the interacting part of the Hamiltonian $H_I$, while the particle-hole symmetry $P$ only commutes with $H_I$ but anti-commutes with $H_0$. In addition to these symmetries, the Hamiltonian also has moir\'e lattice translation symmetry $\hat{T}_{\tilde{\mathbf{a}}_1}$. However, since we fold the moir\'e Brillouin zone along $\tilde{\mathbf{b}}_1$, the moir\'e unit cell will be enlarged along $\tilde{\mathbf{a}}_1$ direction, and it could lead to the spontaneous breaking of $\hat{T}_{\tilde{\mathbf{a}}_1}$. In Table \ref{tab:real_space_sym}, we summarize the commutation properties of these symmetries, and their actions in real space, momentum space, sublattice and layer indices. 

In order to measure the symmetry breaking of a given symmetry $g$, we define the following quantity for a momentum point $\bm{\kappa} \in {\rm FMBZ}$:
\begin{align}
    \mathcal{G}(g, \bm{\kappa}) =& \sum_{bm,b'm'}\Big{|}\langle c^\dagger_{\bm{\kappa}+\mathbf{Q}_b, m,+,\uparrow} c_{\bm{\kappa}+\mathbf{Q}_{b}',n,+,\uparrow} \rangle \nonumber\\ 
    &- \langle g c^\dagger_{\bm{\kappa}+\mathbf{Q}_b, m,+,\uparrow}g^{-1}g c_{\bm{\kappa}+\mathbf{Q}_{b}',n,+,\uparrow}g^{-1} \rangle\Big{|}^2\,,\label{eqn:def_sym_break}
\end{align}
which actually measures how much the order parameter $\Delta(\bm{\kappa})$ changes through certain transformation $g$. We provide a detailed discussion about the transformations of the electron operators in App.~\ref{sec:app_ham_sym}.
Note that the translation symmetry breaking strength defined in Eq.~(\ref{eqn:def_trans_breaking}) can also be written as:
\begin{equation}
    \mathcal{T} = \frac{1}{4N_M}\sum_{\bm{\kappa}\in{\rm FMBZ}}\mathcal{G}(\hat{T}_{\tilde{\mathbf{a}}_1}, \bm{\kappa})\,.
\end{equation}
Thus, $\mathcal{G}(\hat{T}_{\tilde{\mathbf{a}}_1}, \bm{\kappa})$ gives a more detailed description of the translation symmetry breaking than $\mathcal{T}$.

We numerically calculated the symmetry breaking strength $\mathcal{G}(g, \bm{\kappa})$ of the five symmetries mentioned above, {\it i.e.,} $C_{2z}T$, $C_{3z}$, $C_{2x}$, $\hat{T}_{\tilde{\mathbf{a}}_1}$ and $P$, and another combined symmetry $\hat{T}_{\tilde{\mathbf{a}}_1} P$ on a $36\times 36$ lattice at the flat band limit with $w_0/w_1 = 0.8$. 
In Fig.~\ref{fig:flatband_stripe_sym_break}, we provide the values of $\mathcal{G}(C_{3z}, \bm{\kappa})$, $\mathcal{G}(\hat{T}_{\tilde{\mathbf{a}}_1}, \bm{\kappa})$ and $\mathcal{G}(P, \bm{\kappa})$ in the FMBZ. The peak of translation breaking is around $\mu$ point in its FMBZ, showing a strong hybridization between the two $M$ points in the MBZ.
However, the values of $\mathcal{G}(C_{2z}T, \bm{\kappa})$, $\mathcal{G}(C_{2x}, \bm{\kappa})$ and $\mathcal{G}(\hat{T}_{\tilde{\mathbf{a}}_1}P, \bm{\kappa})$ are equal to zero for any $\bm{\kappa} \in {\rm FMBZ}$ up to machine precision ($<10^{-15}$). Thus, the stripe phase at flat band limit does not break $C_{2z}T$, $C_{2x}$ and $\hat{T}_{\tilde{\mathbf{a}}_1}P$ symmetries, although both $\hat{T}_{\tilde{\mathbf{a}}_1}$ and $P$ symmetries are broken. The list of the conserved symmetries of the stripe phase with $t = 0$ can be found in the 8th line of Table \ref{tab:real_space_sym}. As a reference, we also provide the list of conserved symmetries of the stripe phase with kinetic energy ($t=1$), the QAH phase with and without kinetic energy ($t=0$ and $t=1$) in the 9th to 11th lines of Table \ref{tab:real_space_sym}. App.~\ref{sec:add_sym_real} provides a detailed discussion of these solutions.

\subsection{Real space charge distribution}\label{sec:real_space}
\begin{figure}[t]
    \centering
    \includegraphics[width=0.8\linewidth]{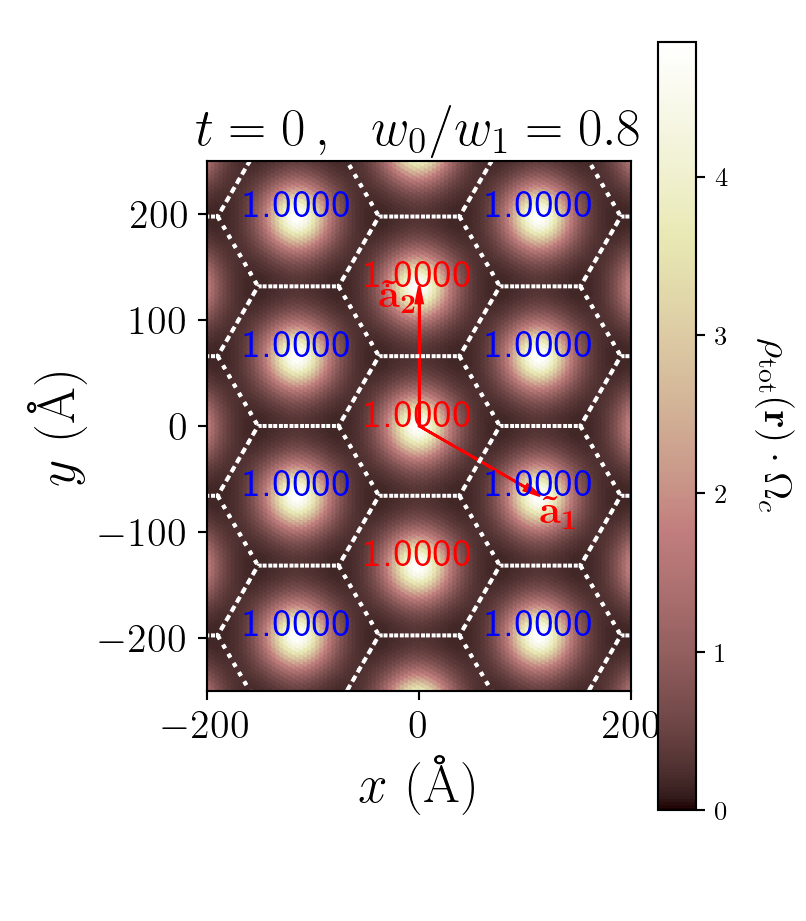}
    \caption{Total charge distribution in real space at flat band limit and $w_0/w_1 = 0.8$. The numbers are the total charge in the corresponding unit cell $Q$ defined in Eq.~(\ref{eqn:def_tot_charge}).}
    \label{fig:flatband_stripe_real_total}
\end{figure}

\begin{figure*}[!htbp]
    \centering
    \includegraphics[width=\linewidth]{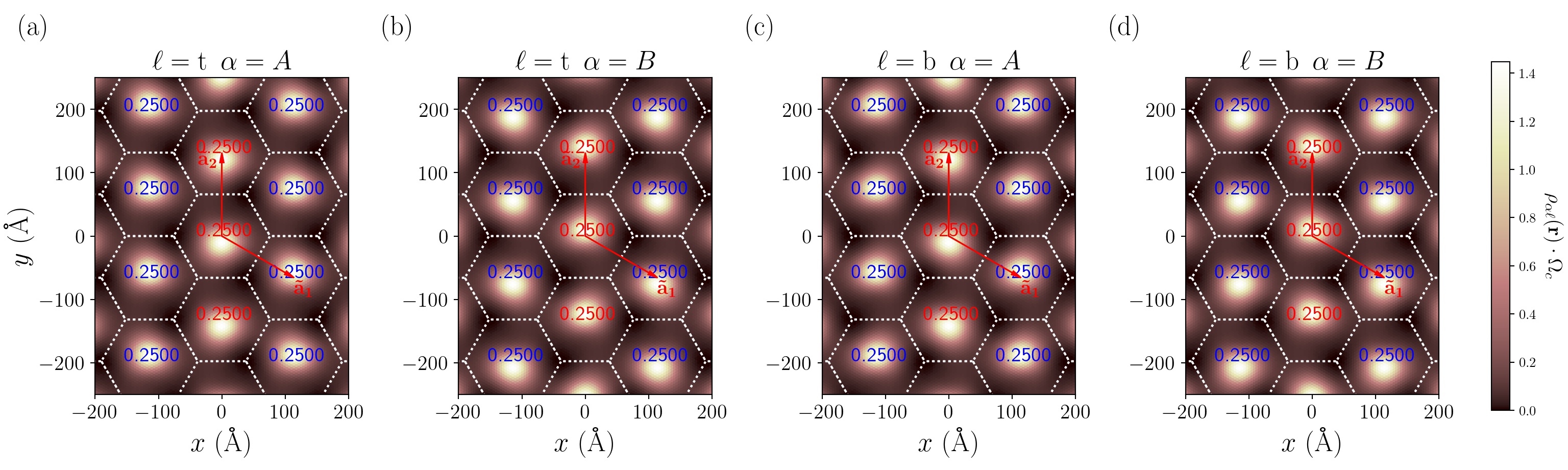}
    \caption{The electron density distribution in real space at flat band limit and $w_0/w_1 = 0.8$, obtained on a $36\times 36$ momentum lattice. $\ell = {\rm t}$ represents the top graphene layer and $\ell = {\rm b}$ represents the bottom layer. The two red arrows represent the Bravais lattice basis of the moir\'e lattice, and white dashed lines depict moir\'e unit cells centered around $AA$ stacking regions, and the red/blue numbers represent the integral of the corresponding component of density in the unit cell $Q_{\alpha\ell}$ (see Eq.~(\ref{eqn:def_charge_sublattice_layer})) in each moir\'e unit cell. The blue and red numbers are only differed by $10^{-9}$ numerically.}
    \label{fig:flatband_stripe_real_sublattice_layer}
\end{figure*}

\begin{figure}[t]
    \centering
    \includegraphics[width=\linewidth]{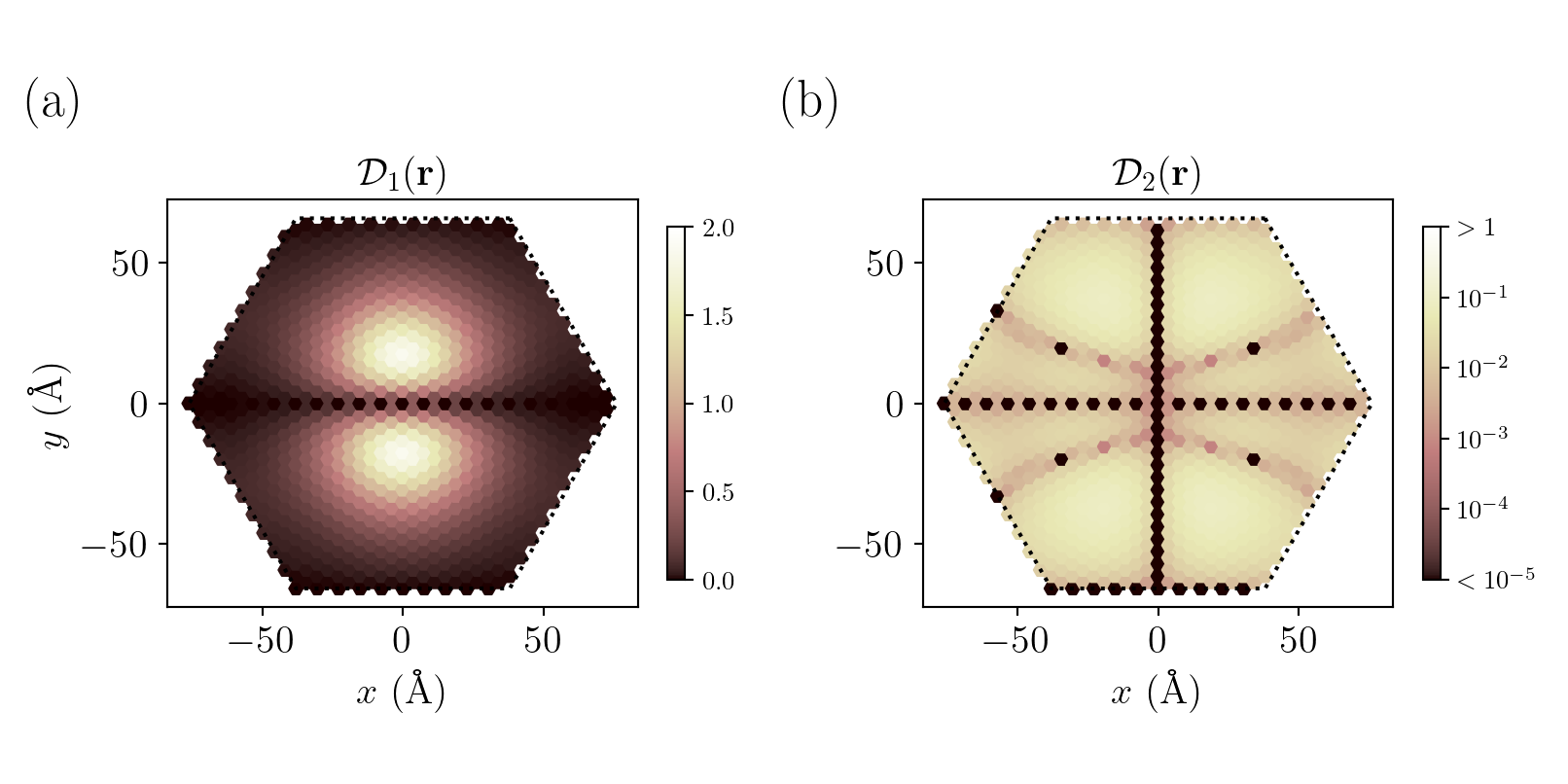}
    \caption{The translation symmetry breaking of the density distribution at flat band limit and $w_0/w_1 = 0.8$. (a) The value of $\mathcal{D}_1(\mathbf{r})$, which equals zero only when all the four sublattice and layer components are invariant under the translation $\hat{T}_{\tilde{\mathbf{a}}_1}$. (b) The value of $\mathcal{D}_2(\mathbf{r})$, which measures the electron charge density change in the {\it top layer} when shifted by $\mathbf{r} \rightarrow \mathbf{r} + \tilde{\mathbf{a}}_1$. The average value of $\mathcal{D}_2(\mathbf{r})$ in a moir\'e unit cell is around $0.026$.}
    \label{fig:flatband_stripe_real_shift}
\end{figure}

We now turn to study the real space distribution of the electron density from the mean field order parameter. The electron operators in real space can be written as:
{\small
    \begin{equation}
    c^\dagger_{\alpha\ell s}(\mathbf{r}) = \frac{1}{\sqrt{\Omega_{\rm tot}}}\sum_{\substack{\vk\in{\rm MBZ}\\\eta, \mathbf{Q}\in\mathcal{Q}_{\eta\ell}\\m}}c^\dagger_{\vk, m, \eta, s} u_{\mathbf{Q}\alpha, m\eta}(\vk)e^{-i(\vk - \mathbf{Q} + \eta \mathbf{K})\cdot \mathbf{r}}\,,
\end{equation}}in which the vector $\mathbf{K}$ is the momentum of $K$ point in the Brillouin zone of single layer graphene. Thus, the real space electron density distribution of a spin and valley polarized state ($\eta = +$, $s = \uparrow$) is given by the following equation:
{\small
\begin{align}
    \rho_{\alpha\ell}(\mathbf{r}) =& \langle c^\dagger_{\alpha\ell \uparrow}(\mathbf{r})c_{\alpha\ell \uparrow}(\mathbf{r})\rangle \nonumber \\
    =& \frac{1}{\Omega_{\rm tot}}\sum_{\substack{\bm{\kappa}\in{\rm FMBZ}\\bb'mm'}}\sum_{\mathbf{Q},\mathbf{Q}'\in\mathcal{Q}_{\ell}}\langle c^\dagger_{\bm{\kappa}+\mathbf{Q}_b, m, +, \uparrow}c_{\bm{\kappa}+\mathbf{Q}_{b'}, m', +, \uparrow} \rangle \nonumber\\
    & \times u^*_{\mathbf{Q}\alpha, m +}(\bm{\kappa} + \mathbf{Q}_b)u_{\mathbf{Q}'\alpha, m' +}(\bm{\kappa} + \mathbf{Q}_{b'})\nonumber\\
    &\times e^{-i\left[(\mathbf{Q}_{b}-\mathbf{Q}_{b'}) - (\mathbf{Q} - \mathbf{Q}')\right]\cdot \mathbf{r}}\,,\label{eqn:density_fourier}
\end{align}
}where the summation over $\bm{\kappa}$ is in the folded moir\'e Brillouin zone. Since the solutions are spin and valley polarized at filling factor $\nu=-3$, we drop the spin indices $s$ for convenience in the following discussion. 

By using the order parameter $\Delta(\bm{\kappa})$ solved at $w_0/w_1 = 0.8$ with flat bands $(t = 0)$ on the $36\times 36$ lattice, we are able to calculate the electron density in real space. Fig.~\ref{fig:flatband_stripe_real_total} provides the {\it total density} in real space over several moir\'e unit cells, which is defined as:
\begin{equation}
    \rho_{\rm tot}(\mathbf{r}) = \sum_{\alpha\ell}\rho_{\alpha\ell}(\mathbf{r})\,.
\end{equation}
The moir\'e unit cells are chosen to be the hexagon region around $AA$ stacking sites, represented by white dashed lines. We can also define the total charge in each unit cell as follows:
\begin{equation}\label{eqn:def_tot_charge}
    Q = \int_c d^2\mathbf{r}\,\rho_{\rm tot}(\mathbf{r})\,,
\end{equation}
and the values of $Q$ in each unit cell is labeled by blue and red numbers in Fig.~\ref{fig:flatband_stripe_real_total}. In Sec.~\ref{sec:c2t_stripe_sym}, we have shown that the order parameter $\Delta(\bm{\kappa})$ has strong translation symmetry breaking along $\tilde{\mathbf{a}}_1$ direction. However, the total electric charge in every moir\'e unit cell $Q$ has the same value $Q = 1$. We also find that the total charge density satisfies $\rho_{\rm tot}(\mathbf{r} + \tilde{\mathbf{a}}_1) = \rho_{\rm tot}(\mathbf{r})$ (up to numerical accuracy). There are still one electron per moir\'e unit cell, thus this state does not modulate the total charge on $AA$ stacking regions \cite{kang_strong_2019, kang_nonabelian_2020}.
From Table \ref{tab:real_space_sym}, we know this translation symmetry breaking solution has $C_{2z}T$, $C_{2x}$, and $\hat{T}_{\tilde{\mathbf{a}}_1} P$ symmetries. Consequently, the wavefunction of the $C_{2z}T$ stripe phase is invariant under the product of $C_{2z}T$ and $\hat{T}_{\tilde{\mathbf{a}}_1}P$. This combined symmetry $C_{2z}T\hat{T}_{\tilde{\mathbf{a}}_1}P$ transforms the real space coordinate as $\mathbf{r} \rightarrow \mathbf{r} + \tilde{\mathbf{a}}_1$, and flips both the graphene layer index $\ell$ and the sublattice index $\alpha$. Thus, the symmetry $C_{2z}T\hat{T}_{\tilde{\mathbf{a}}_1} P$ ensures that the charge density $\rho_{\alpha\ell}(\mathbf{r})$ is invariant under coordinate translation $\mathbf{r} \rightarrow \mathbf{r} + \tilde{\mathbf{a}}_1$ when both $\alpha$ and $\ell$ are flipped, letting the total charge density unchanged under the translation along $\tilde{\mathbf{a}}_1$. 

We also study the charge density components for each sublattice and layer index. We provide the values of $\rho_{\alpha\ell}(\mathbf{r})$ in Fig.~\ref{fig:flatband_stripe_real_sublattice_layer}.
The red/blue numbers represent the charge $Q_{\alpha\ell}$ in the two types of nonequivalent moir\'e unit cells in the enlarged unit cell:
\begin{equation}\label{eqn:def_charge_sublattice_layer}
    Q_{\alpha\ell} = \int_c d^2\mathbf{r} \rho_{\alpha\ell}(\mathbf{r})\,.
\end{equation}
We notice that $Q_{\alpha\ell}$ for a given sublattice $\alpha$ and layer $\ell$ in the unit cell around $\mathbf{r} = 0$ (red) and $\mathbf{r} = \tilde{\mathbf{a}}_1$ (blue) are the same (differ by $10^{-9}$ numerically). However, the charge distributions differ. For example, in the top layer with $\alpha = A$, the charge center in the unit cell around $\mathbf{r} = 0$ is in the lower half of the unit cell, while in the unit cell around $\mathbf{r} = \tilde{\mathbf{a}}_1$, the charge center is in the upper half of the unit cell. 
Moreover, we also numerically confirmed that the charge distribution of layer $\ell = {\rm t}$, sublattice $\alpha = A$ and layer $\ell = {\rm b}$, sublattice $\alpha = B$ are identical with a real space translation $\mathbf{r}\rightarrow \mathbf{r} + \tilde{\mathbf{a}}_1$, as we concluded from $C_{2z}T\hat{T}_{\tilde{\mathbf{a}}_1}P$ symmetry in the last paragraph.

To quantify the charge modulation between two moir\'e unit cells, we first define the following dimensionless quantity:
\begin{equation}\label{eqn:def_D1}
    \mathcal{D}_1(\mathbf{r}) = \Omega_c \sqrt{\sum_{\alpha\ell}|\rho_{\alpha\ell}(\mathbf{r}) - \rho_{\alpha\ell}(\mathbf{r} + \tilde{\mathbf{a}}_1)|^2}\,,
\end{equation}
in which $\Omega_c$ is the volume of one moir\'e unit cell. This quantity equals zero only when all of the four components of $\rho_{\alpha\ell}(\mathbf{r})$ are not changed under translation $\mathcal{\mathbf{r}}\rightarrow\mathbf{r} + \tilde{\mathbf{a}}_1$. It also has the same periodicity as the original moir\'e superlattice by definition, therefore we only have to calculate the values within a single moir\'e unit cell. In Fig.~\ref{fig:flatband_stripe_real_shift} (a), we provide the values of $\mathcal{D}_1(\mathbf{r})$ in a moir\'e unit cell. As can be observed, the charge density components per sublattice and layer are not invariant under the translation. 

Similarly, we can also define the following quantity to quantify the charge density modulation in a single layer (for example, the top layer) under the translation $\mathcal{\mathbf{r}}\rightarrow\mathbf{r} + \tilde{\mathbf{a}}_1$:
\begin{equation}\label{eqn:def_D2}
    \mathcal{D}_2(\mathbf{r}) = \Omega_c \Big{|}\sum_{\alpha}\rho_{\alpha\ell={\rm t}}(\mathbf{r}) - \sum_\alpha \rho_{\alpha\ell={\rm t}}(\mathbf{r} + \tilde{\mathbf{a}}_1)\Big{|}\,.
\end{equation}
$\mathcal{D}_2(\mathbf{r}) = 0$ only when the top layer charge density distributions are the same in two moir\'e unit cells. A plot of $\mathcal{D}_2(\mathbf{r})$ is provided in Fig.~\ref{fig:flatband_stripe_real_shift} (d). It shows that charge distribution for a single layer is not invariant under $\mathbf{r} \rightarrow \mathbf{r} + \tilde{\mathbf{a}}_1$. Therefore, it is still possible to observe a charge density wave by experiments such as scanning tunneling microscope, which mostly detects signals from a single layer, although the total charge density does not have any modulation in $AA$ stacking regions.

We also solved the real space charge distribution of the $C_{2z}T$ stripe phase at $t = 1$, {\it i.e.}, with the kinetic term. As shown in Table \ref{tab:real_space_sym}, this term anti-commutes with $P$, thus the solution does not have the $\hat{T}_{\tilde{\mathbf{a}}_1}P$ symmetry. As a consequence, the total charge no longer has the same periodicity as the moir\'e lattice. However, since the $\hat{T}_{\tilde{\mathbf{a}}_1}P$ symmetry is only weakly broken, the modulation of total charge between different unit cells is less than $0.2\%$. A detailed study of this solution is provided in App.~\ref{sec:add_sym_real}.

\subsection{The motion of Dirac nodes}\label{sec:dirac}
\begin{figure*}[t]
    \centering
    \includegraphics[width=\linewidth]{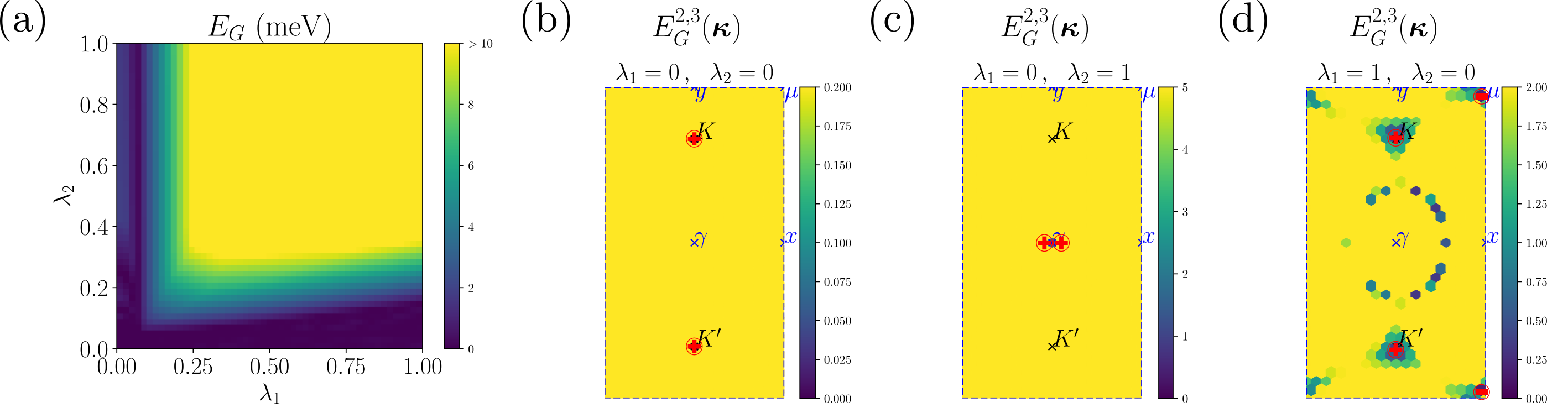}
    \caption{(a) The minimum direct charge gap between the second and the third bands of the hamiltonian $\mathscr{H}(\bm{\kappa}, \lambda_1,\lambda_2)$ as a function of $\lambda_1$ and $\lambda_2$. (b) The direct gap between the second and the third bands in the FMBZ when $\lambda_1 = \lambda_2 = 0$. The direct gaps between the second and the third bands in the FMBZ at $\lambda_1 = 0, \lambda_2 = 1$ and at $\lambda_1 = 1, \lambda_2 = 0$ are also shown in subfigures (c) and (d). The red symbols $\oplus$ and $\ominus$ represent the Dirac nodes and their chiralities in subfigures (b-d). Due to the finite $36\times 36$ mesh in the Brillouin zone, the minimum direct gap is in general not strictly equal to zero (at machine precision). Nevertheless we have checked at several places of the phase diagram by refining the mesh with the non-self-consistent-field method discussed in App.~\ref{sec:hf_app_high_sym_lines} near the nodes that we indeed have direct gap closing, for example, two Dirac nodes can be observed near the $\gamma$ point at $(\lambda_1, \lambda_2) = (0, 1)$. In subfigures (b-d), the values of the direct gap are represented in meV in the colorbars.}
    \label{fig:lam_phase_dirac_position}
\end{figure*}

For any two-band system, the $C_{2z}T$ symmetry can be represented by complex conjugation $\mathcal{K}$ under proper basis choice. Therefore, a $C_{2z}T$ symmetric Hamiltonian will not contain any $\sigma_y$ terms, and a {\it single} Dirac node cannot be gapped locally by any perturbation which respects the $C_{2z}T$ symmetry. Instead, such perturbation can only change the position of the Dirac node in momentum space. The non-interacting TBG flat bands have two Dirac nodes protected by $C_{2z}T$ symmetry with the same chirality, while the $C_{2z}T$ stripe phase does not have any Dirac nodes. However, Dirac nodes can annihilate only when two nodes carry opposite chirality. The gap opening of the $C_{2z}T$ stripe phase is seemingly at odds with the Dirac nodes' chirality of the non-interacting TBG bands.

In this section, we study this process and focus on the $C_{2z}T$ stripe solution with flat band kinetic energy ($t = 1$) at $w_0/w_1 = 0.8$ on a $36\times 36$ momentum lattice. To analyze the gap opening within the $C_{2z}T$ stripe phase, we first introduce the interpolation Hamiltonian with parameters $\lambda_{1}$ and $\lambda_2$:
\begin{align}
    &\mathscr{H}_{bm\eta s;b'n\eta's'}(\bm{k}, \lambda_1, \lambda_2)\nonumber\\ 
    =& \epsilon_{\bm{\vk} + \mathbf{Q}_b,m,\eta}\delta_{bb'}\delta_{mn}\delta_{\eta\eta'}\delta_{ss'}\nonumber\\
    & + \lambda_1 \delta_{bb'}\left(\mathcal{H}^{(H)}(\bm{\kappa}) + \mathcal{H}^{(F)}(\bm{\kappa})\right)_{bm\eta s,b'n\eta's'}\nonumber\\
    & + \lambda_2\left(1 - \delta_{bb'}\right)\left(\mathcal{H}^{(H)}(\bm{\kappa}) + \mathcal{H}^{(F)}(\bm{\kappa})\right)_{bm\eta s,b'n\eta's'}\,,\label{eqn:interpolation_stripe_ham}
\end{align}
in which $\lambda_{1}$ stands for the interpolation coefficients for the translation symmetry preserving part of the self-consistent HF Hamiltonian, and $\lambda_2$ the translation symmetry breaking part of the HF Hamiltonian. Thus, the Hamiltonian at $\lambda_1 = \lambda_2 = 0$ gives us the band structure of the non-interacting bands, while $\lambda_1 = \lambda_2 = 1$ gives us the HF bands of the $C_{2z}T$ stripe phase. In Fig.~\ref{fig:lam_phase_dirac_position}(a), we show the value of the band gap between the second and the third bands of the Hamiltonian $\mathscr{H}(\bm{\kappa}, \lambda_1,\lambda_2)$. Clearly, the gap opens when both the $\lambda_1$ and $\lambda_2$ exceed a critical value. However, different path choices in the $(\lambda_1, \lambda_2)$ space can correspond to different mechanisms of gap opening.  In the following paragraphs, we illustrate how the gapless non-interacting TBG bands become the $C_{2z}T$ stripe phase with a large charge gap along three different paths in this $(\lambda_1, \lambda_2)$ parameter space: one path with non-abelian braiding, one path with annihilation of Dirac nodes from the strong interacting bands, and one path with Dirac nodes annihilation when crossing the Brillouin zone border due to the $\pi$ Berry phase as discussed in Sec.~\ref{sec:c2t_topo_phase_diagram} \cite{ahn_failure_2019}.

\begin{figure}[t]
    \centering
    \includegraphics[width=\linewidth]{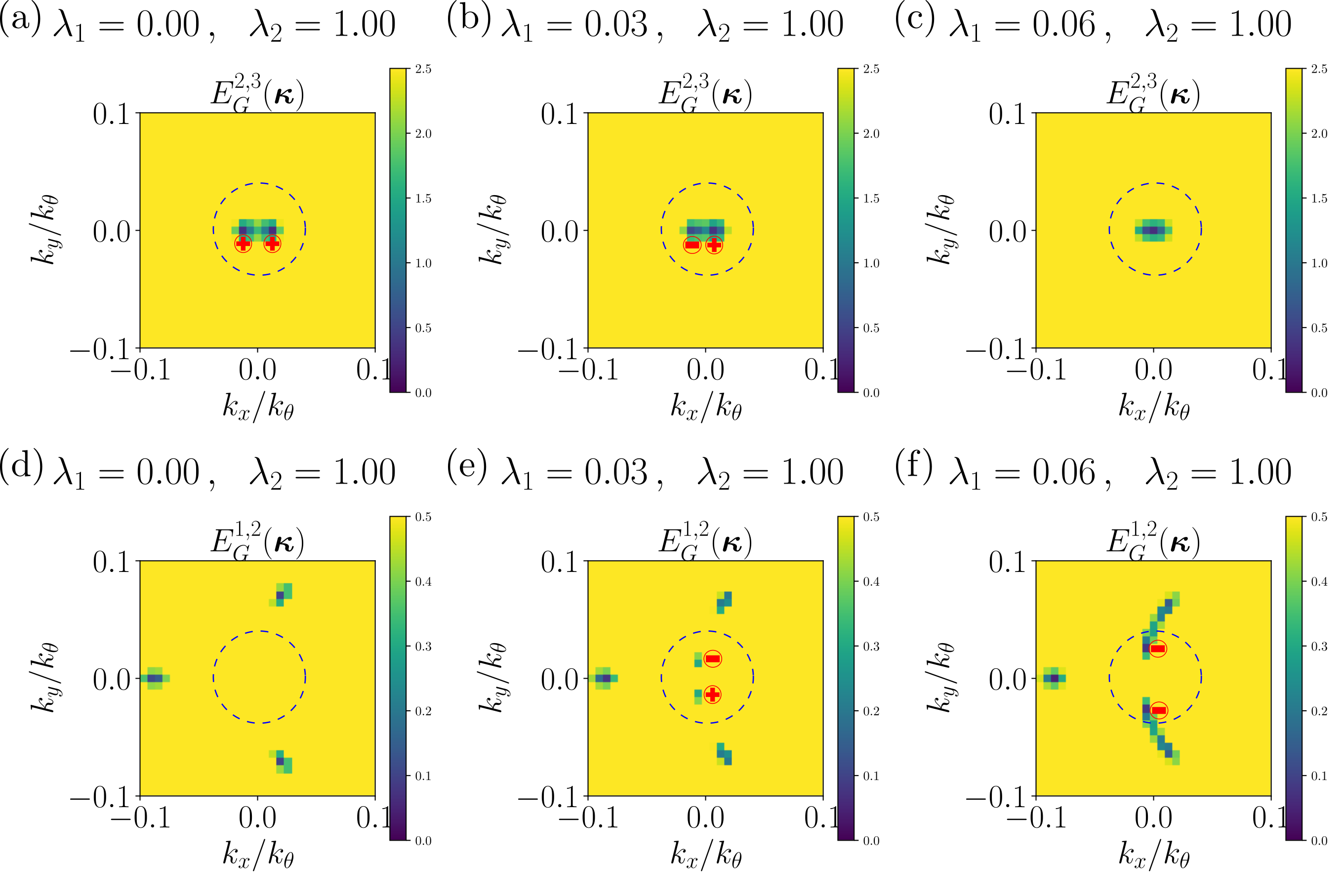}
    \caption{(a-c) The direct gap between the second and the third bands of $\mathscr{H}(\bm{\kappa}, \lambda_1, \lambda_2)$ in the FMBZ patch around $\gamma$ point (see App.~\ref{sec:app_dirac_motion}) with $\lambda_1 = 0, 0.03, 0.06$ and $\lambda_2 = 1$. (d-f) The direct gap between the first and the second bands in same the FMBZ patch. The chiralities of the Dirac nodes in the dashed circles are represented by red symbols $\oplus$ and $\ominus$. The values of the direct gap are represented in meV in the colorbars.}
    \label{fig:non_abelian_nodes}
\end{figure}

\subsubsection{Non-Abelian Dirac node braiding}

The first path we study is along the following direction: $(\lambda_1, \lambda_2) = (0, 0) \rightarrow (0, 1) \rightarrow (1, 1)$. The direct gaps between the second and the third bands in the FMBZ at $(\lambda_1, \lambda_2) = (0,0)$ and $(0, 1)$ are shown in Figs.~\ref{fig:lam_phase_dirac_position}(b) and (c). The Dirac nodes are labeled by red $\oplus$ and $\ominus$ symbols in these figures. Along this first segment of the path, these two Dirac nodes labeled on the figure move to a region around the $\gamma$ point [see Figs.~\ref{fig:lam_phase_dirac_position}(b) and (c)]. 
By using the non-self-consistent-field method discussed in App.~\ref{sec:hf_app_high_sym_lines}, we can solve the band structures of $\mathscr{H}(\bm{\kappa}, \lambda_1, \lambda_2)$ in a small patch around the $\gamma$ point with a $45$ times higher resolution than the original $36\times 36$ lattice, without solving the self-consistent solution on such a dense momentum lattice.
We are also able to evaluate the chirality of Dirac nodes by the method discussed in App.~\ref{sec:chirality_app}, and we provide a detailed numerical study about the chirality of Dirac nodes in App.~\ref{sec:add_lam2_1_results}. 
We now focus on the second segment of the path. In Fig.~\ref{fig:non_abelian_nodes}, we show the position and the chirality of the Dirac nodes between the first and the second bands, and between the second and the third bands at $\lambda_1 = 0, 0.03, 0.06$ and $\lambda_2 = 1$. Fig.~\ref{fig:non_abelian_nodes}(a) shows the zoom-in direct gap plot around the $\gamma$ point of Fig.~\ref{fig:lam_phase_dirac_position}(a), and the two Dirac nodes with the same chirality becomes clearly visible. 
When the value of $\lambda_1$ is increased to $0.03$, one of the nodes flipped its chirality. And these two Dirac nodes annihilate with each other and the charge gap opens when $\lambda_1 > 0.055$, as shown in Figs.~\ref{fig:non_abelian_nodes}(b) and (c). 
Meanwhile, another pair of Dirac nodes are created between the first and the second bands, which can be observed in Figs.~\ref{fig:non_abelian_nodes}(d-f). The two nodes carry opposite chiralities when $\lambda_1 = 0.03$, and one of them flips the chirality when $\lambda_1$ is increased to $0.06$.
The chirality change of Dirac nodes in different bands is a signature of the non-Abelian nature of the braiding between Dirac nodes in multi-band systems \cite{wu2019nonabelian,kang_nonabelian_2020,ahn_failure_2019}. 

\subsubsection{Strong correlated bands}

The second path is along the direction: $(\lambda_1, \lambda_2) = (0, 0) \rightarrow (1,0) \rightarrow (1, 1)$. When $\lambda_1$ continuously increases from $0$ to $1$, the Hamiltonian does not break the translation symmetry, and thus we can still study the bands in the MBZ. Since the Coulomb interaction dominates over the kinetic energy of the narrow bands, the Hamiltonian is in the strong coupling limit when $\lambda_1$ is large enough, especially at $\lambda_1 = 1$.  As discussed in Ref.~\cite{kang_cascade_2021} and App.~\ref{sec:add_lam1_1_results},  the bands are degenerate at the high symmetry points, $\Gamma$, $M$, and $K$. The degeneracy at $\Gamma$ is protected by the  $C_{2z} T$ and the particle-hole symmetry, carrying the winding number of $+3$. The MBZ contains three different $M$ points, related by $C_{3z}$ symmetry. Similar to the $\Gamma$ point, the degeneracy at $M$ is also protected by $C_{2z} T$ and particle-hole, but carries the winding number of $-1$. The degeneracy at $K$ and $K'$ points, however, is protected by $C_{2z} T$ and $C_{3z}$ symmetry, and carries the winding number of $1$. So the total winding number is $2$, reflecting the nontrivial topological properties of the flat bands around the charge neutral point.

For the second part of the path, i.e., $(1,0) \rightarrow (1, 1)$, the starting point is the previously described strong interacting band structure of $\mathscr{H}(\bm{\kappa}, 1, 0)$, but folded into the FMBZ. There, the two Dirac nodes originally at different $M$ points are moved to the $\mu$ point. In contrast, the third $M$ point will be moved to the $\gamma$ point, and it becomes a Dirac node between the first and the second bands. 
Therefore, there are four Dirac nodes between the second and the third bands. 
The two nodes at the $\mu$ point carry opposite chirality from the nodes at $K$ and $K'$ point. When increasing $\lambda_2$, these four nodes move towards $y$ point in FMBZ and annihilate with each other. 
Thus, the Brillouin zone folding is also necessary along the second path for gap opening between the second and the third bands, although there is no non-Abelian braiding involved.
We also provide detailed discussion of the motion and chirality of these nodes in App.~\ref{sec:add_lam1_1_results}. 

\subsubsection{Brillouin zone border}

\begin{figure}[t]
    \centering
    \includegraphics[width=\linewidth]{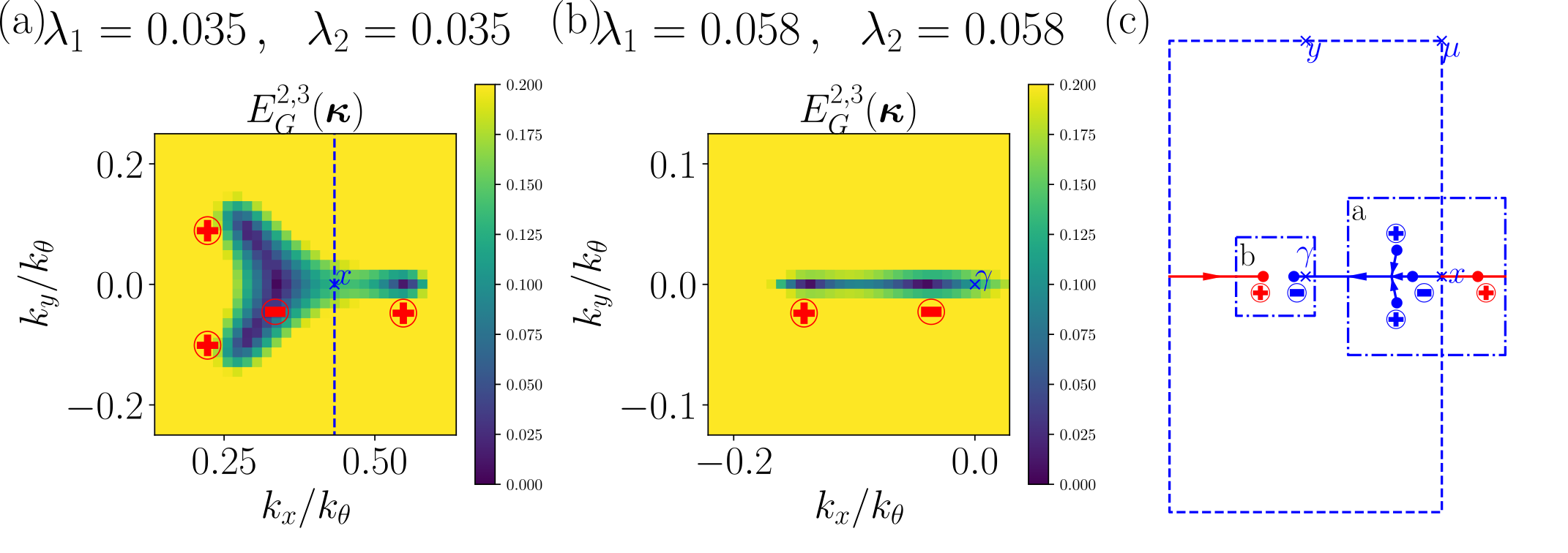}
    \caption{(a) The direct gap between the second and the third bands of $\mathscr{H}(\bm{\kappa}, \lambda_1, \lambda_2)$ in the FMBZ patch around the $x$ point at $\lambda_1 = \lambda_2 = 0.035$. (b) The direct gap between the second and the third bands in the FMBZ patch around $\gamma$ point at $\lambda_1 = \lambda_2 = 0.058$. Dirac nodes and their chiralities are represented by the red $\oplus$ and $\ominus$ symbols in subfigures (a-b). (c) The motion of Dirac nodes in the FMBZ along the third path in the $(\lambda_1, \lambda_2)$ plane. The three Dirac nodes represented by blue dots will merge into one node and move leftward, while the Dirac node represented by red dot moves rightward. The two blue dashed squares represent the FMBZ patches shown in subfigures (a-b). The values of the direct gap are represented in meV in the colorbars.}
    \label{fig:diag_path}
\end{figure}

Our third path is a linear interpolation along the direction $(\lambda_1, \lambda_2) = (0, 0) \rightarrow (1, 1)$. As soon as we move away from $(0, 0)$, the two Dirac nodes of the non-interacting Hamiltonian between the second and third bands start moving in the FMBZ. 
Around $\lambda_1 = \lambda_2 = 0.03$, the two nodes with the same chirality move to the proximity of $x$ point of the FMBZ (see Fig.~\ref{fig:folded_brillouin_zone}). Another pair of Dirac nodes with opposite chiralities are also created in this region.
As shown in Fig.~\ref{fig:diag_path}~(a), there are four Dirac nodes around the $x$ point when $\lambda_1 = \lambda_2 = 0.035$. 
By using the method discussed in App.~\ref{sec:chirality_app}, we are able to evaluate the chiralities of these Dirac nodes.  
The three nodes on the left will merge into one node once the values of $\lambda_{1}$ and $\lambda_2$ are increased to $0.04$ (see App.~\ref{sec:add_lam1_lam2_results}).
Thus, there will be two nodes with both $+1$ chirality moving leftward and rightward from the $x$ point when increasing the values of $\lambda_1$ and $\lambda_2$. 
The path of these nodes wrap around the FMBZ along the axis $\tilde{\mathbf{b}}_1$, and they move towards the proximity of $\gamma$ point around $\lambda_1 = \lambda_2 = 0.05$. In Fig.~\ref{fig:diag_path}~(b), we observe these nodes in the FMBZ patch near the $\gamma$ point at $\lambda_1 = \lambda_2 = 0.058$. 
The relative chirality of different Dirac nodes is well-defined on local patches in the FMBZ. For nodes far apart from each other, finding such a single large patch is problematic (see App.~\ref{sec:chirality_app}), which is why we resort only to local patches once they contain the two nodes. Interestingly, as implied by the analysis in App.~\ref{sec:app_wilson_loop}, the relative chirality of a Dirac node can flip once it encircles the FMBZ (see also Ref.\cite{ahn_failure_2019} for a simple example of a checkerboard lattice with $C_2T$ symmetry and unobstructed single quadratic band touching).
As shown in Fig.~\ref{fig:diag_path}~(b), the two Dirac nodes carry the opposite chiralities when they meet near the $\gamma$ point in FMBZ, which is different from the $+2$ chirality when they were in the proximity of $x$ point. 
The nodes annihilate with each other at around $\lambda_1 = \lambda_2 = 0.06$, and the gap between the second and the third bands is opened. 
We also demonstrate the paths of these nodes wrapping around the FMBZ in Fig.~\ref{fig:diag_path}~(c), where the blue (red) dots and arrows stand for the motion of left (right) moving Dirac nodes. 
Detailed numerical results about the Dirac nodes and chiralities along this path can also be found in App.~\ref{sec:add_lam1_lam2_results}.

\section{Conclusion}\label{sec:conclusion}
Using the translation symmetry breaking Hartree-Fock calculation, we have mapped the phase diagram of TBG at filling factor $\nu=-3$ (or $\nu=+3$ thanks to the particle-hole symmetry) as a function of $w_0/w_1$. Our results show that the quantum anomalous Hall state obtained at the chiral limit is still the self-consistent solution when the interlayer hopping ratio $w_0/w_1$ is smaller than a critical value of $0.5$. 
Around the more experimentally realistic value $w_0/w_1 \approx 0.8$, a translation symmetry breaking phase with $C_{2z}T$ symmetry, a doubled moir\'e unit cell and a large charge gap, which we dub as \emph{$C_{2z}T$ stripe phase}, becomes energetically preferred. By computing its Wilson loop, we also found the $C_{2z}T$ stripe phase carries zero Chern number, which is different from the quantum anomalous Hall phase. 
The vanishing Chern number and the large charge gap imply that this $C_{2z}T$ stripe phase could depict the insulating state at $\nu = +3$ filling observed in experiments \cite{lu2019superconductors,yankowitz2019tuning}.
In the region between the quantum anomalous Hall and the $C_{2z}T$ stripe phases with an intermediate value  $0.5\lesssim w_0/w_1 \lesssim 0.65$, we also find that these states and another phase with a tripling of the moir\'e unit cell all have competitive energy. The candidate states in this intermediate region have small charge gaps, whereas large charge gaps can be observed away from the intermediate region.

Compared to the states proposed in previous studies, the $C_{2z}T$ stripe phase we obtained does not require any strain~\cite{KWAN2021}. This $C_{2z}T$ stripe phase is invariant under $\hat{T}_{\tilde{\mathbf{a}}_1}P$ transformation, and, although similar, it is different from the translation breaking phase in Refs.~\cite{kang_nonabelian_2020}, which has the $\hat{T}_{\tilde{\mathbf{a}}_1}C_{2x}$ symmetry that does not enforce the invariance of the total charge density $\rho_{\rm tot}(\mathbf{r})$ at each $\mathbf{r}$ when translating by a moir\'e unit cell.
The real space charge distribution in this $C_{2z}T$ stripe phase is also evaluated from the mean field order parameter. 
We discovered that the total charge density in the flat band limit does not have modulation in different moir\'e unit cells because of a new non-symmorphic symmetry $\hat{T}_{\tilde{\mathbf{a}}_1}P$ symmetry, although the $C_{2z}T$ stripe phase itself strongly breaks the translation symmetry $\hat{T}_{\tilde{\mathbf{a}}_1}$. 
This non-symmorphic symmetry is no longer fulfilled when the flat band kinetic terms are considered, yet it is only weakly violated.
Meanwhile, the charge density in a single layer still has a clear modulation even in the flat band limit, and it is experimentally testable by scanning tunneling microscope, which only detects the electron states from a single layer. 
We also analyze how the non-interacting TBG flat bands with two Dirac nodes with the same chirality are deformed into the $C_{2z}T$ stripe phase with a large charge gap. The gap opening mechanism depends on the path selected to connect these two extreme cases. In particular, moving to the strongly correlated bands regime first and then breaking the translation symmetry unveils the non-Abelian nature of Dirac nodes' charge in multi-band systems.

The existence of the $C_{2z}T$ stripe phase at $\nu=-3$ naturally raises the question of a similar phase at integer filling $\nu=-1$. 
Indeed, this filling factor shares similarities with $\nu=-3$, with only quantum anomalous Hall states in the chiral flat band limit, as opposed to even integer fillings which have exact eigenstates with zero Chern number~\cite{kang_strong_2019, bultinck_ground_2020, ourpaper4}.
We did solve the self-consistent equation at another odd integer filling $\nu = -1$ and $w_0/w_1 = 0.8$, and translation symmetry breaking is not observed. We leave the search for possible symmetry breaking phases at $\nu = -1$ filling and perturbations which would stabilize them to further works.

\begin{acknowledgments}

We are grateful to Zhi-Da Song for valuable discussions and suggestions in the early stage of this work. We would also like to thank Dumitru C\u{a}lug\u{a}ru, Biao Lian and Run Hou for helpful discussions. 
B.~A.~B. and N.~R. were supported by the DOE Grant No. DE-SC0016239. B.~A.~B. was also supported by the Gordon and Betty Moore Foundation through  Grant No. GBMF11070 towards the EPiQS Initiative. N.~R.acknowledges support from the Princeton Global Network Funds, and the QuantERA II Programme that has received funding from the European Union’s Horizon 2020 research and innovation programme under Grant Agreement No. 101017733. 
This project has also received funding from the European Union's Horizon 2020 Research and Innovation Programme under Grant Agreement No. 731473 and No. 101017733.
This work is also partly supported by a project that has received funding from the European Research Council (ERC) under the European Union’s Horizon 2020 Research and Innovation Programme (Grant Agreement No. 101020833). 
 J.~K.~acknowledges the support from the NSFC Grant No.~12074276 and the start-up grant of ShanghaiTech University. O.~V. was supported by NSF Grant No.~DMR-1916958 and is partially funded by the Gordon and Betty Moore Foundation's EPiQS Initiative Grant GBMF11070, National High Magnetic Field Laboratory through NSF Grant No.~DMR-1157490 and the State of Florida.

\end{acknowledgments}

\bibliographystyle{apsrev4-2}
\bibliography{TBG.bib}

\onecolumngrid
\tableofcontents
\appendix

\section{Hartree-Fock Method}\label{sec:hf_app}
In this appendix, we discuss the details of Hartree-Fock mean field theory with folded moir\'e Brillouin zones in App.~\ref{sec:hf_app_expression}. We also show a method to obtain a smooth visualization of mean field band structure along high symmetry lines in App.~\ref{sec:hf_app_high_sym_lines}.

\subsection{Hartree-Fock Hamiltonian with folded moir\'e Brillouin zone}\label{sec:hf_app_expression}
Here, we provide the explicit expression for the Hartree-Fock Hamiltonian with folded moir\'e Brillouin zones that was sketched in Sec.~\ref{sec:hf_main_text}. We first rewrite the projected interacting Hamiltonian using the following alternative form:
\begin{equation}
    H_I = \frac{1}{2\Omega_{\rm tot}}\sum_{\vk,\vk',\vq\in{\rm MBZ}}\sum_{\substack{\eta\eta',ss'\\mnm'n'}}U^{(\eta\eta')}_{mn;m'n'}(\vq;\vk, \vk')\left(c^\dagger_{\vk+\vq,m\eta s}c_{\vk, n \eta s} - \frac{1}{2}\delta_{\vq, 0}\delta_{mn}\right)\left(c^\dagger_{\vk'-\vq, m'\eta's'}c_{\vk', n'\eta's'} - \frac12 \delta_{\vq, 0}\delta_{m'n'}\right)\,,
\end{equation}
in which the interacting elements $U^{(\eta\eta')}_{mn'm'n'}(\vq;, \vk, \vk')$ are defined as:
\begin{equation}
    U^{(\eta\eta')}_{mn;m'n'}(\vq;\vk,\vk') = \sum_{\mathbf{G}\in\mathcal{Q}_0}V(\vq + \mathbf{G})M^{(\eta)}_{mn}(\vk, \vq + \mathbf{G}) M^{(\eta')}_{m'n'}(\vk', -\vq - \mathbf{G})\,.
\end{equation}
Here $M^{(\eta)}(\vk, \vq + \mathbf{G})$ is the form factor defined in Eq.~(\ref{eqn:def_form_factor}) in the main text. By using the mean field approximation, the interacting Hamiltonian can be written into the Hartree and Fock terms:
\begin{align}
    H^{(H)} &= \sum_{\bm{\kappa}\in{\rm FMBZ}}\sum_{bb',mn,\eta s}\mathcal{H}^{(H)}_{bm\eta s;bn\eta s}(\bm{\kappa})\left(c^\dagger_{\bm{\kappa}+\mathbf{Q}_b, m\eta s} c_{\bm{\kappa} + \mathbf{Q}_{b'}, n\eta s} - \frac12 \delta_{bb'}\delta_{mn}\right)\,,\\
    H^{(F)} &= \sum_{\bm{\kappa}\in{\rm FMBZ}}\sum_{bb',\eta\eta',mn,ss'}\mathcal{H}^{(F)}_{bm\eta s;bn\eta's'}(\bm{\kappa})\left(c^\dagger_{\bm{\kappa}+\mathbf{Q}_b, m\eta s} c_{\bm{\kappa} + \mathbf{Q}_{b'}, n\eta's'} - \frac12 \delta_{bb'}\delta_{mn}\delta_{\eta\eta'}\delta_{ss'}\right)\,.
\end{align}
The matrices $\mathcal{H}^{(H)}(\bm{\kappa})$ and $\mathcal{H}^{(F)}(\bm{\kappa})$ can be written as:
\begin{align}
    \mathcal{H}^{(H)}_{bm\eta s;b'n\eta's'}(\bm{\kappa}) &= \frac{1}{\Omega_
    {\rm tot}}\sum_{\bm{\kappa}'\in{\rm FMBZ}}\sum_{bb'b''b'''}\sum_{\eta''s''}\sum_{mnm'n'}U^H_{bm\eta s, b'n\eta s; b''m'\eta''s''; b'''n'\eta''s''}(\bm{\kappa}, \bm{\kappa}')\Delta_{b''m'\eta''s''; b'''n'\eta''s''}(\bm{\kappa}')\delta_{\eta\eta'}\delta_{ss'}\label{eqn:hartree_sb}\\
    \mathcal{H}^{(F)}_{bm\eta s;b'n\eta's'}(\bm{\kappa}) &= -\frac{1}{\Omega_{\rm tot}}\sum_{\bm{\kappa}'\in{\rm FMBZ}}\sum_{bb'b''b'''}\sum_{mnm'n'}U^F_{bm\eta s, b'n\eta's';b''m'\eta's';b'''n'\eta s}(\bm{\kappa}, \bm{\kappa}')\Delta_{b''m'\eta's';b'''n'\eta s}(\bm{\kappa}')\,,\label{eqn:fock_sb}
\end{align}
in which the matrices $\Delta(\bm{\kappa})$ is the order parameter defined in Eq.~(\ref{eqn:def_order_parameter}) in the main text. We can also use the interaction elements $U^{(\eta\eta')}_{mn;m'n'}(\vq;\vk,\vk')$ to represent the coefficients $U^H(\bm{\kappa}, \bm{\kappa}')$ and $U^F(\bm{\kappa}, \bm{\kappa}')$ as follows:
\begin{align}
    U^H_{bm\eta s, b'n\eta s; b''m'\eta's'; b'''n'\eta's'}(\bm{\kappa}, \bm{\kappa}') &= U^{(\eta\eta')}_{mn;m'n'}(\mathbf{Q}_b - \mathbf{Q}_{b'}; \bm{\kappa}+\mathbf{Q}_{b'}, \bm{\kappa}' + \mathbf{Q}_{b'''})\sum_{\mathbf{G}\in\mathcal{Q}_0}\delta_{\mathbf{Q}_b-\mathbf{Q}_{b'} + \mathbf{Q}_{b''} - \mathbf{Q}_{b'''}, \mathbf{G}}\\
    U^F_{bm\eta s, b'n\eta's';b''m'\eta's';b'''n'\eta s}(\bm{\kappa},\bm{\kappa}') &= U^{(\eta'\eta)}_{m'n;mn'}(\bm{\kappa}'-\bm{\kappa} + \mathbf{Q}_{b''} - \mathbf{Q}_{b'}; \bm{\kappa} + \mathbf{Q}_{b'}, \bm{\kappa}' + \mathbf{Q}_{b'''})\sum_{\mathbf{G}\in\mathcal{Q}_0}\delta_{\mathbf{Q}_b-\mathbf{Q}_{b'} + \mathbf{Q}_{b''} - \mathbf{Q}_{b'''}, \mathbf{G}}\,.
\end{align}
We can also write down the total mean field Hamiltonian by adding the kinetic term:
\begin{align}
    &\mathcal{H}^{(0)}_{b m \eta s; b'n\eta' s'}(\bm{\kappa}) = \epsilon_{\bm{\kappa} + \mathbf{Q}_{b}, m, \eta}~\delta_{bb'}\delta_{mn}\delta_{\eta\eta'}\delta_{ss'}\,,\\
    &\mathcal{H}^{HF}(\bm{\kappa}) = t \mathcal{H}^{(0)}(\bm{\kappa}) + \mathcal{H}^{(H)}(\bm{\kappa}) + \mathcal{H}^{(F)}(\bm{\kappa})\,.
\end{align}
For convenience, we have introduced a parameter $t$ to go from the flat band limit $(t=0)$ to the full fledged kinetic term $(t = 1)$. As discussed in Sec.~\ref{sec:hf_main_text}, we use $\phi_{bm\eta s, i}(\bm{\kappa})$ to represent the eigenstates of the Hamiltonian $\mathcal{H}^{HF}(\bm{\kappa})$. By using $\phi_{bm\eta s, i}(\bm{\kappa})$, we can also obtain the self-consistent condition for the order parameter:
\begin{equation}\label{eqn:self_consistent}
    \Delta_{bm\eta s; b'n\eta's'}(\bm{\kappa}) = \sum_{i\in{\rm occupied}}\left(\phi^*_{bm\eta s, i}(\bm{\kappa})\phi_{b'n\eta's', i}(\bm{\kappa})\right) - \frac12 \delta_{bb'}\delta_{mn}\delta_{\eta\eta'}\delta_{ss'}\,.
\end{equation}
Since we solely focus on the filling factor $\nu=-3$, only the $N_M$ states with the lowest eigenvalues $E_i(\bm{\kappa})$ among all eigenstates are counted as occupied states. We start from a randomized initial order parameter, and we build $\mathcal{H}^{HF}(\bm{\kappa})$ from this order parameter. We can then solve the new order parameter from the self-consistent condition Eq.~(\ref{eqn:self_consistent}) until both the order parameter and Hamiltonian converge. For a given self-consistent solution, the total energy can be evaluated by the following equation:
\begin{equation}
    E_{\rm tot} = \sum_{\bm{\kappa}\in{\rm FMBZ}}{\rm Tr}\left[\left(\mathcal{H}^{(0)}(\bm{\kappa}) + \frac12 \left(\mathcal{H}^{(H)}(\bm{\kappa}) + \mathcal{H}^{(F)}(\bm{\kappa})\right)\right)\Delta^{\rm T}(\bm{\kappa})\right]\,.
\end{equation}

\subsection{Band structure along high symmetry lines}\label{sec:hf_app_high_sym_lines}
In this subsection, we discuss the non-self-consistent-field method we use to obtain a smooth visualization of HF band structure without solving the self-consistent equation on a dense momentum lattice. Solving the self-consistent equation on a dense momentum lattice discretizing the (folded) moir\'e Brillouin zone requires a large amount of computing resources. The storage requirement for saving the coefficients $U^{H,F}(\bm{\kappa}, \bm{\kappa}')$ also grows quadratically with the lattice size. Thus, our mean field solutions are obtained on relatively small lattices, such as $12\times 12$, $18\times 18$ up to $36\times 36$. However, there are only a few points of this discretized mesh that are along the high symmetry lines on such small momentum lattice. These points are not dense enough to obtain a smooth visualization of the mean field band structure along these high symmetry lines. 

\begin{figure}[!htbp]
    \centering
    \includegraphics[width=0.7\linewidth]{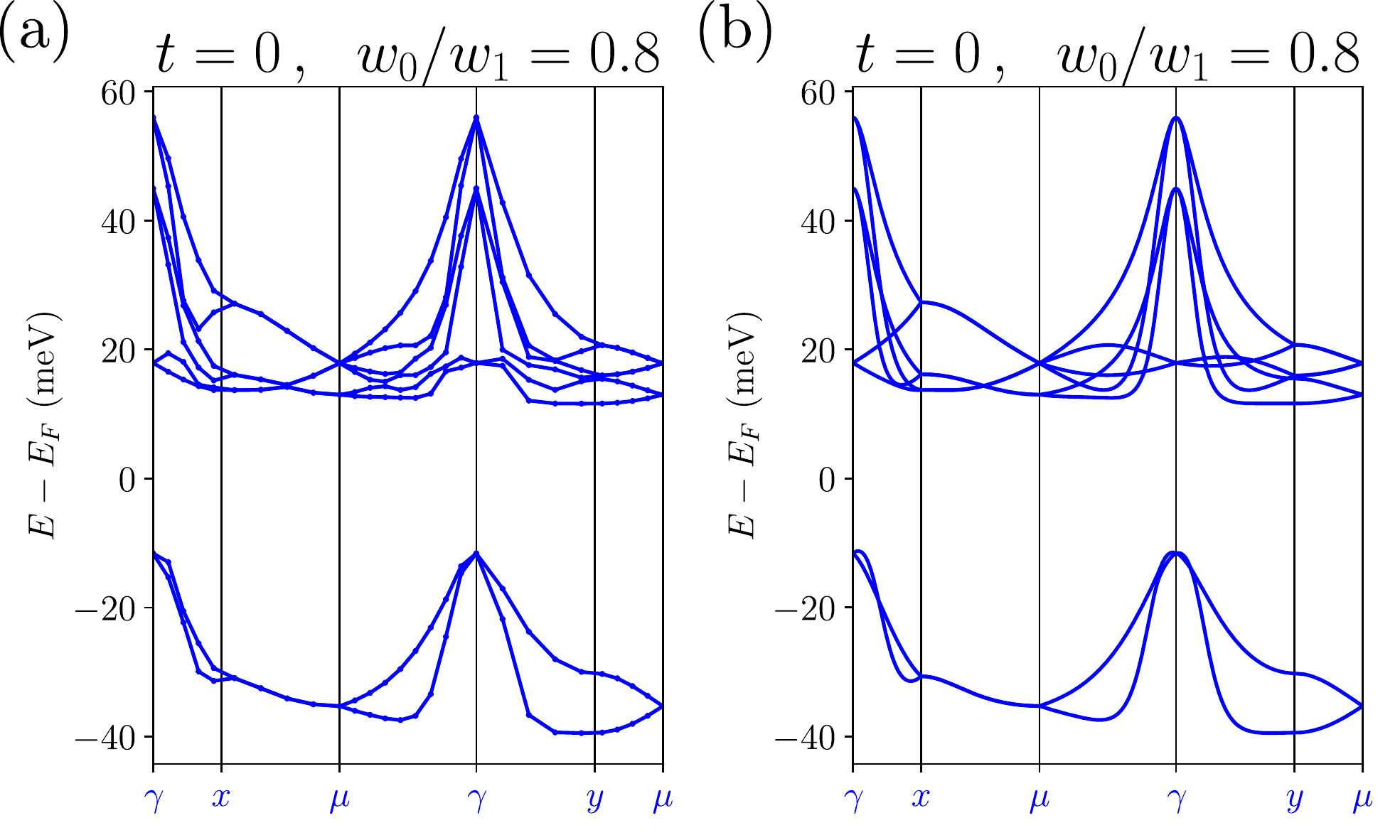}
    \caption{The Hartree-Fock band structure of the $C_{2z}T$ stripe phase along the high symmetry lines. (a) The HF band structure obtained directly from the solution on $18\times 18$ momentum lattice. (b) The ``continuum'' HF band structure along the high symmetry lines. We used the order parameter on the same $18\times 18$ momentum lattice as in subfigure (a) to calculate the HF Hamiltonians along these high symmetry lines.}
    \label{fig:continuum_plot}
\end{figure}

We choose our $C_{2z}T$ stripe phase solution at $w_0/w_1 = 0.8$ on $18\times 18$ momentum lattice as an example. As shown in Fig.~\ref{fig:continuum_plot}~(a), we simply diagonalize the Hamiltonian $\mathcal{H}^{HF}(\bm{\kappa})$ on this momentum lattice, and we show the energy spectra for $\bm{\kappa}$ along the high symmetry lines. Albeit the shape of the bands and the charge gap can be roughly observed in this plot, the details, such as band crossing points, cannot be easily identified due to the large distances between these momentum points. 

In order to solve the energy spectra for any given momentum $\bm{\kappa}$ along the high symmetry lines, we can still use Eqs. (\ref{eqn:hartree_sb}) and (\ref{eqn:fock_sb}). These equations show that the Hartree Fock Hamiltonian $\mathcal{H}^{HF}(\bm{\kappa})$ has a summation for $\bm{\kappa}' \in {\rm FMBZ}$. For an arbitrary value of $\bm{\kappa}$, we can enforce that the summation of $\bm{\kappa}'$ is always on the sparse momentum lattice. Therefore, for the purpose of building the HF Hamiltonian along a dense high symmetry line, we have to know the order parameter $\Delta(\bm{\kappa}')$, and the HF coefficients $U^{H,F}(\bm{\kappa}, \bm{\kappa}')$ with $\bm{\kappa}$ along these dense high symmetry lines and $\bm{\kappa}'$ on this sparse lattice. To obtain all of these coefficients, we only need to solve the single body wavefunctions of the BM model on the sparse lattice and along the dense high symmetry line, instead of the wavefunctions on a dense momentum lattice. The storage requirement for saving these HF coefficients becomes linear with the lattice size, and thus we are able to enhance the point density along the high symmetry lines at a moderate cost. Once we obtain the HF Hamiltonian, we are also able to get its eigenstate $\phi_{bm\eta s, i}(\bm{\kappa})$ and order parameter $\Delta(\bm{\kappa})$ along the dense high symmetry line. Fig.~\ref{fig:continuum_plot} (b) shows the HF band structure we calculated using the order parameter on the $18\times 18$ momentum lattice, which is the same as the one we used in Fig.~\ref{fig:continuum_plot} (a). The band structure plot in Fig.~\ref{fig:continuum_plot} (b) has the same shape qualitatively as in Fig.~\ref{fig:continuum_plot} (a), while it also provides more details like the band crossing point along the $\gamma$-$x$ line. The plots in Fig.~\ref{fig:mean_field_bands_18x18} in the main text have also been obtained by this method.

This method is not limited to studying the band structure along high symmetry lines. We can also replace the momentum points $\bm{\kappa}$ along the high symmetry lines by momentum points on a small patch in the FMBZ. Thus, we are also able to study the band structure and the HF wavefunctions in a small region of the FMBZ without solving the self-consistent equation on a dense lattice. This technique was used in Sec.~\ref{sec:dirac} to get an accurate result for the Dirac nodes motions around the $\gamma$ point in the $C_{2z}T$ stripe phase. The HF Hamiltonians $\mathcal{H}^{HF}(\bm{\kappa})$ built in the small FMBZ patch retain the symmetries of the order parameter $\Delta(\bm{\kappa}')$ on the sparse momentum lattice. Therefore, the Dirac nodes in the FMBZ patch, which are protected by the $C_{2z}T$ symmetry, should be well-captured by the non-self-consistent-field method. 

\section{Symmetries and sewing matrices}\label{sec:app_ham_sym}
In this appendix, we review the representation of several symmetries of the single valley TBG Hamiltonian, which is mentioned in Sec.~\ref{sec:model} and used in Sec.~\ref{sec:c2t_stripe_sym} in the main text. As discussed in Refs.~\cite{ahn_failure_2019,song_all_2019, ourpaper3}, the single valley TBG Hamiltonian has $C_{2z}T$, $C_{3z}$, $C_{2x}$ and $P$ symmetries. The representations of these symmetries are given by:
\begin{align}
    D_{\mathbf{Q},\mathbf{Q}'}(C_{2z}T) &= \sigma_{x} \delta_{\mathbf{Q}, \mathbf{Q}'}\,,\\
    D_{\mathbf{Q}, \mathbf{Q}'}(C_{3z}) &= \exp\left(i\frac{2\pi\eta}{3}\sigma_z\right)\delta_{\mathbf{Q},C_{3z} \mathbf{Q}'}\,,\\
    D_{\mathbf{Q},\mathbf{Q}'}(C_{2x}) &= \sigma_x\delta_{\mathbf{Q}, C_{2x}\mathbf{Q}'}\,,\\
    D_{\mathbf{Q}, \mathbf{Q}'}(P) &= \zeta_{\mathbf{Q}}\delta_{\mathbf{Q}, -\mathbf{Q}'}\,,
\end{align}
where $\zeta_\mathbf{Q} = 1$ when $\mathbf{Q}\in\mathcal{Q}_+$, and $\zeta_\mathbf{Q} = -1$ when $\mathbf{Q} \in \mathcal{Q}_-$. Both $C_{3z}$ and $C_{2x}$ are unitary symmetries which commute with the non-interacting Hamiltonian. In contrast, particle hole symmetry $P$ is a unitary symmetry which anti-commute with the non-interacting Hamiltonian. Unlike the other three, $C_{2z}T$ is an anti-unitary symmetry, which also contains a complex conjugation operation. These representation matrices describe the transformation of electron operators in plane wave basis under these transformations:
\begin{equation}
    g^{-1}c^\dagger_{\vk,\mathbf{Q},\eta,\alpha, s} g = \sum_{\mathbf{Q}'\beta} D^*_{\mathbf{Q}\alpha,\mathbf{Q}'\beta}(g) c^\dagger_{g\vk, \mathbf{Q}', \eta, \beta, s}
\end{equation}
Therefore, the non-interacting Hamiltonian will transform as follows under these symmetries:
\begin{align}
    D(C_{2z}T)^{-1}h^{(\eta)*}(\vk)D(C_{2z}T) &= h^{(\eta)}(\vk)\,,\label{eqn:c2t_transformation_ham}\\
    D(C_{3z})^{-1}h^{(\eta)}(\vk)D(C_{3z}) &= h^{(\eta)}(C_{3z}\vk)\,,\\
    D(C_{2x})^{-1}h^{(\eta)}(\vk)D(C_{2x}) &= h^{(\eta)}(C_{2x}\vk)\,,\\
    D(P)^{-1}h^{(\eta)}(\vk)D(P) &= -h^{(\eta)}(-\vk)\,.
\end{align}
Notice that on the left hand side of Eq.~(\ref{eqn:c2t_transformation_ham}), we transform the {\it complex conjugation} of the Hamiltonian by matrix $D(C_{2z}T)$ because of the anti-unitary nature of $C_{2z}T$ transformation. For each unitary symmetry $g$, we can define the sewing matrix $B^g(\vk)$ as:
\begin{equation}
    B^g_{mn}(\vk) = \sum_{\mathbf{Q}\alpha, \mathbf{Q}'\beta} u^*_{\mathbf{Q}\alpha, m\eta}(\vk) D_{\mathbf{Q}\alpha,\mathbf{Q}'\beta}(g)u_{\mathbf{Q}'\beta, n\eta}(g\vk)\,,
\end{equation}
and this matrix has the following property:
\begin{equation}
    \sum_{m}u_{\mathbf{Q}\alpha, m\eta}(\vk) B^g_{mn}(\vk) = \sum_{\mathbf{Q}'\beta}  D_{\mathbf{Q}\alpha,\mathbf{Q}'\beta}(g)u_{\mathbf{Q}'\beta, n\eta}(g\vk)\,.
\end{equation}
Similarly, for anti-unitary symmetry $C_{2z}T$, we can define its sewing matrix as:
\begin{equation}
    B^{C_{2z}T}_{mn}(\vk) = \sum_{\mathbf{Q}\alpha, \mathbf{Q}'\beta} u^*_{\mathbf{Q}\alpha, m\eta}(\vk) D_{\mathbf{Q}\alpha,\mathbf{Q}'\beta}(C_{2z}T)u^*_{\mathbf{Q}'\beta, n\eta}(\vk)\,.
\end{equation}
For each given unitary symmetry, the electron operators will transform as their sewing matrices:
\begin{align}
    g^{-1} c^\dagger_{\vk, m\eta s} g &= \sum_{\mathbf{Q}\alpha}u_{\mathbf{Q}\alpha, m\eta}(\vk)g^{-1}c^\dagger_{\vk, \mathbf{Q}, \eta, \alpha, s}g\nonumber\\
    &= \sum_{\mathbf{Q}\alpha}u_{\mathbf{Q}\alpha, m\eta}(\vk)\sum_{\mathbf{Q}'\beta}D^*_{\mathbf{Q}\alpha,\mathbf{Q}'\beta}(g)c^\dagger_{g\vk,\mathbf{Q}',\eta,\beta,s}\nonumber\\
    &= \sum_{\mathbf{Q'}\beta}\sum_n u_{\mathbf{Q}'\beta, n\eta}(g\vk)B^{g*}_{mn}(\vk)c^\dagger_{g\vk,\mathbf{Q}', \eta, \beta, s}\nonumber\\
    &= \sum_{n}B^{g*}_{mn}(\vk)c^\dagger_{g\vk, n, \eta, s}\,.\label{eqn:transform_sewing}
\end{align}
Similar result can also be derived for anti-unitary symmetry $C_{2z}T$. As mentioned in Refs.~\cite{ourpaper3, ourpaper4, ourpaper5, ourpaper6}, we fix the gauge choice of the wavefunctions such that the sewing matrices $B^{C_{2z}T}_{mn}(\vk) = \delta_{mn}$. Thus, the electron operators are transformed as:
\begin{equation}\label{eqn:c2t_transform}
    (C_{2z}T)^{-1}c^\dagger_{\vk, m\eta s} (C_{2z}T) = c^\dagger_{\vk, m \eta s}\,.
\end{equation}

Except for these symmetries, this Hamiltonian also has the translation symmetries along the basis vectors $\tilde{\mathbf{a}}_{1,2}$ of the moir\'e superlattice, which we denote by $\hat{T}_{\tilde{\mathbf{a}}_{1,2}}$. For each electron operator, it will gain a phase factor under such translation transformation as shown:
\begin{equation}\label{eqn:translation_transform}
    \hat{T}^{-1}_{\tilde{\mathbf{a}}_{1,2}} c^\dagger_{\vk,m,\eta,s} \hat{T}_{\tilde{\mathbf{a}}_{1,2}} = e^{i\vk\cdot \tilde{\mathbf{a}}_{1,2}}c^\dagger_{\vk,m,\eta,s}\,.
\end{equation}
We obtained the values of $\mathcal{G}(g, \bm{\kappa})$ shown in Fig.~\ref{fig:flatband_stripe_sym_break} in the main text by applying Eqs.~(\ref{eqn:transform_sewing}), (\ref{eqn:c2t_transform}) and (\ref{eqn:translation_transform}) to its definition Eq.~(\ref{eqn:def_sym_break}).

\section{Wilson loops in folded MBZ}\label{sec:wilson_app}
In this appendix, we discuss the method to represent the HF wavefunctions using the plane wave basis and we derive the expression of the non-Abelian Wilson loops, which is used in Sec.~\ref{sec:c2t_topo_phase_diagram} in the main text.
We start with the mean field Hamiltonian $\mathcal{H}^{HF}(\bm{\kappa}) = \mathcal{H}^{(H)}(\bm{\kappa}) + \mathcal{H}^{(F)}(\bm{\kappa})$. By diagonalizing this Hamiltonian, we can obtain the eigenvectors:
\begin{equation}
    \sum_{b'm'\eta's'}\mathcal{H}^{(HF)}_{bm\eta s; b'm'\eta's'}(\bm{\kappa}) \phi_{b' m' \eta' s', i}(\bm{\kappa}) = E_{i}(\bm{\kappa}) \phi_{b m \eta s, i}(\bm{\kappa})\,.
\end{equation}
where $\phi_{bm\eta s, i}(\bm{\kappa})$ is the Hartree-Fock band wavefunction of the $i$-th mean field band. We can write the wavefunction into the following format:
\begin{equation}
    |\phi_i(\bm{\kappa})\rangle = \sum_{bm\eta s}\phi_{bm\eta s, i}(\bm{\kappa})c^\dagger_{\bm{\kappa} + \mathbf{Q}_b m \eta s}|0\rangle.
\end{equation}
In order to compute the Wilson loops or Berry connection of these Bloch wavefunctions, we have to rewrite these states using the plane wave basis of the continuum model \cite{bistritzer_moire_2011}:
\begin{equation}
    |\phi_i(\bm{\kappa})\rangle = \sum_{bm\eta s}\phi_{bm\eta s, i}(\bm{\kappa}) \sum_{\mathbf{Q}\in\mathcal{Q}_\pm ,\alpha} u_{\mathbf{Q}\alpha, m\eta}(\bm{\kappa} + \mathbf{Q}_b)c^\dagger_{\bm{\kappa} + \mathbf{Q}_b, \mathbf{Q},\eta, \alpha, s}|0\rangle\,.
\end{equation}
Therefore, to represent the Bloch wavefunction of the mean field bands by the plane wave basis, we introduce the coefficients $\Phi_{\mathbf{Q}\alpha,b,\eta,s; i}(\bm{\kappa})$:
\begin{align}
    \Phi_{\mathbf{Q}\alpha, b, \eta, s; i}(\bm{\kappa}) &= \sum_{m}\phi_{bm\eta s, i}(\bm{\kappa}) u_{\mathbf{Q}\alpha,m \eta}(\bm{\kappa} + \mathbf{Q}_b)\,,\\
    |\phi_i(\bm{\kappa})\rangle &= \sum_{\mathbf{Q}\in\mathcal{Q}_{\pm}, \alpha}\sum_{b\eta s} \Phi_{\mathbf{Q}\alpha, b,\eta,s;i}(\bm{\kappa})c^\dagger_{\bm{\kappa} + \mathbf{Q}_b, \mathbf{Q}, \eta, \alpha, s}|0\rangle\,.
\end{align}
The eigenvectors $\phi_{bm\eta s, i}(\bm{\kappa})$ are not periodic in the FMBZ. When the momentum $\bm{\kappa}\in {\rm FMBZ}$ is shifted by a reciprocal vector of FMBZ ($\bm{\kappa} \rightarrow \bm{\kappa} + \mathbf{g}$), the subband index $b$ of $\phi_{bm\eta s, i}(\bm{\kappa})$ will be transformed by the embedding matrix $\mathcal{V}_{\mathbf{g}}$ as follows:
\begin{equation}
    \phi_{bm\eta s, i}(\bm{\kappa} + \mathbf{g}) = \sum_{b'}(\mathcal{V}_{\mathbf{g}})_{bb'}\phi_{b'm\eta s}(\bm{\kappa})\,,
\end{equation}
in which the matrix $\mathcal{V}_{\mathbf{g}}$ is given by:
\begin{equation}
    (\mathcal{V}_{\mathbf{g}})_{bb'} = \sum_{\mathbf{G}\in\mathcal{Q}_0}\delta_{\mathbf{g} + \mathbf{Q}_b, \mathbf{Q}_{b'} + \mathbf{G}}\,.
\end{equation}
Therefore, the subband index $b$ in the Bloch wavefunction coefficients $\Phi_{\mathbf{Q}\alpha, b, \eta, s; i}(\bm{\kappa})$ also has to be shifted accordingly:
\begin{equation}\label{eqn:hf_wf_plane}
    \Phi_{\mathbf{Q}\alpha,b,\eta, s;i}(\bm{\kappa} + \mathbf{g}) = \sum_{m,b'}\sum_{\mathbf{G}\in\mathcal{Q}_0}\delta_{\mathbf{g} + \mathbf{Q}_b, \mathbf{Q}_{b'} + \mathbf{G}} \phi_{b' m \eta s, i}(\bm{\kappa}) u_{\mathbf{Q}\alpha, m\eta}(\bm{\kappa} + \mathbf{g} + \mathbf{Q}_b)\,.
\end{equation}
We also use $\Phi_i(\bm{\kappa})$ to denote the vector made of the coefficients $\Phi_{\mathbf{Q}\alpha, b, \eta, s; i}(\bm{\kappa})$. Thus, we are able to define the non-Abelian Wilson loop of the mean field bands. For the two types of favored foldings $(\sqrt3\times \sqrt3)$ and $(2\times 1)$, we represent the momentum in FMBZ by $\bm{\kappa} = \frac{\kappa_1}{2\pi}\mathbf{Q}_1+ \frac{\kappa_2}{2\pi}\tilde{\mathbf{b}}_2$, in which $\mathbf{Q}_1$ is the basis vectors of the reciprocal lattices of the folded Brillouin zones defined in Table. \ref{tab:unit_cell_choices} and $\tilde{\mathbf{b}}_2$ is the reciprocal vector of the original moir\'e Brillouin zone. We evaluate the Wilson loops along the direction of $\mathbf{Q}_1$ with the lowest $N_F$ bands:
\begin{align}
    W_{ij}(\kappa_2) =&\sum_{i_1,i_2,\cdots i_{n-1}=1}^{N_F}\left[\Phi^\dagger_i(\kappa_1=0, \kappa_2) \Phi_{i_1}(\kappa_1=\delta\kappa, \kappa_2) \Phi^\dagger_{i_1}(\kappa_1=\delta\kappa, \kappa_2) \cdots \right.\nonumber\\
    &\left.\times \Phi_{i_{n-1}}(\kappa_1 = 2\pi-\delta\kappa, \kappa_2)\Phi_{i_{n-1}}^\dagger(\kappa_1=2\pi-\delta\kappa, \kappa_2)\Phi_j(\kappa_1=2\pi,\kappa_2)\right]\,, \delta\kappa = \frac{2\pi}{n}.\label{eqn:def_wilson}
\end{align}
Here the integer $n$ is the number of points along the direction of $\mathbf{Q}_1$ on the discretized momentum lattice in the FMBZ. The winding of Wilson loop eigenvalue exponent, computed by this expression, contains the information of the band topology, as discussed for the plots in Figs.~\ref{fig:wilson} and \ref{fig:wilson_extra} in the main text.

\section{Wilson loop of the stripe phase}\label{sec:app_wilson_loop}

In this Appendix, we derive the properties of the Wilson loop for the $C_{2z}T$ symmetric stripe phase that was numerically evaluated in Sec.~\ref{sec:c2t_topo_phase_diagram}. For that purpose, we focus on the simple limiting case of a gapped $C_{2z}T$ stripe with effective Hamiltonian given in Eq.~(69) of Ref.~\cite{kang_nonabelian_2020}.
This Hamiltonian has a gap between the band 2 and band 3, and because it is a special case, band 1 is degenerate with band 2, as is band 3 with band 4. 
This model is enough to understand the topology of these special subspaces, because, as long as the gap between 1-2 and 3-4 does not close, the topology must remain, i.e., we cannot change the sign of the determinant of the Wilson loops $\det W$ under continuous deformations which do not close the 2-3 gap.

\subsection{Simple effective Hamiltonian for the \texorpdfstring{$C_{2z}T$}{C2T} stripe}

For convenience, the Chern states in this Appendix (App.~\ref{sec:app_wilson_loop}) are chosen to be the Bloch state basis and their gauge is fixed as in Ref.~\cite{kang_nonabelian_2020}, i.e. the constructed  Chern states are continuous in momentum, but not periodic. To be more specific, these  Chern  states satisfy the following boundary conditions:
\begin{align}
	| \psi_{\mu, \eta, s}(\bm{\vk}) \rangle = | \psi_{\mu, \eta, s}(\bm{\vk} + \tilde{\mathbf{b}}_1) \rangle \ , \quad | \psi_{\mu, \eta, s}(\bm{\vk}) \rangle = e^{- i \mu \bm{\vk} \cdot \tilde{\mathbf{a}}_1}| \psi_{\mu, \eta, s}(\bm{\vk} + \tilde{\mathbf{b}}_2) \rangle  \label{eq:ChernBC}
\end{align}
where the subscript $\mu = \pm 1$ is the Chern number of the associated states, $\eta$ and $s$ are the indices for valley and spin respectively, and $\tilde{a}_i$ are the moir\'e lattice vectors defined in Fig.~\ref{fig:flatband_stripe_real_total}. In addition, these Chern states transform under $C_{2z}T$ as follows, 
\begin{align} \label{eq:ChernStatesC2T}
	C_{2z} T | \psi_{\mu, \eta, s}(\bm{\vk}) \rangle = | \psi_{-\mu, \eta, s}(\bm{\vk}) \rangle  \ . 
\end{align}
Note that the gauge choice in Eq.~(\ref{eq:ChernBC}) is different from the gauge choice that we used for numerical calculations, which was discussed in Refs.~\cite{ourpaper3,ourpaper4,ourpaper5,ourpaper6}.

Since the $C_{2z}T$ stripe phase is both spin and valley polarized, we can focus only on a particular spin and valley, and thus drop the the spin and valley indices in the rest of this appendix. In the $C_{2z} T$ phase, with the four-component Chern basis  $ \left\{ |\psi_{+1}(\bm{\kappa}) \rangle, |\psi_{+1}(\bm{\kappa} + \tilde{\mathbf{b}}_1/2)\rangle, |\psi_{-1}(\bm{\kappa})\rangle, |\psi_{-1}(\bm{\kappa} + \tilde{\mathbf{b}}_1/2) \rangle \right\}$, the effective Hamiltonian Eq.~(69) of Ref.~\cite{kang_nonabelian_2020} can be written as:
\begin{eqnarray}
H^{C_{2z}T}_{\rm eff}(\bm{\kappa})&=&\left(\begin{array}{cccc} \epsilon(\bm{\kappa}) & 0 & 0 & \Delta_2(\bm{\kappa}) \\ 
0 & \epsilon(\bm{\kappa} + \tilde{\mathbf{b}}_1/2) & \Delta_2(\bm{\kappa}) & 0 \\ 
0 & \Delta^*_2(\bm{\kappa}) & \epsilon(\bm{\kappa}) & 0 \\ 
\Delta^*_2(\bm{\kappa}) & 0 & 0 & \epsilon(\bm{\kappa} + \tilde{\mathbf{b}}_1/2) 
\end{array}\right)\,,\label{eq:effectivehamiltonianc2zt}
\end{eqnarray}
where the matrix elements satisfy the non-trivial periodicity conditions (with the proper gauge choice listed in Eq.~(\ref{eq:ChernBC}) and Ref.~\cite{kang_nonabelian_2020}):
\begin{eqnarray}
\Delta_2(\bm{\kappa} + \tilde{\mathbf{b}}_1/2) & = & \Delta_2(\bm{\kappa})\,, \label{eq:effectivec2tperiodicitydelta2}\\
\Delta_2(\bm{\kappa} + \tilde{\mathbf{b}_2}) & = & -e^{2i\bm{\kappa} \cdot \tilde{\mathbf{a}}_1  }\Delta_2(\bm{\kappa})\,,\label{eq:effectivec2tperiodicitydelta2_y}\\
\epsilon(\bm{\kappa} + \tilde{\mathbf{b}}_1) & = & \epsilon(\bm{\kappa})\,. \label{eq:effectivec2tperiodicityepsilon}
\end{eqnarray}
The corresponding single particle states for the smooth Chern gauge states built in Eq.~(13) of Ref.~\cite{kang_nonabelian_2020} are:
\begin{eqnarray}
\mbox{\bf upper doublet:}\;\;\; &&|\phi_4(\bm{\kappa})\rangle =\cos\frac{\theta(\bm{\kappa})}{2}|\psi_{+1}(\bm{\kappa})\rangle+\sin\frac{\theta(\bm{\kappa})}{2}e^{-i\varphi_2(\bm{\kappa})}|\psi_{-1}(\bm{\kappa} + \tilde{\mathbf{b}}_1/2)\rangle\,,\\
&&|\phi_3(\bm{\kappa})\rangle =\sin\frac{\theta(\bm{\kappa})}{2}e^{i\varphi_2(\bm{\kappa})}|\psi_{+1}(\bm{\kappa} + \tilde{\mathbf{b}}_1/2)\rangle+\cos\frac{\theta(\bm{\kappa})}{2}|\psi_{-1}(\bm{\kappa})\rangle\,,\\
\mbox{\bf lower doublet:}\;\;\; &&|\phi_2(\bm{\kappa})\rangle  =-\sin\frac{\theta(\bm{\kappa})}{2}e^{i\varphi_2(\bm{\kappa})}|\psi_{+1}(\bm{\kappa})\rangle+\cos\frac{\theta(\bm{\kappa})}{2}|\psi_{-1}(\bm{\kappa} + \tilde{\mathbf{b}}_1/2)\rangle\,,\\
&&|\phi_1(\bm{\kappa})\rangle=\cos\frac{\theta(\bm{\kappa})}{2}|\psi_{+1}(\bm{\kappa} + \tilde{\mathbf{b}}_1/2)\rangle-\sin\frac{\theta(\bm{\kappa})}{2}e^{-i\varphi_2(\bm{\kappa})}|\psi_{-1}(\bm{\kappa})\rangle\,.
\end{eqnarray}  
The quantities $\theta(\bm{\kappa})$ and $\varphi_2(\bm{\kappa})$ are defined by:
\begin{eqnarray}
e^{i\varphi_2(\bm{\kappa})}&=&\frac{\Delta_2(\bm{\kappa})}{|\Delta_2(\bm{\kappa})|}\,,\label{eq:defphi2}\\
\cos\theta(\bm{\kappa}) &=& \frac{\epsilon'(\bm{\kappa})}{\sqrt{\epsilon'(\bm{\kappa})^2+|\Delta_2(\bm{\kappa})|^2}}\,,\label{eq:deftheta}\\
\epsilon'(\bm{\kappa})&=&\frac{1}{2}\left(\epsilon(\bm{\kappa})-\epsilon(\bm{\kappa} + \tilde{\mathbf{b}}_1/2)\right)\,.\label{eq:defepsilonprime}
\end{eqnarray}

Under $C_{2z}T$, we have $|\psi_\pm(\vk)\rangle\rightarrow |\psi_\mp(\vk)\rangle$ which follows from Eq.~(\ref{eq:ChernStatesC2T}). Thus, $|\phi_4(\bm{\kappa})\rangle$ and $|\phi_3(\bm{\kappa})\rangle$ get interchanged by $C_{2z}T$ (and similarly for $|\phi_1(\bm{\kappa})\rangle$ and $|\phi_2(\bm{\kappa})\rangle$). Now, we are able to construct states which have a diagonal $C_{2z}T$ sewing matrix:
{\small{\begin{eqnarray}
&&\mbox{\bf upper doublet:}\nonumber\\
&&|\phi_4'(\bm{\kappa})\rangle = \frac{|\phi_4(\bm{\kappa})\rangle+e^{-i\varphi_2(\bm{\kappa})}|\phi_3(\bm{\kappa})\rangle}{\sqrt{2}}=\\
&&\frac{1}{\sqrt{2}}\left(\cos\frac{\theta(\bm{\kappa})}{2}|\psi_{+1}(\bm{\kappa})\rangle+\sin\frac{\theta(\bm{\kappa})}{2}|\psi_{+1}(\bm{\kappa} + \tilde{\mathbf{b}}_1/2)\rangle
+e^{-i\varphi_2(\bm{\kappa})}\left(\sin\frac{\theta(\bm{\kappa})}{2}|\psi_{-1}(\bm{\kappa} + \tilde{\mathbf{b}}_1/2)\rangle+\cos\frac{\theta(\bm{\kappa})}{2}|\psi_{-1}(\bm{\kappa})\rangle
\right)\right)\,,
\\
&&|\phi_3'(\bm{\kappa})\rangle =\frac{|\phi_4(\bm{\kappa})\rangle-e^{-i\varphi_2(\bm{\kappa})}|\phi_3(\bm{\kappa})\rangle}{\sqrt{2}}=\\
&&\frac{1}{\sqrt{2}}\left(\cos\frac{\theta(\bm{\kappa})}{2}|\psi_{+1}(\bm{\kappa})\rangle-\sin\frac{\theta(\bm{\kappa})}{2}|\psi_{+1}(\bm{\kappa} + \tilde{\mathbf{b}}_1/2)\rangle
+e^{-i\varphi_2(\bm{\kappa})}\left(\sin\frac{\theta(\bm{\kappa})}{2}|\psi_{-1}(\bm{\kappa} + \tilde{\mathbf{b}}_1/2)\rangle-\cos\frac{\theta(\bm{\kappa})}{2}|\psi_{-1}(\bm{\kappa})\rangle\right)\right)\,,\\
&&\mbox{\bf lower doublet:}\nonumber\\
&&|\phi_2'(\bm{\kappa})\rangle =\frac{|\phi_1(\bm{\kappa})\rangle+e^{-i\varphi_2(\bm{\kappa})}|\phi_2(\bm{\kappa})\rangle}{\sqrt{2}}=\\
&&\frac{1}{\sqrt{2}}\left(\cos\frac{\theta(\bm{\kappa})}{2}|\psi_{+1}(\bm{\kappa} + \tilde{\mathbf{b}}_1/2)\rangle-\sin\frac{\theta(\bm{\kappa})}{2}|\psi_{+1}(\bm{\kappa})\rangle
+e^{-i\varphi_2(\bm{\kappa})}\left(-\sin\frac{\theta(\bm{\kappa})}{2}|\psi_{-1}(\bm{\kappa})\rangle
+\cos\frac{\theta(\bm{\kappa})}{2}|\psi_{-1}(\bm{\kappa} + \tilde{\mathbf{b}}_1/2)\rangle\right)\right)\,,\\
&&|\phi_1'(\bm{\kappa})\rangle =\frac{|\phi_1(\bm{\kappa})\rangle-e^{-i\varphi_2(\bm{\kappa})}|\phi_2(\bm{\kappa})\rangle}{\sqrt{2}}=\\
&&\frac{1}{\sqrt{2}}\left(\cos\frac{\theta(\bm{\kappa})}{2}|\psi_{+1}(\bm{\kappa} + \tilde{\mathbf{b}}_1/2)\rangle+\sin\frac{\theta(\bm{\kappa})}{2}|\psi_{+1}(\bm{\kappa})\rangle
+e^{-i\varphi_2(\bm{\kappa})}\left(-\sin\frac{\theta(\bm{\kappa})}{2}|\psi_{-1}(\bm{\kappa})\rangle
-\cos\frac{\theta(\bm{\kappa})}{2}|\psi_{-1}(\bm{\kappa} + \tilde{\mathbf{b}}_1/2)\rangle\right)\right)\,.
\end{eqnarray}}}

\subsection{Periodicity of wavefunctions \texorpdfstring{$|\phi_i'(\text{{\boldmath$\kappa$}})\rangle$}{phi'i(kappa)}}
We now study the periodicity of the eigenstates $|\phi'_i(\bm{\kappa})\rangle$ along the both directions of the FMBZ. We start our discussion with the direction along $\tilde{\mathbf{b}}_1/2$ axis.
In order to understand what happens to $|\phi_i'(\bm{\kappa})\rangle$ under $\bm{\kappa} \rightarrow \bm{\kappa} + \tilde{\mathbf{b}}_1/2$, where $i=1,2,3,4$, we first note that 
\begin{eqnarray}
e^{i\varphi_2(\bm{\kappa} + \tilde{\mathbf{b}}_1/2)}&=&e^{i\varphi_2(\bm{\kappa})}\,,
\end{eqnarray}
which follows from the definition of $\varphi_2$ in Eq.~(\ref{eq:defphi2}) and the property of $\Delta_2(\bm{\kappa})$ given by Eq.~(\ref{eq:effectivec2tperiodicitydelta2}). Now, from Eq.~(\ref{eq:defepsilonprime}) we clearly have
\begin{eqnarray}
\epsilon'(\bm{\kappa} + \tilde{\mathbf{b}}_1/2)=-\epsilon'(\bm{\kappa})&
\Rightarrow& \cos\theta(\bm{\kappa} + \tilde{\mathbf{b}}_1/2)=-\cos\theta(\bm{\kappa})\,.
\end{eqnarray}
The spherical polar coordinate is defined in $\theta\in[0,\pi)$. 
Therefore, the angle $\theta(\bm{\kappa})$ will transform as:
\begin{eqnarray}
\theta(\bm{\kappa} + \tilde{\mathbf{b}}_1/2)&=&\pi-\theta(\bm{\kappa})
\end{eqnarray}
\begin{equation}
\cos\left[\frac{\theta(\bm{\kappa} + \tilde{\mathbf{b}}_1/2)}{2}\right]=\sin\left[\frac{\theta(\bm{\kappa})}{2}\right], ~~~~~~ \sin\left[\frac{\theta(\bm{\kappa} + \tilde{\mathbf{b}}_1/2)}{2}\right]=\cos\left[\frac{\theta(\bm{\kappa})}{2}\right].
\end{equation}
The wavefunctions $|\phi'_i(\bm{\kappa})\rangle$ will transform accordingly:
\begin{eqnarray}
\mbox{\bf upper doublet:}\;\;\;&&
|\phi_4'(\bm{\kappa} + \tilde{\mathbf{b}}_1/2)\rangle = |\phi'_4(\bm{\kappa})\rangle\,,
\\
&&|\phi_3'(\bm{\kappa} + \tilde{\mathbf{b}}_1/2)\rangle = -|\phi'_3(\bm{\kappa})\rangle\,,\\
\mbox{\bf lower doublet:}\;\;\;&&|\phi_2'(\bm{\kappa} + \tilde{\mathbf{b}}_1/2)\rangle = -|\phi'_2(\bm{\kappa})\rangle\,,
\\
&&|\phi_1'(\bm{\kappa} + \tilde{\mathbf{b}}_1/2)\rangle = |\phi'_1(\bm{\kappa})\rangle\,.
\end{eqnarray}
We see that one component is periodic ($|\phi'_1(\bm{\kappa})\rangle$ and $|\phi'_4(\bm{\kappa})\rangle$) and one anti-periodic ($|\phi'_2(\kappa)\rangle$ and $|\phi'_3(\bm{\kappa})\rangle$) in each doublet. 
To adopt the same boundary conditions, say periodic, we multiply the anti-periodic state with a smooth phase $e^{i\chi(\bm{\kappa})}$ such that $\chi(0)=0$ and $\chi(\bm{\kappa} + \tilde{\mathbf{b}}_1/2) = \chi(\bm{\kappa}) + \pi$. For example, we can choose this gauge phase factor as:
\begin{equation}
    \chi(\bm{\kappa}) = \bm{\kappa} \cdot \tilde{\mathbf{a}}_1  \,.
\end{equation}
So, finally, we define the periodic single particle states along $\bm{\kappa} \rightarrow \bm{\kappa} + \tilde{\mathbf{b}}_1/2$ as follows:
\begin{eqnarray}
\mbox{\bf upper doublet:}\;\;\;&&|\phi''_4(\bm{\kappa})\rangle = |\phi'_4(\bm{\kappa})\rangle\,,
\\
&&|\phi''_3(\bm{\kappa})\rangle = e^{i\chi(\bm{\kappa})}|\phi'_3(\bm{\kappa})\rangle\,,\\
\mbox{\bf lower doublet:}\;\;\;&&
|\phi''_2(\bm{\kappa})\rangle = e^{i\chi(\bm{\kappa})}|\phi'_2(\bm{\kappa})\rangle\,,
\\
&&|\phi''_1(\bm{\kappa})\rangle = |\phi'_1(\bm{\kappa})\rangle\,.
\end{eqnarray}
These states transform as follows under $C_{2z}T$ transformation:
\begin{eqnarray}
\mbox{\bf upper doublet:}\;\;\;&& C_{2z}T |\phi''_4(\bm{\kappa})\rangle = e^{i\varphi_2(\bm{\kappa})}|\phi''_4(\bm{\kappa})\rangle\,,
\\
&&C_{2z}T|\phi''_3(\bm{\kappa})\rangle = e^{-2i\chi(\bm{\kappa})}e^{i(\varphi_2(\bm{\kappa})+\pi)}|\phi''_3(\bm{\kappa})\rangle\,,\\
\mbox{\bf lower doublet:}\;\;\;&&
C_{2z}T|\phi''_2(\bm{\kappa})\rangle = e^{-2i\chi(\bm{\kappa})}e^{i\varphi_2(\bm{\kappa})}|\phi''_2(\bm{\kappa})\rangle
\,,\\
&&C_{2z}T|\phi''_1(\bm{\kappa})\rangle =e^{i(\varphi_2(\bm{\kappa})+\pi)} |\phi''_1(\bm{\kappa})\rangle\,.
\end{eqnarray}
The $C_{2z}T$ sewing matrices for the upper and lower doublets can be written as:
\begin{align}
    B^{C_{2z}T}_{\textrm{upper}}(\bm{\kappa}) &= \left(\begin{array}{cc}
        e^{i\varphi_2(\bm{\kappa})} & 0 \\
        0 & e^{-2i\chi(\bm{\kappa})}e^{i(\varphi_2(\bm{\kappa}) + \pi)}
    \end{array}
    \right)\,,\label{eqn:model_c2t_sewing_x1}\\
    B^{C_{2z}T}_{\textrm{lower}}(\bm{\kappa}) &= \left(\begin{array}{cc}
        e^{-2i\chi(\bm{\kappa})}e^{i\varphi_2(\bm{\kappa})} & 0 \\
        0 & e^{i(\varphi_2(\bm{\kappa}) + \pi)}
    \end{array}
    \right)\,.\label{eqn:model_c2t_sewing_x2}
\end{align}
These sewing matrices with this gauge choice can be used to calculate the determinant of the Wilson loop operator along the $\tilde{\mathbf{b}}_1/2$ direction.

Next, we study the periodicity of the wavefunctions $|\phi'_i(\bm{\kappa})\rangle$ under transformation $\bm{\kappa} \rightarrow \bm{\kappa} + \tilde{\mathbf{b}}_2$. The states $|\phi''_i(\bm{\kappa})\rangle$ defined in previous paragraphs are periodic along $\bm{\kappa}\rightarrow\bm{\kappa} + \tilde{\mathbf{b}}_1/2$, but they are not periodic along $\bm{\kappa} \rightarrow \bm{\kappa} + \tilde{\mathbf{b}}_2$. In order to evaluate the Wilson loop along $\tilde{\mathbf{b}}_2$ direction, we have to find the states which are periodic along $\bm{\kappa} \rightarrow \bm{\kappa} + \tilde{\mathbf{b}}_2$, which are not necessary to be periodic along $\bm{\kappa}\rightarrow\bm{\kappa} + \tilde{\mathbf{b}}_1/2$. From Eq.~(\ref{eq:effectivec2tperiodicitydelta2_y}), we notice that the phase $\varphi_2(\bm{\kappa})$ will transform as follows:
\begin{equation}
    e^{i\varphi_2(\bm{\kappa} + \tilde{\mathbf{b}}_2)} = -e^{i\varphi_2(\bm{\kappa})}e^{2i\bm{\kappa}\cdot \tilde{\mathbf{a}}_1}\,,
\end{equation}
and the angle $\theta(\bm{\kappa})$ is not changed when $\bm{\kappa} \rightarrow \bm{\kappa} + \tilde{\mathbf{b}}_2$. Then, by using Eq.~(\ref{eq:ChernBC}), we can obtain the wavefunctions $|\phi'_{i}(\bm{\kappa} + \tilde{\mathbf{b}}_2)\rangle$:
\begin{eqnarray}
\mbox{\bf upper doublet:}\;\;\;&& |\phi'_4(\bm{\kappa} + \tilde{\mathbf{b}}_2)\rangle = e^{-i\bm{\kappa}\cdot \tilde{\mathbf{a}}_1 }|\phi'_3(\bm{\kappa})\rangle\,,
\\
&&|\phi'_3(\bm{\kappa} + \tilde{\mathbf{b}}_2)\rangle = e^{-i\bm{\kappa}\cdot \tilde{\mathbf{a}}_1}|\phi'_4(\bm{\kappa})\rangle\,,\\
\mbox{\bf lower doublet:}\;\;\;&&
|\phi'_2(\bm{\kappa} + \tilde{\mathbf{b}}_2)\rangle = -e^{-i\bm{\kappa}\cdot \tilde{\mathbf{a}}_1}|\phi'_1(\bm{\kappa})\rangle\,,\\
&&|\phi'_1(\bm{\kappa} + \tilde{\mathbf{b}}_2)\rangle = -e^{-i\bm{\kappa}\cdot \tilde{\mathbf{a}}_1}|\phi'_2(\bm{\kappa})\rangle\,.
\end{eqnarray}
Similar to the case along $\tilde{\mathbf{b}}_1/2$ direction, these wavefunctions are not periodic along $\tilde{\mathbf{b}}_2$ as well. In order to obtain periodic wavefunctions, we define the following superposition states for both the upper and lower doublet states:
\begin{eqnarray}
\mbox{\bf upper doublet:}\;\;\;&& |\phi'''_4(\bm{\kappa})\rangle = e^{i\bm{\kappa}\cdot \tilde{\mathbf{a}}_2\left(\frac14 + \frac{\bm{\kappa}\cdot \tilde{\mathbf{a}}_1}{2\pi}\right)}\left(\cos\frac{\bm{\kappa} \cdot \tilde{\mathbf{a}}_2}{4}|\phi'_4(\bm{\kappa})\rangle - i\sin\frac{\bm{\kappa}\cdot \tilde{\mathbf{a}}_2}{4}|\phi'_3(\bm{\kappa})\rangle\right)\,,
\\
&&|\phi'''_3(\bm{\kappa})\rangle = e^{i\bm{\kappa}\cdot \tilde{\mathbf{a}}_2 \left(\frac14 + \frac{\bm{\kappa}\cdot \tilde{\mathbf{a}}_1  }{2\pi}\right)}\left(-i\sin\frac{\bm{\kappa} \cdot \tilde{\mathbf{a}}_2}{4}|\phi'_4(\bm{\kappa})\rangle + \cos\frac{\bm{\kappa}\cdot  \tilde{\mathbf{a}}_2}{4}|\phi'_3(\bm{\kappa})\rangle\right)\,,\\
\mbox{\bf lower doublet:}\;\;\;&&
|\phi'''_2(\bm{\kappa})\rangle = e^{i\bm{\kappa}\cdot \tilde{\mathbf{a}}_2\left(-\frac14 + \frac{\bm{\kappa}\cdot \tilde{\mathbf{a}}_1}{2\pi}\right)}\left(\cos\frac{\bm{\kappa} \cdot \tilde{\mathbf{a}}_2}{4}|\phi'_2(\bm{\kappa})\rangle - i\sin\frac{\bm{\kappa}\cdot  \tilde{\mathbf{a}}_2}{4}|\phi'_1(\bm{\kappa})\rangle\right)\,,\\
&&|\phi'''_1(\bm{\kappa})\rangle = e^{i\bm{\kappa}\cdot \tilde{\mathbf{a}}_2\left(-\frac14 + \frac{\bm{\kappa}\cdot \tilde{\mathbf{a}}_1}{2\pi}\right)}\left(-i\sin\frac{\bm{\kappa} \cdot \tilde{\mathbf{a}}_2}{4}|\phi'_2(\bm{\kappa})\rangle +\cos\frac{\bm{\kappa}\cdot \tilde{\mathbf{a}}_2}{4}|\phi'_1(\bm{\kappa})\rangle\right)\,.
\end{eqnarray}
It can be easily proved that these states satisfy the periodic condition $|\phi'''_i(\bm{\kappa} + \tilde{\mathbf{b}}_2)\rangle = |\phi'''_i(\bm{\kappa}) \rangle$. Thus, by applying the $C_{2z}T$ operator, these states will transform as the following equations:
\begin{eqnarray}
\mbox{\bf upper doublet:}\;\;\;&& C_{2z}T |\phi'''_4(\bm{\kappa})\rangle = e^{i\varphi_2(\bm{\kappa})}e^{-2i\bm{\kappa}\cdot\tilde{\mathbf{a}}_2\left(\frac14 + \frac{\bm{\kappa}\cdot\tilde{\mathbf{a}}_1}{2\pi}\right)}|\phi'''_4(\bm{\kappa})\rangle\,,
\\
&&C_{2z}T|\phi'''_3(\bm{\kappa})\rangle = e^{i(\varphi_2(\bm{\kappa}) + \pi)}e^{-2i\bm{\kappa}\cdot\tilde{\mathbf{a}}_2\left(\frac14 + \frac{\bm{\kappa}\cdot\tilde{\mathbf{a}}_1}{2\pi}\right)}|\phi'''_3(\bm{\kappa})\rangle\,,\\
\mbox{\bf lower doublet:}\;\;\;&&
C_{2z}T|\phi'''_2(\bm{\kappa})\rangle = e^{i\varphi_2(\bm{\kappa})}e^{-2i\bm{\kappa}\cdot\tilde{\mathbf{a}}_2\left(-\frac14 + \frac{\bm{\kappa}\cdot\tilde{\mathbf{a}}_1}{2\pi}\right)}|\phi'''_2(\bm{\kappa})\rangle
\,,\\
&&C_{2z}T|\phi'''_1(\bm{\kappa})\rangle =e^{i(\varphi_2(\bm{\kappa}) + \pi)}e^{-2i\bm{\kappa}\cdot\tilde{\mathbf{a}}_2\left(-\frac14 + \frac{\bm{\kappa}\cdot\tilde{\mathbf{a}}_1}{2\pi}\right)} |\phi'''_1(\bm{\kappa})\rangle\,.
\end{eqnarray}
And consequently, the $C_{2z}T$ sewing matrices can be written as follows:
\begin{align}
    B^{C_{2z}T}_{\textrm{upper}}(\bm{\kappa}) &= \left(
    \begin{array}{cc}
        e^{i\varphi_2(\bm{\kappa})}e^{-2i\bm{\kappa}\cdot\tilde{\mathbf{a}}_2\left(\frac14 + \frac{\bm{\kappa}\cdot\tilde{\mathbf{a}}_1}{2\pi}\right)} & 0 \\
        0 & e^{i(\varphi_2(\bm{\kappa}) + \pi)}e^{-2i\bm{\kappa}\cdot\tilde{\mathbf{a}}_2\left(\frac14 + \frac{\bm{\kappa}\cdot\tilde{\mathbf{a}}_1}{2\pi}\right)}
    \end{array}
    \right)\,,\label{eqn:model_c2t_sewing_y1}\\
    B^{C_{2z}T}_{\textrm{lower}}(\bm{\kappa}) &= \left(
    \begin{array}{cc}
        e^{i\varphi_2(\bm{\kappa})}e^{-2i\bm{\kappa}\cdot\tilde{\mathbf{a}}_2\left(-\frac14 + \frac{\bm{\kappa}\cdot\tilde{\mathbf{a}}_1}{2\pi}\right)} & 0 \\
        0 & e^{i(\varphi_2(\bm{\kappa}) + \pi)}e^{-2i\bm{\kappa}\cdot\tilde{\mathbf{a}}_2\left(-\frac14 + \frac{\bm{\kappa}\cdot\tilde{\mathbf{a}}_1}{2\pi}\right)}
    \end{array}
    \right)\,.\label{eqn:model_c2t_sewing_y2}
\end{align}
Similar to Eqs.~(\ref{eqn:model_c2t_sewing_x1}) and (\ref{eqn:model_c2t_sewing_x2}), we will use these sewing matrices when evaluating the determinant of the Wilson loop operators along the $\tilde{\mathbf{b}}_2$ direction.

\subsection{Computing the determinant of the Wilson loop operator \texorpdfstring{$W$}{W}}\label{app:c2twilsonloopdet}

\begin{figure}[!htbp]
    \centering
    \includegraphics[width=0.9\linewidth]{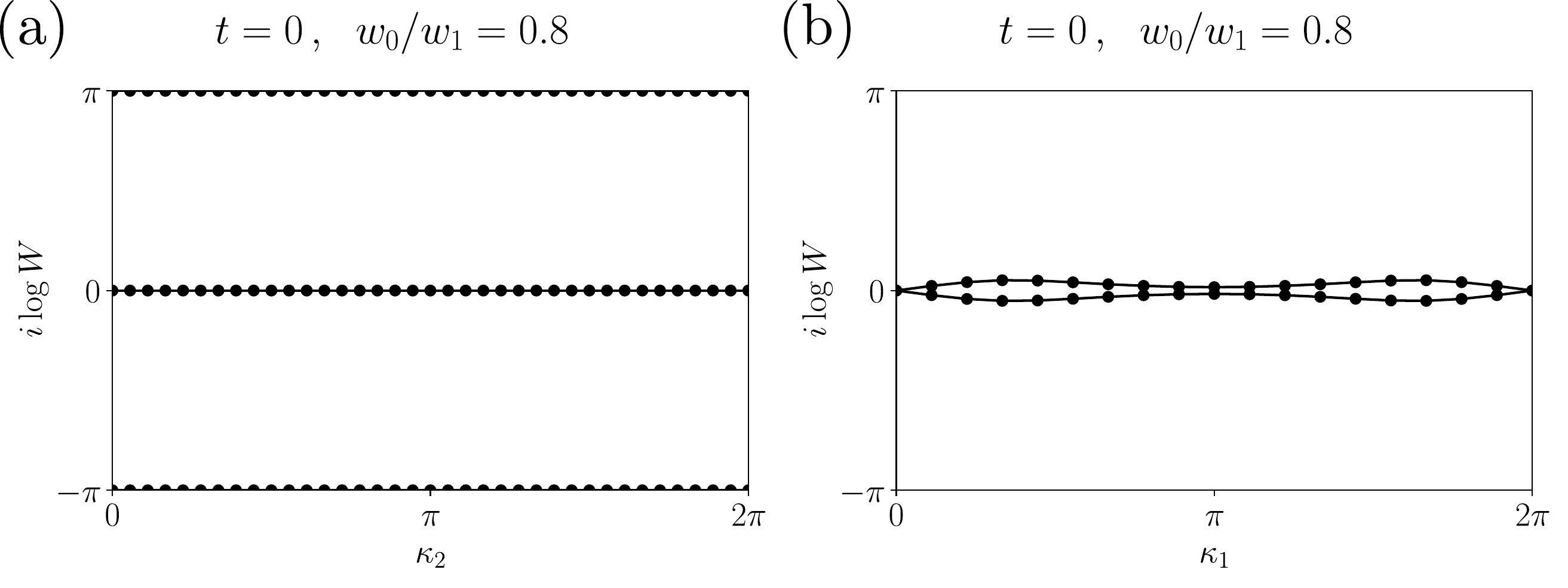}
    \caption{The Wilson loops of the two occupied HF bands evaluated from the $C_{2z}T$ stripe phase obtained at $w_0/w_1=0.8$ at $t=0$. (a) Wilson loop along the $\tilde{\mathbf{b}}_1/2$ direction. This figure is similar to Fig.~\ref{fig:wilson}~(d) but with a mesh in momentum space doubled in both directions ($36\times 36$ as opposed to $18 \times 18$) and assuming spin and valley polarization. (b) Wilson loop along the $\tilde{\mathbf{b}}_2$ direction. The results are as expected from the derivation of presented in App.~\ref{app:c2twilsonloopdet}.}
    \label{fig:app_wilson_two_direction}
\end{figure}

The determinant of the Wilson loop operator can be evaluated following the derivation provided in Sec.~V~C of Ref.~\cite{xie_superfluid_2020}. We first evaluate the Wilson loop along $\tilde{\mathbf{b}}_1/2$. As defined previously, the basis $|\phi_4''(\bm{\kappa})\rangle,|\phi_3''(\bm{\kappa})\rangle$ and $|\phi_2''(\bm{\kappa})\rangle,|\phi_1''(\bm{\kappa})\rangle$ is periodic under $\bm{\kappa}\rightarrow\bm{\kappa} + \tilde{\mathbf{b}}_1/2$. It can been proved that the $C_{2z}T$ sewing matrix of a two band system is deeply related tp its non-Abelian Berry connection. If the sewing matrix $B^{C_{2z}T}(\bm{\kappa})$ can be written as:
\begin{equation}
    B^{C_{2z}T}(\bm{\kappa}) = \left(
    \begin{array}{cc}
        e^{i\vartheta_1(\bm{\kappa})} & 0 \\
        0 & e^{i\vartheta_2(\bm{\kappa})}
    \end{array}
    \right)\,,
\end{equation}
then the non-Abelian Berry connection has the following form:
\begin{equation}\label{eq:c2tnonabelianconnection}
\mathbf{A}(\bm{\kappa}) = \left(\begin{array}{cc}\frac{1}{2}\partial_{\bm{\kappa}} \vartheta_{1}(\bm{\kappa}) & i \mathbf{a}(\bm{\kappa})e^{i\frac12 (\vartheta_1(\bm{\kappa}) - \vartheta_2(\bm{\kappa)})} \\ -i \mathbf{a}(\bm{\kappa})e^{-i\frac12 (\vartheta_1(\bm{\kappa}) - \vartheta_2(\bm{\kappa}))} & \frac12 \partial_{\bm{\kappa}} \vartheta_{2}(\bm{\kappa})\end{array}\right)\,.
\end{equation}
As we have mentioned in the main text, we use $\kappa_1$ and $\kappa_2$ to parameterize the momentum $\bm{\kappa}$ in the FMBZ as $\bm{\kappa} = \frac{\kappa_1}{2\pi}\frac{\tilde{\mathbf{b}}_1}{2} + \frac{\kappa_2}{2\pi}\tilde{\mathbf{b}}_2$.
We choose the path of the Wilson loop $c$ along the direction of $\tilde{\mathbf{b}}_1/2$ with a fixed value of $\kappa_2$. The non-Abelian Wilson loop can be written as:
\begin{equation}
W(\kappa_2) = \mathcal{P}\exp\left(-i\oint_c d\bm{\kappa}\cdot \mathbf{A}(\bm{\kappa})\right) = \mathcal{P}\exp\left(-i\int_0^{2\pi} d\kappa_1 A_1(\kappa_1,\kappa_2)\right).
\end{equation}
Therefore, the determinant of the lower doublet Wilson loop will be given by:
\begin{align}
\det W(\kappa_2) &= \exp\left(-i\int_0^{2\pi} d\kappa_1~{\rm Tr}{A_1}(\kappa_1, \kappa_2)\right)\nonumber\\
&=e^{-\frac{i}{2}\left(\varphi_2(\tilde{\mathbf{b}}_1/2)-\varphi_2(0)+\varphi_2(\tilde{\mathbf{b}}_1/2)-\varphi_2(0)-2\chi(\tilde{\mathbf{b}}_1/2)+2\chi(0)\right)}\nonumber\\
&=e^{i\left(\chi(\tilde{\mathbf{b}}_1/2)-\chi(0)\right)}\nonumber\\
&=-1.
\end{align}
The $C_{2z}T$ symmetry also requires $W(\kappa_2)$ and $W^*(\kappa_2)$ to have the same eigenvalue spectrum \cite{xie_superfluid_2020}. Therefore, the only possible eigenvalues of $W$ are $+1$ and $-1$ independent of $\kappa_2$ for the lower doublet. This is identical to the numerical evaluation of the Wilson loop of the $C_{2z}T$ stripe phase shown in Fig.~\ref{fig:wilson}~(d) of Sec.~\ref{sec:c2t_topo_phase_diagram} (see also Fig.~\ref{fig:app_wilson_two_direction}~(a)). Similarly, we are also able to show that the two eigenvalues of the the upper doublet Wilson loop are $\pm 1$, leading to flat Wilson loops for both the occupied and unoccupied bands.
  
We can also evaluate the Wilson loop along the other direction. The path of the Wilson loop $c$ is chosen along $\tilde{\mathbf{b}}_2$ with a fixed value of $\kappa_1$, and we choose the basis $|\phi_i'''(\bm{\kappa})\rangle$, which is periodic along $\tilde{\mathbf{b}}_2$ direction. Thus, the determinant of the lower doublet Wilson loop can be expressed as:
\begin{align}
    \det W(\kappa_1) &= \exp\left(-i\int_0^{2\pi} d\kappa_2\,{\rm Tr}A_2(\kappa_1, \kappa_2) \right)\nonumber\\
    &= e^{-i(\varphi_2(\tilde{\mathbf{b}}_2) - \varphi_2(0))}e^{i4\pi\left(-\frac{1}{4}+\frac{\kappa\cdot \tilde{\mathbf{a}}_1}{2\pi}\right)}\nonumber\\
    &= -e^{-2i\bm{\kappa}\cdot \tilde{\mathbf{a}}_1} e^{-i\pi} e^{2i\bm{\kappa}\cdot \tilde{\mathbf{a}}_1} \nonumber\\
    &= +1\,.
\end{align}
Thus, the two eigenvalues of $W(\kappa_1)$ are complex conjugation of each other at each $\kappa_1$ value. This also agrees with the numerical result we obtained from the self-consistent HF state, as shown in Fig.~\ref{fig:app_wilson_two_direction}~(b). We can also show that the determinant of upper doublet Wilson loop is $+1$ as well.

\section{Dirac nodes in the \texorpdfstring{$C_{2z}T$}{C2T} stripe phase}\label{sec:dirac_app}
\subsection{Chirality of Dirac nodes}\label{sec:chirality_app}
For a two band system, $C_{2z}T$ transformation can be represented by complex conjugation operator $\mathcal{K}$ with proper basis choice. With this basis choice, the $\sigma_y$ terms are forbidden in the Hamiltonian. Thus, the chirality of a Dirac node can be defined by the winding number of the state on the $xz$ plane of the Bloch sphere, along a circle surrounding the Dirac node.

To identify the chirality of several Dirac nodes between the $i$-th and $(i + 1)$-th Hartree-Fock bands, we have to study the wavefunctions of the HF states around these nodes. We start with finding a patch $\Pi$ in the Brillouin zone, in which the two bands are completely gapped from other bands, while also containing all the Dirac nodes between them. Writing these two bands as an effective two-band Hamiltonian which satisfies $C_{2z}T$ symmetry will help us determine the chirality of the Dirac nodes. We first represent the HF states wavefunctions as vectors $\Phi_{i}(\bm{\kappa})$ in the plane wave basis defined in Eq.~(\ref{eqn:hf_wf_plane}) of App.~\ref{sec:wilson_app}. Next, we choose a point $\bm{\kappa}_0$ in $\Pi$ as the reference point, and we use the two band wavefunctions at $\bm{\kappa}_0$ as the basis for our effective two-band model on $\Pi$. To determine whether $\bm{\kappa}_0$ is a good choice as the reference point, we can define the following quantity $\mathcal{N}_i(\bm{\kappa})$ to quantify the wavefunction overlap between the momentum point $\bm{\kappa}$ and the reference point $\bm{\kappa}_0$ in the $i$-th and $(i + 1)$-th bands:
\begin{equation}
    \mathcal{N}_i(\bm{\kappa}) = \frac12 \sum_{j,k = 0, 1}|\Phi^\dagger_{i + j}(\bm{\kappa}) \Phi_{i+k}(\bm{\kappa}_0)|^2\,.
\end{equation}
If the value of $\mathcal{N}_i(\bm{\kappa})$ is close to $1$ on the chosen patch $\Pi$, the Hilbert space spanned by the two bands over $\Pi$ will also be close to the Hilbert space spanned by the two bands at the reference point $\bm{\kappa}_0$. Therefore, it is reasonable to choose $\Phi_{i}(\bm{\kappa}_0)$ and $\Phi_{i+1}(\bm{\kappa}_0)$ as momentum independent basis, and we can project the wavefunction $\Phi_i(\bm{\kappa})$ onto these two states. The two coefficients $\alpha$ and $\beta$ for such a projection are given by the following equations:
\begin{align}
    \alpha(\bm{\kappa}) &= \Phi^\dagger_i(\bm{\kappa}) \Phi_{i}(\bm{\kappa}_0)\,\\
    \beta(\bm{\kappa}) &= \Phi^\dagger_i(\bm{\kappa}) \Phi_{i+1}(\bm{\kappa}_0)\,.
\end{align}
Because of the $C_{2z}T$ symmetry, both coefficients are real. Thus, the vector $\left(\alpha, \beta\right)$ will lie in the $xz$ plane of the Bloch sphere if $|\alpha(\bm{\kappa})|^2 + |\beta(\bm{\kappa})|^2 = 1$. This is usually a reasonable assumption when $\bm{\kappa}$ is not far from the reference point $\bm{\kappa}_0$, and it will be checked numerically in the following calculation. Under this assumption, the effective two-band model can be captured by the Hamiltonian $h(\bm{\kappa}) = (\alpha^2(\bm{\kappa}) - \beta^2(\bm{\kappa}))\sigma_z + 2\alpha(\bm{\kappa})\beta(\bm{\kappa})\sigma_x$. We now use the angle $\varphi^{i,i+1}_{xz}(\bm{\kappa})$ to describe the direction of this Hamiltonian:
\begin{equation}
    \varphi^{i,i+1}_{xz}(\bm{\kappa}) = {\rm arg}\left[(\alpha(\bm{\kappa}) + i\beta(\bm{\kappa}))^2\right]\,.
\end{equation}
This quantity measures the angle between the $+z$ axis and the direction on Bloch sphere. The winding number of this angle around a Dirac point measures the ``chirality'' of this node between $i$-th band and $(i+1)$-th band.

In the following Apps.~\ref{sec:add_lam1_1_results} and \ref{sec:add_lam2_1_results}, we use the method discussed in this subsection to study the chiralities of the Dirac nodes.

\subsection{Motion of Dirac nodes}\label{sec:app_dirac_motion}
In this subsection, we provide additional information about the motion and the chirality of the Dirac nodes in the $C_{2z}T$ stripe phase obtained with flat band kinetic energy ($t=1$) and at $w_0/w_1 = 0.8$ as completing our discussion in Sec.~\ref{sec:dirac} of the main text. 

\subsubsection{Non-Abelian braiding of Dirac nodes}\label{sec:add_lam2_1_results}

\begin{figure}[t]
    \centering
    \includegraphics[width=0.25\linewidth]{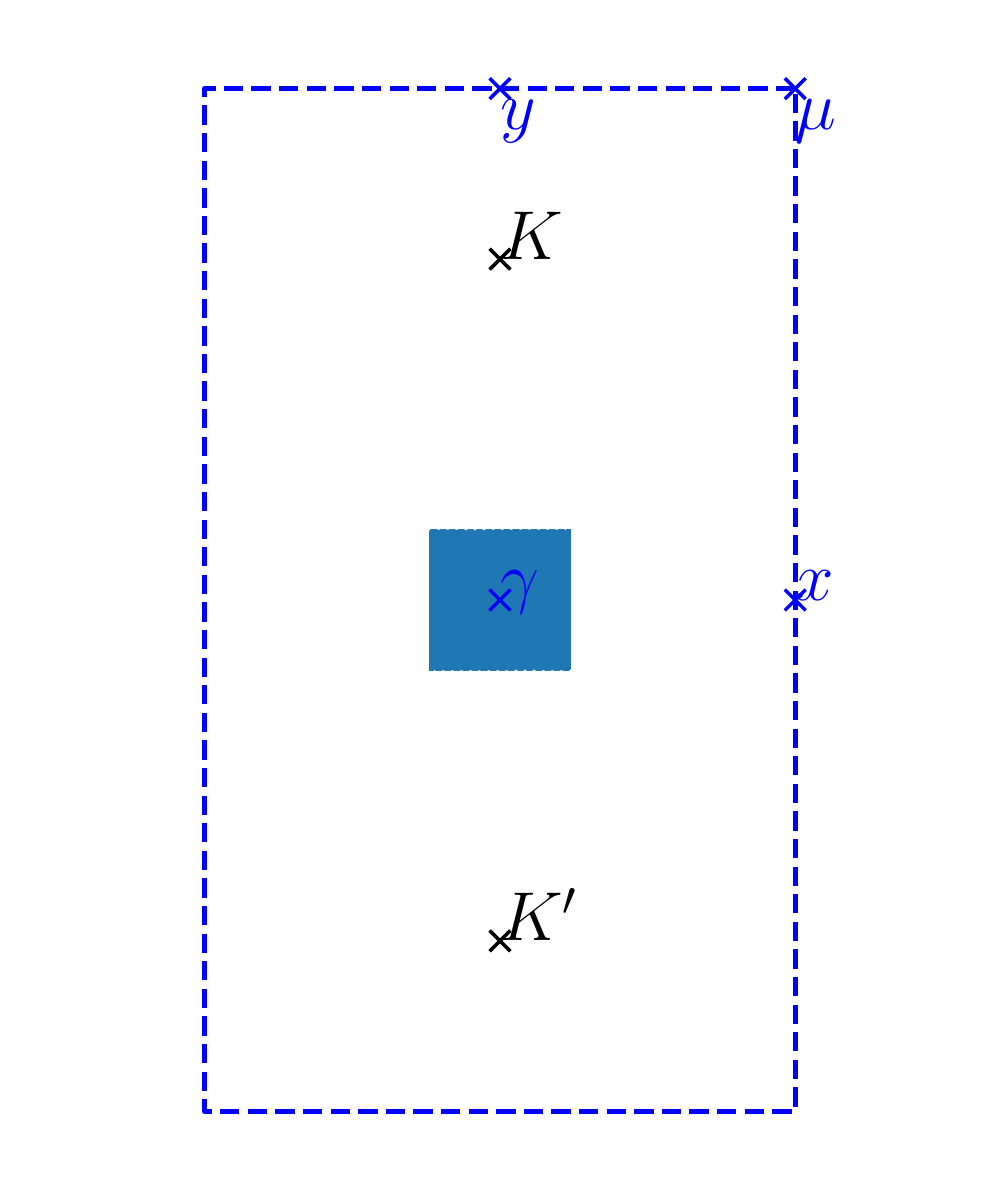}
    \caption{The patch (blue square) around the $\gamma$ point used to evaluate the Dirac nodes chirality. Data shown in Figs.~\ref{fig:non_abelian_nodes}, \ref{fig:add_chirality_12}-\ref{fig:add_chirality_34} are computed in this patch.}
    \label{fig:patch}
\end{figure}

We first study the Dirac nodes motion in detail along the first path introduced in Sec.~\ref{sec:dirac} [$(\lambda_1, \lambda_2) = (0, 0)\rightarrow (0, 1)\rightarrow (1, 1)$], where the parameters $\lambda_1$ and $\lambda_2$ are defined in Eq.~(\ref{eqn:interpolation_stripe_ham}) in the main text. As mentioned in the main text, we start from the non-interacting bands at $\lambda_1 = \lambda_2 = 0$, which has two Dirac nodes connecting the second and the third bands with the same chirality at $K$ and $K'$ points. When $\lambda_1 = 0$ and $\lambda_2$ is increased from $0$ to $1$, the two Dirac nodes are moved into a small region around $\gamma$ point in the FMBZ, and no other Dirac node between the second and the third bands are generated.
Hence, we choose a patch around $\gamma$ point as shown in Fig.~\ref{fig:patch}, and we use the method described in App.~\ref{sec:hf_app_high_sym_lines} to compute the HF bands and wavefunctions in this patch with a higher resolution. 
With the increasing value of $\lambda_1$, the two Dirac nodes move towards each other and annihilated at around $\lambda_1 \approx 0.05$. Meanwhile, we also observed that there are other pairs of Dirac nodes between the first and second bands, and between the third and fourth bands, which can be observed in Figs.~\ref{fig:add_chirality_12} and \ref{fig:add_chirality_34}.

\begin{figure}[!htbp]
    \centering
    \includegraphics[width=0.7\linewidth]{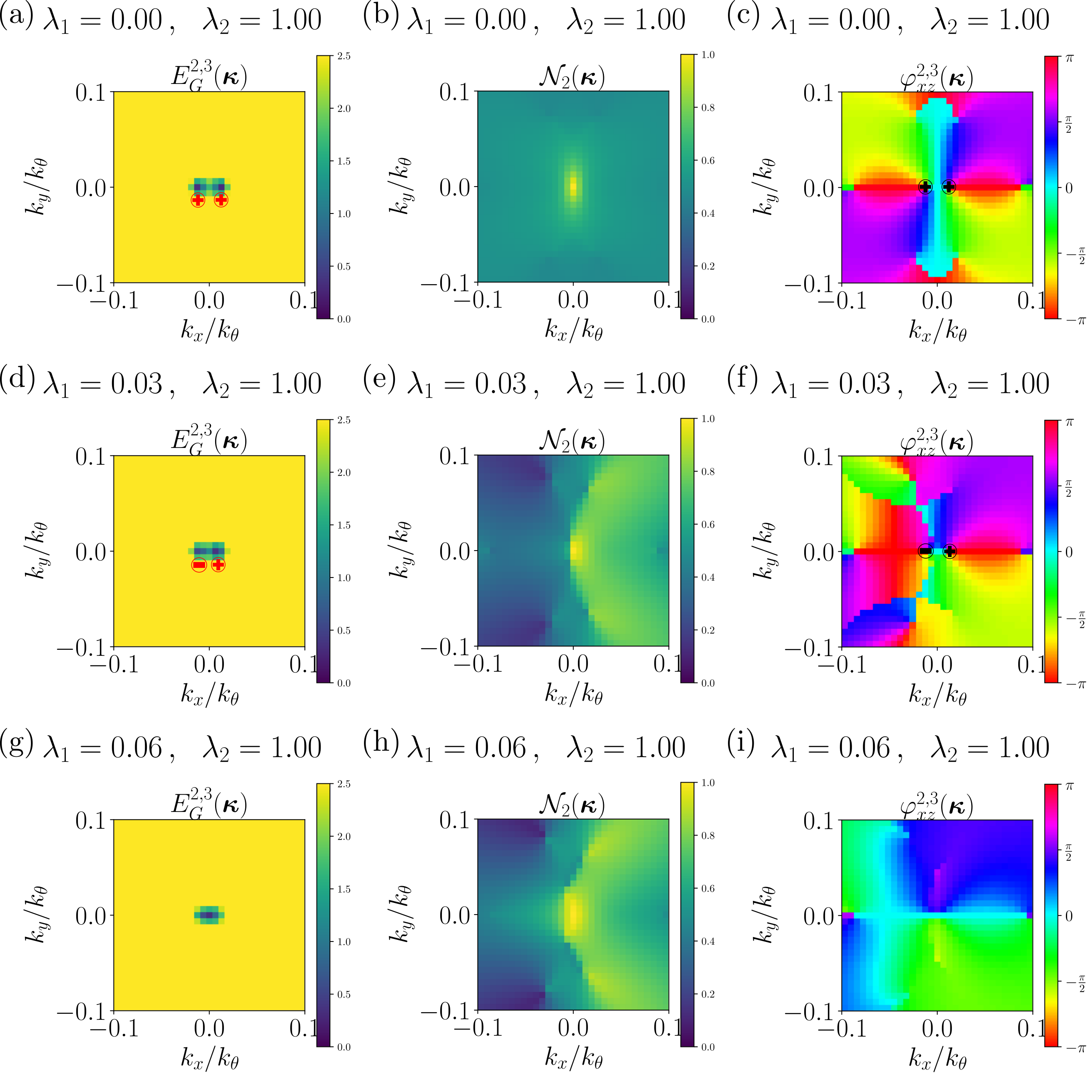}
    \caption{(a) The direct gap between the second and the third HF bands at $\lambda_1 = 0$ and $\lambda_2 = 1$. The symbols $\oplus$ and $\ominus$ in red represent the Dirac nodes with opposite chiralities. (b) The wavefunction overlap of the first and second HF bands between a given momentum $\bm{\kappa}$ and the reference point $\bm{\kappa}_0$ at $\lambda_1 = 0$ and $\lambda_2 = 1$. (c) The angle $\varphi_{xz}^{2,3}(\bm{\kappa})$ computed in the same FMBZ patch. The black symbols $\oplus$ and $\ominus$ stand for the chirality of the corresponding Dirac nodes. Similarly, subfigures (d-f) are computed at $\lambda_1 = 0.03$ and $\lambda_2 = 1$, and subfigures (g-i) are computed at $\lambda_1 = 0.06$ and $\lambda_2 = 1$.}
    \label{fig:add_chirality_23}
\end{figure}

To determine the chiralities of these Dirac nodes between different bands, we choose $\bm{\kappa}_0 = \gamma$ as the reference point, and we computed the values of $\mathcal{N}_{i}(\bm{\kappa})$ and $\varphi_{xz}^{i,i+1}(\bm{\kappa})$ with $i=1,2,3$ at $\lambda_1 = 0, 0.03, 0.06$, and $\lambda_2 = 1$. The results are shown in Figs.~\ref{fig:add_chirality_12}, \ref{fig:add_chirality_23} and \ref{fig:add_chirality_34}. As shown in Fig.~\ref{fig:add_chirality_23}(c), the two Dirac nodes between the second and third bands carry the same chirality when $\lambda_2 = 0$. However, one of these two Dirac nodes flipped its chirality if $\lambda_2$ is raised to $0.03$ as shown in Figs.~\ref{fig:add_chirality_23}(f). Finally, the two Dirac nodes annihilate with each other as shown in Figs.~\ref{fig:add_chirality_23}(g-i), and thus the charge gap can be created between the second and the third HF bands.
In the meantime, another pair of Dirac nodes with opposite chiralities are created between the first and the second bands, which are shown in Figs.~\ref{fig:add_chirality_12}(a) and (d). And in Figs.~\ref{fig:add_chirality_12}(g-i), we can also observe that one of the nodes also flipped its chirality with the increased value of $\lambda_1$. 
Another pairs of nodes can also be observed between the first and the second bands. They carry opposite chiralities when $\lambda_1 = 0$, and one of them also changed the chirality when $\lambda_1$ is increased to $0.06$, as can be seen in Figs.~\ref{fig:add_chirality_12}(a), (d) and (g).
We also notice similar phenomenon between the third and four bands. In Fig.~\ref{fig:add_chirality_34}(a), two nodes with opposite chiralities can be found when $\lambda_1 = 0$. And in Fig.~\ref{fig:add_chirality_34}(g), they carry the same chirality when the value of $\lambda_1$ increases to $0.06$.
The change of chiralities demonstrated the non-Abelian nature of the Dirac node charges in $C_{2z}T$ symmetric multi-band systems.

\begin{figure}[!htbp]
    \centering
    \includegraphics[width=0.7\linewidth]{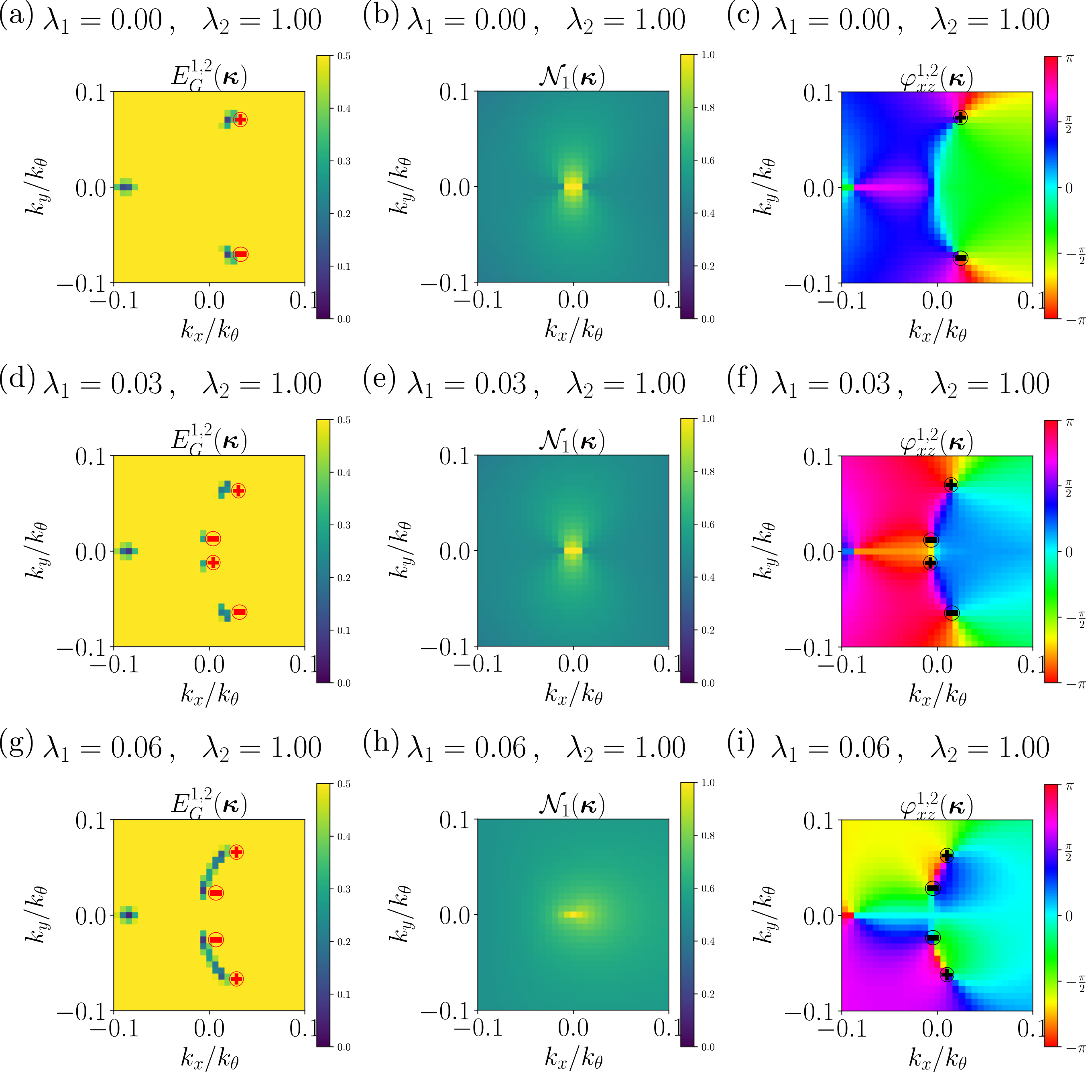}
    \caption{Similar to Fig.~\ref{fig:add_chirality_12}, subfigure (a-c) provide the direct gap $E_G^{1,2}(\bm{\kappa})$, the wavefunction overlaps $\mathcal{N}_1(\bm{\kappa})$ and the angle $\varphi_{xz}^{1,2}(\bm{\kappa})$ between the second and the third HF bands at $\lambda_1 = 0$ and $\lambda_2 = 1$. Subfigures (d-f) are obtained at $\lambda_1 = 0.03$, $\lambda_2 = 1$, and subfigure (g-i) are obtained at $\lambda_1 = 0.06$, $\lambda_2 = 1$.}
    \label{fig:add_chirality_12}
\end{figure}

\begin{figure}[!htbp]
    \centering
    \includegraphics[width=0.7\linewidth]{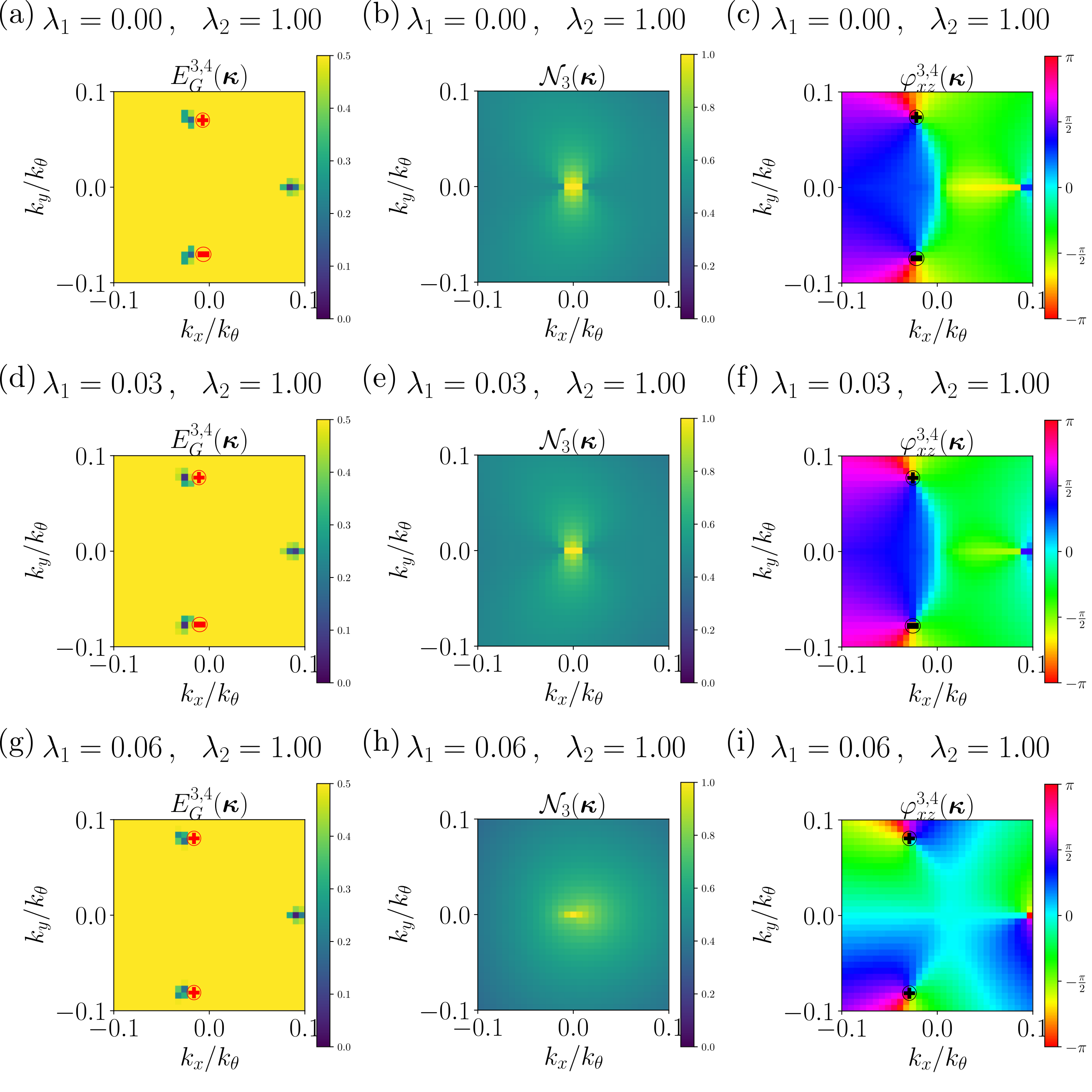}
    \caption{Subfigure (a-c) provide the direct gap $E_G^{3,4}(\bm{\kappa})$, the wavefunction overlaps $\mathcal{N}_3(\bm{\kappa})$ and the angle $\varphi_{xz}^{3,4}(\bm{\kappa})$ between the third and the fourth HF bands at $\lambda_1 = 0$ and $\lambda_2 = 1$. Subfigures (d-f) are obtained at $\lambda_1 = 0.03$, $\lambda_2 = 1$, and subfigure (g-i) are obtained at $\lambda_1 = 0.06$, $\lambda_2 = 1$.}
    \label{fig:add_chirality_34}
\end{figure}

\subsubsection{Four Dirac nodes annihilation}\label{sec:add_lam1_1_results}

We also study the Dirac nodes motion and annihilation along the second path introduced in Sec.~\ref{sec:dirac} [$(\lambda_1, \lambda_2) = (0, 0)\rightarrow (1, 0)\rightarrow (1, 1)$]. As mentioned in the main text, when the value of $\lambda_1$ is increased from $0$ to $1$, the non-interacting TBG bands is turned into the ``strong interacting bands''.
In this case,  we will show that the Dirac nodes are located at the high symmetry momenta $\Gamma$, $M$, and $K$ (and $K'$)~\cite{kang_cascade_2021} and calculate the associated winding numbers.
For convenience, the Chern states $| \psi_{\mu, \eta, s}(\bm{\vk}) \rangle$ are chosen to be the Bloch state basis and their gauge is fixed as in Ref.~\cite{kang_nonabelian_2020} (also see Sec.~\ref{sec:app_wilson_loop}). 
Under the transformation $C_{2z} T$, 
\begin{align}
	C_{2z} T | \psi_{\mu, \eta, s}(\bm{\vk}) \rangle = | \psi_{-\mu, \eta, s}(\bm{\vk}) \rangle  \ . 
\end{align}
Under the particle-hole transformation,
\begin{align}
	& P | \psi_{\mu, \eta, s}(\bm{\vk}) \rangle = e^{i \theta_{\mu}(\mk)} | \psi_{\mu, \eta, s}(-\bm{\vk}) \rangle \ ,
\end{align}
Since $P$ commutes with $C_{2z} T$, we have $\theta_{+1}(\mk) = - \theta_{-1}(\mk)$. In addition, because $P^2 = -1$, $\theta_{\mu}(\mk) = \pi - \theta_{\mu}(-\mk)$. When $\mk = \Gamma$ or $M$, $\mk = - \mk$, and these constraints lead to $e^{i \theta_{\mu}(\mk)} = - e^{- i\theta_{\mu}(-\mk)}$. Numerically, we found $e^{i \theta_{+1}(\Gamma)} = -i$ and $e^{i \theta_{+1}(M)} = i$. Similar to the main text, we introduce the creation and annihilation operators $d^{\dagger}_{\mk, \mu, \eta, s}$ ($d_{\mk, \mu, \eta, s}$) so that $d^{\dagger}_{\mk, \mu, \eta, s}|\emptyset \rangle = | \psi_{\mu, \eta, s}(\mk) \rangle$. Notice that they can be expressed in terms of the creation/annihilation operators of the eigenstate basis introduced in Eq.~\ref{eqn:h0}: 
\begin{align}
	d^{\dagger}_{\mk, +1, \eta, s} = \frac1{\sqrt{2}} e^{i \vartheta(\mk)} \left( c^{\dagger}_{\mk, 1, \eta, s} +   i c^{\dagger}_{\mk, 2, \eta, s} \right) \ , \quad \mbox{and} \qquad d^{\dagger}_{\mk, -1, \eta, s} = \frac1{\sqrt{2}} e^{-i \vartheta(\mk)} \left( c^{\dagger}_{\mk, 1, \eta, s} -  i c^{\dagger}_{\mk, 2, \eta, s} \right) 
\end{align}
where the phase factor $e^{i \vartheta(\mk)}$ is inserted to guarantee the constructed Chern states are smooth in $\mk$ and satisfy the boundary conditions Eq.~(\ref{eq:ChernBC}).

Next, we consider the properties of the Dirac nodes when the system is in the strong coupling limit without breaking the translation symmetry. 
As we have mentioned in the main text, when  $\lambda_1$ in Eq.~(\ref{eqn:interpolation_stripe_ham}) increases from $0$ to $1$ and $\lambda_2$ is kept to be $0$, the non-interacting TBG bands are turned into the ``strong interacting bands'' without breaking the translation symmetry. Because both the spin and valley are polarized for the $C_{2z}T$ stripe phase,  $\mathcal{H}^{(H)}$ and $\mathcal{H}^{(F)}$ in Eq.~(\ref{eqn:interpolation_stripe_ham}) are diagonal in spin and valley indices. Furthermore, in the strong coupling limit, we neglect the dispersion of the narrow bands, i.e.~$\epsilon_{\bm{\vk}} = 0$. 

Now, we focus on the translationally invariant part of the Hamiltonian in the strong coupling limit, i.e.~$\delta_{bb'}\left(\mathcal{H}^{(H)}(\bm{\kappa}) + \mathcal{H}^{(F)}(\bm{\kappa})\right)_{bm\eta s,b'n\eta's'}$  in Eq.~(\ref{eqn:interpolation_stripe_ham}). As opposed to Sec.~\ref{sec:model} in the main text, here we use the Chern basis described above to study the properties of this part of the Hamiltonian. Since it does not break the translation symmetry, we can remove the $b$ subscript and replace the label $\bm{\kappa}$ for momentum  in FMBZ by $\mk$ that ranges over the whole MBZ.  For particular spin $s$ and valley $\eta$, $\left(\mathcal{H}^{(H)}(\mk) + \mathcal{H}^{(F)}(\mk)\right)_{m\eta s, n\eta s}$ becomes a $2 \times 2$ matrix. Thus, in the basis of Chern states, the effective Hamiltonian containing only this translationally invariant part can be written as 
\begin{align}
	H_{\rm eff} = \sum_{\mk \in MBZ, \eta, s} \sum_{\mu\nu} d^{\dagger}_{\mk, \mu, \eta, s} \vec{n}(\mk) \cdot \vec{\sigma}_{\mu\nu} d_{\mk, \nu, \eta, s}
\end{align}
For simplicity, we drop the valley and spin indices from the vector $\vec{n}$. Since the Chern basis is continuous but not periodic in $\mk$, the same is true for the vector $\vec{n}(\mk)$. Because $H_{\rm eff}$ is invariant under $C_{2z} T$ transformation, $\vec n(\mk) = (n_{1}(\mk), n_{2}(\mk), 0)$, i.e., the third component of $\vec{n}(\mk)$ must vanish. 

Here, we outline the argument for the winding numbers at the high symmetry momenta, while the details can be found in the appendix of Ref.~\cite{kang_cascade_2021}. 

When $\mk$ is near $\Gamma$, we expand the two components of $\vec{n}(\mk)$ to the powers of $\mk$ :
\begin{align}
	n_i(\mk \approx \Gamma) = n_i^{(0)} + \sum_{a} n_{i, a}^{(1)} k_a +  \sum_{a, b} n_{i, a b}^{(2)} k_a k_b + \sum_{a b c} n_{i, a b c}^{(3)} k_a k_b k_c + O(\mk^4)
\end{align}
Since $H_{\rm eff}$ is the effective Hamiltonian in the strong coupling limit, it is particle-hole symmetric. Under the particle-hole transformation $P$, $d_{\Gamma, \mu, \eta, s} \longrightarrow i (\sigma_3)_{\mu\nu} d_{\Gamma, \nu, \eta, s}$ and $\mk \longrightarrow - \mk$. As a consequence, $n_i^{(0)} = 0$ and thus a Dirac node appears at $\Gamma$. In addition, the particle-hole symmetry also gives $n_{i, a b}^{(2)} = 0$. Furthermore, the Bloch states at $\Gamma$ are invariant under $C_3$ transformation, leading to $n^{(1)}_{i, a} = 0$. This implies that the effective Hamiltonian close to $\Gamma$ is dominated by $\mk^3$ terms, giving the winding of $\pm 3$ for the Dirac node at $\Gamma$.

\begin{figure}[htb]
	\centering
    \includegraphics[width=0.9\linewidth]{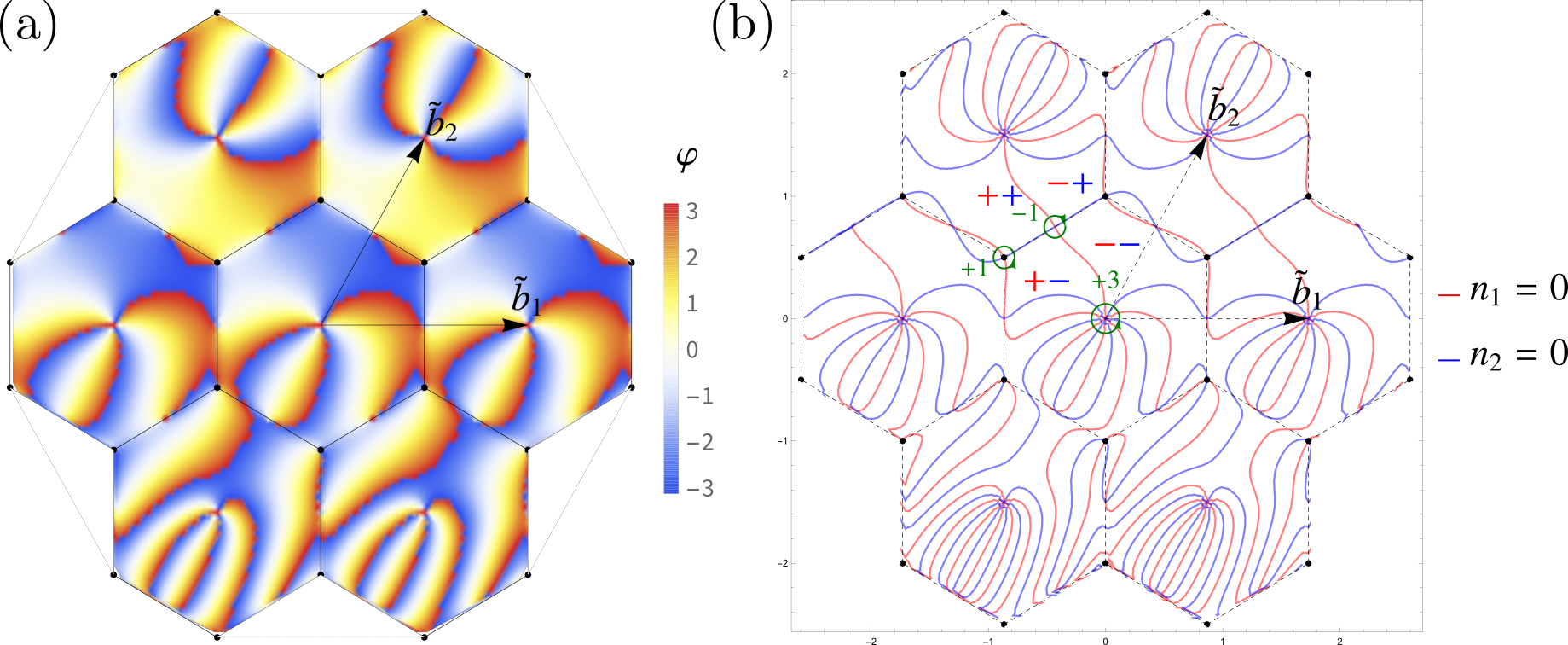}
	\caption{(a) The phase $\varphi(\mk)$ of $n_1(\mk) + i n_2(\mk)$ over the first and extended MBZs.  (b) the contours where (red) $n_1(\mk)$ and (blue) $n_2(\mk)$ vanish and their signs in different region. The two bands become degenerate and a Dirac node appears at the intersection point of the ed and blue curves. The winding numbers of the Dirac nodes are shown in dark green. }
	\label{Fig:Vort}
\end{figure}

Similarly, due to the particle-hole symmetry, the winding number of the Dirac node at $M$ can be shown to be $\pm 1$. Additionally, the Dirac node at $K$ also has a winding number of $\pm 1$ by $C_3$ symmetry. However, we emphasize that this argument does not give the sign of the winding numbers at these high symmetry momenta. With the Chern basis constructed in Ref.~\cite{kang_nonabelian_2020}, the winding numbers of Dirac nodes can be numerically obtained. Since the detailed calculation has already been presented in Ref.~\cite{kang_nonabelian_2020}, here we will only summarize the results in Fig.~\ref{Fig:Vort}. In Fig.~\ref{Fig:Vort} (a), we plotted the phase $\varphi(\mk) = \mathrm{arg}(n_1(\mk) + i n_2(\mk))$ with $-\pi < \varphi(\mk) \leq \pi$. Fig.~\ref{Fig:Vort} (b) shows the colored curves along which $n_1(\mk) = 0$ and $n_2(\mk) = 0$.  Fig.~\ref{Fig:Vort} has demonstrated clearly that the the Dirac nodes at $\Gamma$, $M$, and $K$ have the winding numbers of $+3$, $-1$, and $+1$ respectively.

\begin{figure}[t]
    \centering
    \includegraphics[width=\linewidth]{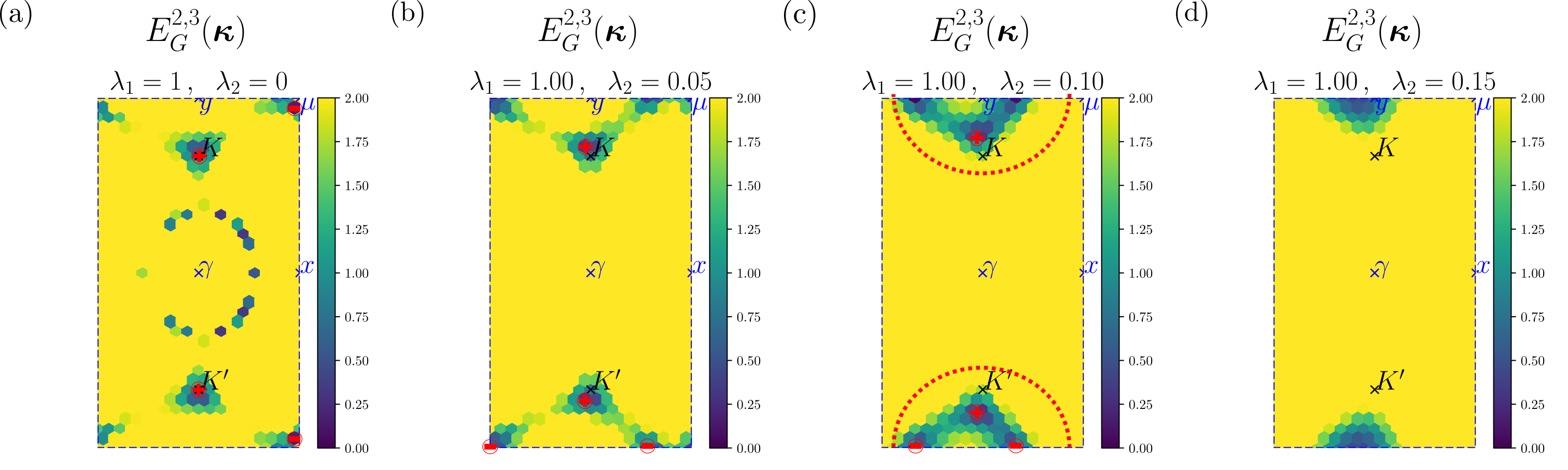}
    \caption{The band gap between the second and the third bands of the interpolated Hamiltonian $\mathscr{H}(\bm{\kappa})$ for $\lambda_1 = 1$ and $\lambda_2 = 0, 0.05, 0.1$ and $0.15$. The color code represents the gap between the second and the third bands. Red symbols $\oplus$ and $\ominus$ stand for the Dirac nodes with opposite chiralities. As shown by the red dashed circle in subfigure (c), the four Dirac nodes are in a patch around $y$ point, and they move towards a point close to $y$ with an increasing $\lambda_2$.}
    \label{fig:add_four_dirac_gap}
\end{figure}

As we have just shown, the band structure at $\lambda_1 = 1$ and $\lambda_2 = 0$ resembles these strong interacting bands. 
By folding this band structure into the FMBZ, we can get four Dirac nodes between the second and the third bands. Nodes at $K$ and $K'$ points with the same chirality in the original MBZ are folded into the FMBZ, which are also labeled by $K$ and $K'$ in Fig.~\ref{fig:add_four_dirac_gap}(a). The other two nodes are close to the $\mu$ point on the corner of the FMBZ, which comes from the two different $M$ points in the original MBZ. The third $M$ point is folded together with the $\Gamma$ point, which becomes a Dirac node between the first and the second bands.
In Figs.~\ref{fig:add_four_dirac_gap}(b-d), we provide the band gap between the second and third bands in the FMBZ when $\lambda_1 = 1$ and $\lambda_2 = 0.05, 0.1$ and $0.15$. 
It can be observed that the four Dirac nodes move away from $K$, $K'$ and $\mu$ points towards a point close to (but not exactly at) $y$ point when the value of $\lambda_2$ is increased.
The four Dirac nodes will meet with each other and annihilate, and the band gap between the two will open at around $\lambda_2 = 0.14$.  

\begin{figure}[t]
    \centering
    \includegraphics[width=\linewidth]{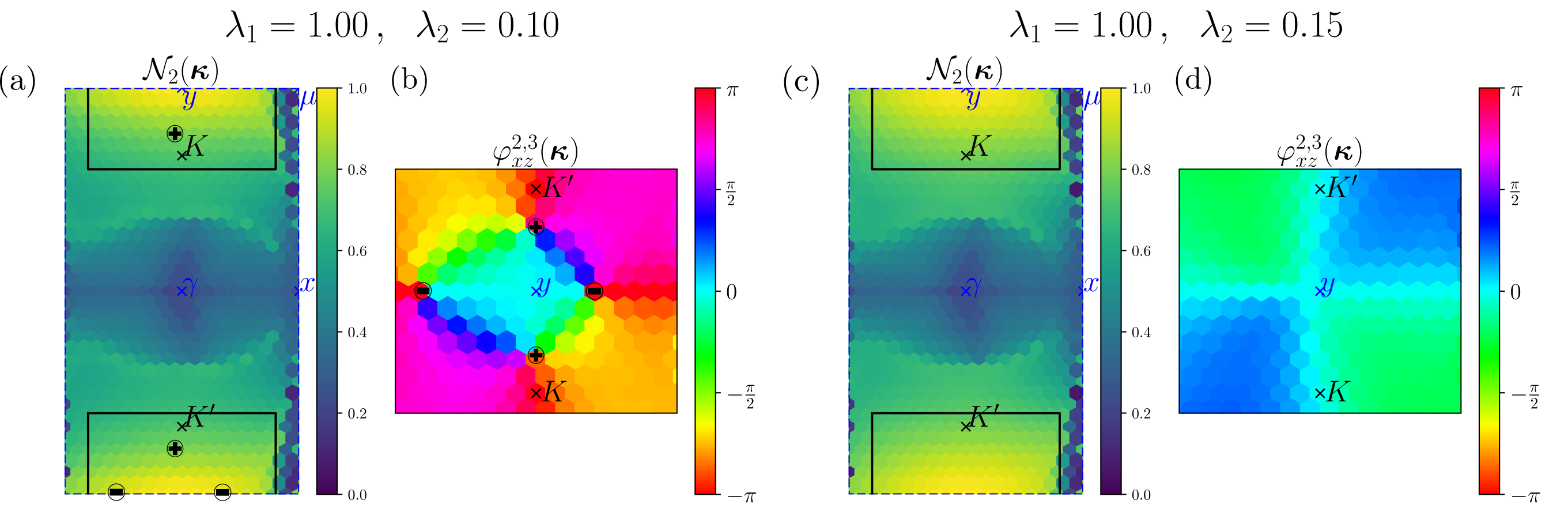}
    \caption{(a) The middle two bands' wavefunction overlap between a given momentum $\bm{\kappa}$ and the reference point $\bm{\kappa}_0 = y$ at $\lambda_1 = 1$ and $\lambda_2 = 0.1$. (b) The angle $\varphi_{xz}^{2,3}(\bm{\kappa})$ at $\lambda_1 = 1, \lambda_2 = 0.1$ in a rectangular patch around $y$ in the FMBZ. The patch $\Pi$ is represented by the black rectangle in subfigure (a). Black symbols $\oplus$ and $\ominus$ represent the Dirac nodes. (c-d) are calculated at $\lambda_1 = 1, \lambda_2 = 0.15$.}
    \label{fig:add_four_dirac_angle}
\end{figure}

We also used the method discussed in App.~\ref{sec:chirality_app} to determine the chirality of these Dirac nodes. 
Since the nodes annihilate around the point $y$, it is natural to choose $\bm{\kappa}_0 = y$ as the reference point of the basis. 
In Fig.~\ref{fig:add_four_dirac_angle} (a), we computed the values of $\mathcal{N}_2(\bm{\kappa})$ over the FMBZ using $y$ as the reference point at $\lambda_2 = 0.1$. 
It can be seen that the value of $\mathcal{N}_2(\bm{\kappa})$ is relatively large around $y$, including the four Dirac nodes represented by black crosses. 
Therefore, as we have discussed in App.~\ref{sec:chirality_app}, we can represent the wavefunctions of the states around these Dirac nodes by the wavefunctions at $\bm{\kappa}_0$. 
In Fig.~\ref{fig:add_four_dirac_angle} (b), we calculated the direction of the wavefunctions on the Bloch sphere $\varphi_{xz}^{2,3}(\bm{\kappa})$ on a rectangular patch around the reference point. 
This patch is also represented by the black solid line in Fig.~\ref{fig:add_four_dirac_angle} (a). 
As shown by the winding direction of $\varphi_{xz}^{2,3}(\bm{\kappa})$, we find that the two Dirac nodes along $\gamma$-$x$ carry the same chirality, while the other two nodes close to $K$, $K'$ points carry the opposite chirality. This result agrees with the strong interacting bands picture we discussed previously in this Appendix.

Thus, the four nodes can annihilate when they meet with each other. 
We also computed the value of $\mathcal{N}_2(\bm{\kappa})$ and $\varphi_{xz}^{2,3}(\bm{\kappa})$ at $\lambda_2 = 0.15$, which can be found in Figs.~\ref{fig:add_four_dirac_angle} (c-d). All the four Dirac nodes are gapped at this point, and there is no winding of $\varphi_{xz}^{2,3}(\bm{\kappa})$.

\subsubsection{Brillouin zone border}\label{sec:add_lam1_lam2_results}

\begin{figure}
    \centering
    \includegraphics[width=0.7\linewidth]{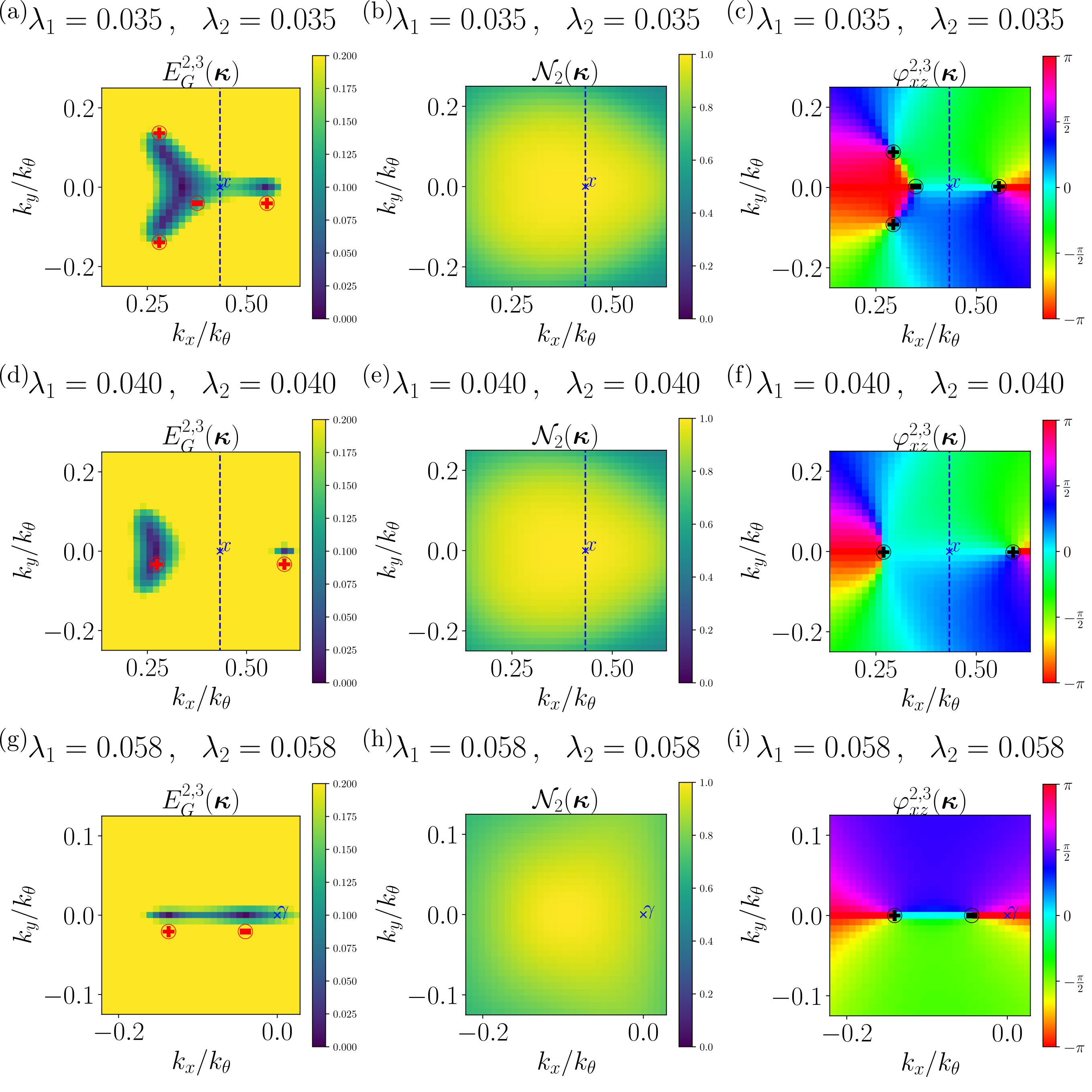}
    \caption{Subfigures (a-c) provide the direct gap $E^{2,3}_G(\bm{\kappa})$, the wavefunction overlaps $\mathcal{N}_2(\bm{\kappa})$ and the angle parameter $\varphi^{2,3}_{xz}(\bm{\kappa})$ between the second and the third bands at $\lambda_1 = \lambda_2 = 0.035$ in the proximity region of $x$ point. Subfigures (d-f) are obtained at $\lambda_1 = \lambda_2 = 0.04$ in the same FMBZ patch around the $x$ point. The results in subfigures (g-i) are obtained in another FMBZ patch around the $\gamma$ point at $\lambda_1 = \lambda_2 = 0.058$. The FMBZ patch choices are shown in Fig.~\ref{fig:diag_path}~(c) in the main text.}
    \label{fig:app_diag_path}
\end{figure}

In this subsection, we provide detailed numerical results about the Dirac nodes motion along the third path introduced in Sec.~\ref{sec:dirac}, namely the direct path [$(\lambda_1, \lambda_2) = (0, 0)\rightarrow (1, 1)$]. We have mentioned in the main text that the two nodes of the non-interacting bands move into the proximity of the $x$ point in the FMBZ around $\lambda_1 = \lambda_2 = 0.035$, while another pair of nodes are also created in this region at the same time. Similar to App.~\ref{sec:add_lam2_1_results}, we use the method described in App.~\ref{sec:chirality_app} to evaluate the chiralities of these nodes. As shown in Figs.~\ref{fig:app_diag_path}~(c), we observe this pair of nodes created around the $x$ point with opposite chiralities along the $\gamma$-$x$ line. One of these two nodes with $-1$ chirality is close to the other nodes with the same chirality $+1$. The other node with $+1$ chirality is on the right of $x$ point. With increasing values of $\lambda_1$ and $\lambda_2$, the three left-moving nodes merge into one node with chirality $+1$, and the right moving mode also carries $+1$ chirality, which can be seen in Fig.~\ref{fig:app_diag_path}~(f). These two nodes move apart from each other as $\lambda_1$ and $\lambda_2$ get larger. Their path wrap around the FMBZ and they approach each other again around $\gamma$ point when $\lambda_1 = \lambda_2 \approx 0.055$. In Fig.~\ref{fig:app_diag_path}~(i), we provide the value of $\varphi^{2,3}_{xz}(\bm{\kappa})$ in a FMBZ patch around the $\gamma$ point at $\lambda_1 = \lambda_2 = 0.058$, and we find that these two Dirac nodes carry opposite chiralities after they went across the FMBZ along the $\tilde{\mathbf{b}}_1$ axis. Thus, band gap between the second and the third bands can be opened after these two nodes annihilate with each other.

\section{Additional numerical results}\label{sec:additional_results}
In this appendix, we provide additional numerical results that were mentioned in Secs.~\ref{sec:phase_diagram} and \ref{sec:c2t_stripe}. In particular, we discuss the spin and valley polarization in various phases in App.~\ref{sec:add_polarization}, and symmetries and real space charge distributions in $C_{2z}T$ stripe and QAH phases in App.~\ref{sec:add_sym_real}.

\subsection{The effect of kinetic energy \& spin and valley polarization}\label{sec:add_polarization}

\begin{figure}[t]
    \centering
    \includegraphics[width=\linewidth]{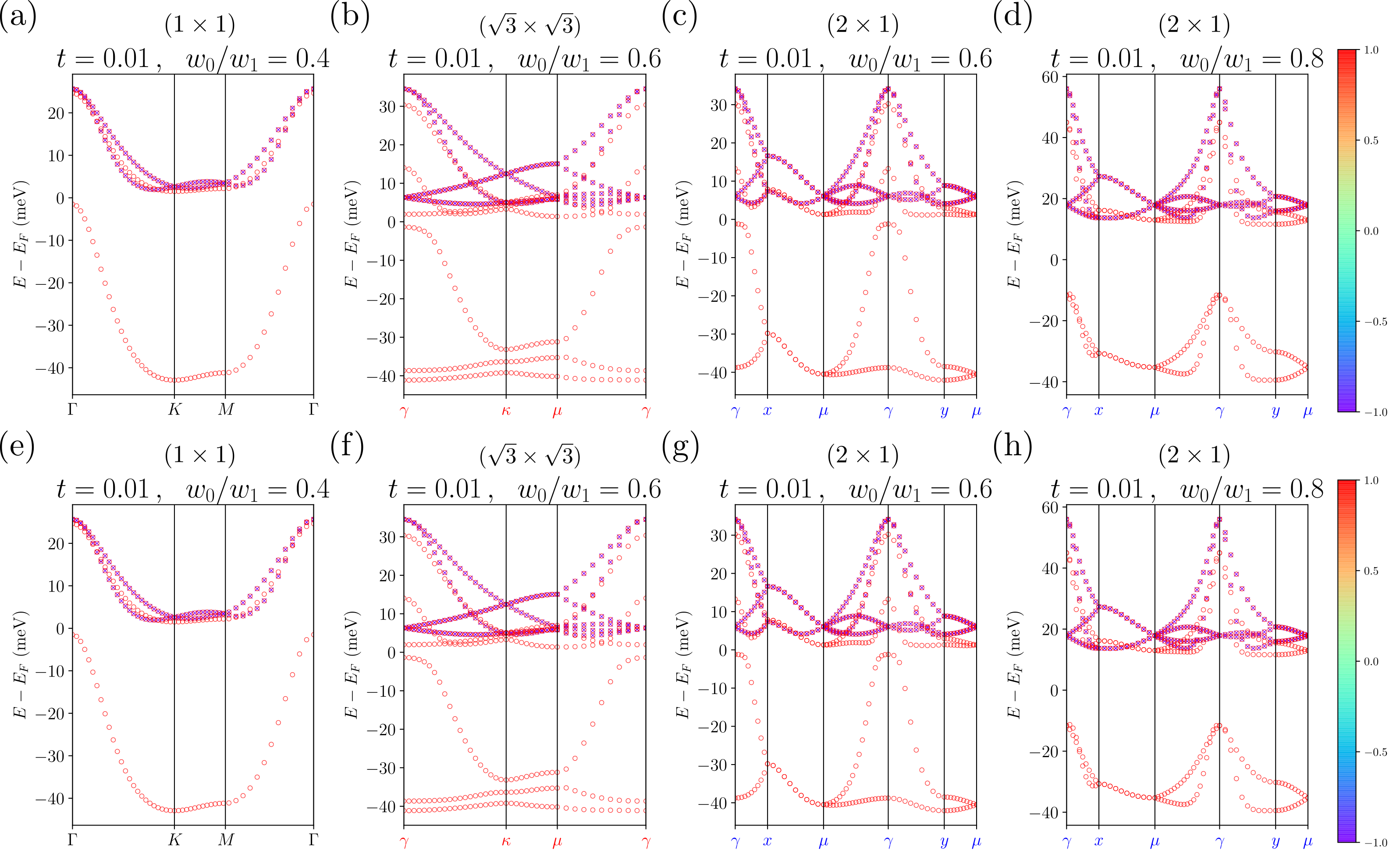}
    \caption{The HF band structure and spin valley polarizations close to flat band limit $(t = 0.01)$. (a-d) The spin polarization $s_{i,z}(\bm{\kappa})$ of each HF band state. (e-h) The valley polarization $v_i(\bm{\kappa})$ of each HF band states.
    We use ``$\circ$'' to represent the states with $s_{z,i}(\bm{\kappa})$ or $v_i(\bm{\kappa}) > 0$ and ``$\times$'' to represent states with $s_{z,i}(\bm{\kappa})$ or $v_i(\bm{\kappa}) < 0$, such that the degenerate states with opposite spin or valley indices become visible.}
    \label{fig:add_band_spinvalley_t_0}
\end{figure}

\begin{figure}[t]
    \centering
    \includegraphics[width=\linewidth]{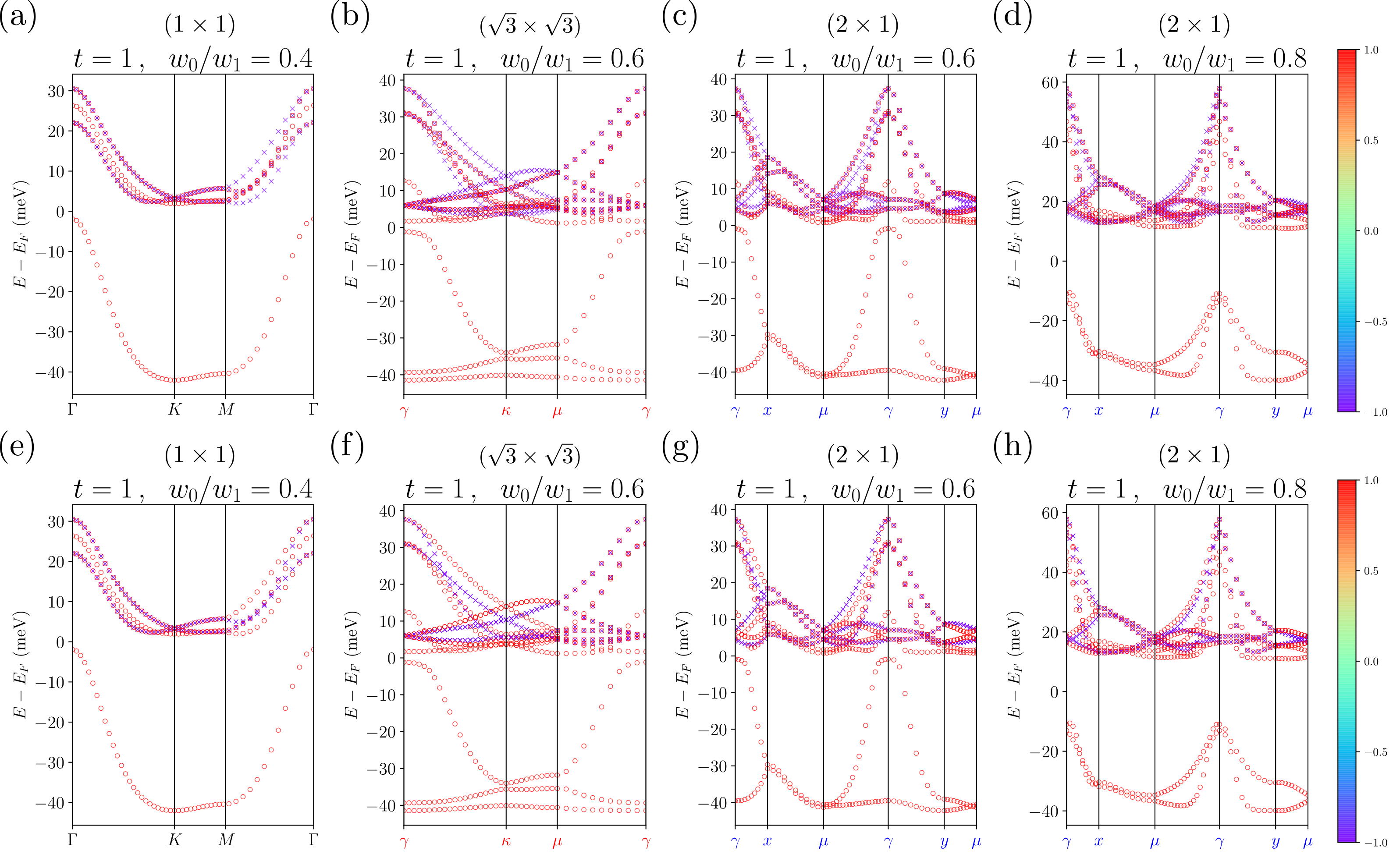}
    \caption{The HF band structure and spin valley polarizations with kinetic energy considered ($t = 1$). (a-d) The spin polarization $s_{i,z}(\bm{\kappa})$ of each HF band state. (e-h) The valley polarization $v_i(\bm{\kappa})$ of each HF band states.
    We use ``$\circ$'' to represent the states with $s_{z,i}(\bm{\kappa})$ or $v_i(\bm{\kappa}) > 0$ and ``$\times$'' to represent states with $s_{z,i}(\bm{\kappa})$ or $v_i(\bm{\kappa}) < 0$, such that the degenerate states with opposite spin or valley indices become visible.}
    \label{fig:add_band_spinvalley_t_1}
\end{figure}

In Sec.~\ref{sec:phase_diagram} of the main text, we have shown the phase diagram and the representative Hartree-Fock band structures by assuming valley polarization and flat band limit ($t = 0$).
In this subsection, we will provide the self-consistent solutions of various phases without assuming flat band limit or any spin or valley polarization, albeit these solutions are obtained on a slightly smaller $12\times 12$ momentum lattice. The spin and valley polarization of these states can also be computed. We found that all these states are spin and valley fully polarized, including the QAH phase, the $C_{2z}T$ stripe phase, and the competing states with intermediate values of $w_0/w_1$.

For each state in the HF bands $\phi_{bm\eta s; i}(\bm{\kappa})$, we can calculate its spin and valley polarization. The spin vector is given by: 
\begin{equation}
    \vec{s}_i(\bm{\kappa}) = \sum_{bm\eta}\sum_{ss'} (\vec{s})_{ss'}\phi^*_{bm\eta s; i}(\bm{\kappa})\phi_{bm\eta s'; i}(\bm{\kappa})\,,
\end{equation}
in which $\vec{s} = (s_x, s_y, s_z)$ are the Pauli matrices of spin indices. Similarly, we can also define the valley polarization as follows:
\begin{equation}
    v_i(\bm{\kappa}) = \sum_{bm s}\sum_{\eta\eta'}(\tau_z)_{\eta\eta'} \phi^*_{bm\eta s; i}(\bm{\kappa})\phi_{bm\eta' s; i}(\bm{\kappa})\,,
\end{equation}
in which $\tau_z$ is the Pauli $z$ matrix acting in valley indices. By studying the values of $\vec{s}_i(\bm{\kappa})$ and $v_i(\bm{\kappa})$ of the occupied states, we can determine whether the solution is spin and valley fully polarized or not, and validate the polarization assumptions in Secs.~\ref{sec:phase_diagram} and \ref{sec:c2t_stripe}.

Since the interacting Hamiltonian of TBG has the spin $SU(2)$ symmetry, the spins of the self-consistent solutions could be along any direction due to the random initial conditions. For that reason, we rotate the direction of the spin of the lowest energy band at $\Gamma$ (when considering the MBZ) or $\gamma$ (for the FMBZ) point to $+z$ direction. We also add a small term $\Delta \mathcal{H}_{bm\eta s;b'n\eta's'}(\bm{\kappa}) = \varepsilon\, \delta_{bb'}\delta_{mn}\delta_{\eta\eta'} (s_z)_{ss'}$ with $\varepsilon \approx 10^{-6}$ to lift the degenerate bands with opposite spins when evaluating the values of $s_{i,z}(\bm{\kappa})$ for each band. Similarly, we also add a term $\Delta \mathcal{H}_{bm\eta s; b'n\eta's'} = \varepsilon\, \delta_{bb'}\delta_{mn}(\tau_z)_{\eta\eta'}\delta_{ss'}$ to lift the degeneracy of bands from opposite valleys when solving the values of $v_i(\bm{\kappa})$.

In the flat band limit, the symmetry of the interacting Hamiltonian is enhanced to $U(4)$ \cite{kang_strong_2019,bultinck_ground_2020,ourpaper3}. Thus, the spin and valley indices could be mixed together due to the randomized initial condition. We solve the self-consistent equation without assuming spin and valley polarization at both $t = 0$ and $t=0.01$. With the kinetic energy being slightly turned on, we can lift the spin and valley degeneracy due to the $U(4)$ symmetry, while the band structures are not strongly affected. Numerical solutions also shows that the energy of the HF energy bands are only changed by $0.014$ meV at most. In Fig.~\ref{fig:add_band_spinvalley_t_0}, we provide the HF band structures at $w_0/w_1 = 0.4$, $0.6$ and $0.8$ with $t = 0.01$. More precisely, the color code represents the spin of each state $s_{z, i}(\bm{\kappa})$ in Figs.~\ref{fig:add_band_spinvalley_t_0} (a-d). It can be observed that the $N_F$ occupied bands are fully spin polarized in the QAH phase, $C_{2z}T$ stripe phase and intermediate states. Similarly, the values of valley polarization $v_i(\bm{\kappa})$ are represented by the color code in Figs.~\ref{fig:add_band_spinvalley_t_0} (e-h), and we also found that the valley is fully polarized in all these three phases.

We also solved the self-consistent solutions with the kinetic energy considered ($t = 1$) at $w_0/w_1 = 0.4, 0.6$ and $0.8$. The HF band structures and the spin valley polarization of these states are shown in Fig.~\ref{fig:add_band_spinvalley_t_1}. Similar to the solutions in the flat band limit, the $N_F$ occupied bands are all spin and valley fully polarized as can be seen by the color code. Moreover, the energy of the states with $(\sqrt{3}\times\sqrt{3})$ enlarged unit cell is still slightly lower than the state with $(2\times 1)$ enlarge unit cell by $\sim 0.013~\rm meV$ per moir\'e unit cell at $w_0/w_1 = 0.6$, which is comparable to the results in the competing region at flat band limit. Furthermore, the energy (per moir\'e unit cell) of $C_{2z}T$ stripe phase at $w_0/w_1 = 0.8$ and $t = 1$ is $\sim 0.21~\rm meV$ lower than the translation symmetric solution, and $\sim 0.13~\rm meV$ lower than the $(\sqrt{3}\times\sqrt{3})$ enlarged unit cell state, which also echo the values shown in Fig.~\ref{fig:mean_field_phase_diag}~(a). In conclusion, the self-consistent solutions at $t = 1$ demonstrate the stability of the $C_{2z}T$ stripe phase against the perturbation from the kinetic energy.

\subsection{Symmetries and real space charge distributions of  \texorpdfstring{$C_{2z}T$}{C2zT} stripe and QAH phases}\label{sec:add_sym_real}
\subsubsection{\texorpdfstring{$C_{2z}T$}{C2T} stripe phase}
As shown in App.~\ref{sec:add_polarization}, the $C_{2z}T$ stripe phase is spin and valley polarized. Therefore, we can perform the self-consistent mean field solution in the presence of kinetic term $H_0$ ({\it i.e.}, when $t = 1$) at $w_0/w_1 = 0.8$ on a much larger $36\times 36$ momentum lattice by assuming spin and valley polarization to study the properties of the $C_{2z}T$ stripe phase.  

\begin{figure}[!htbp]
    \centering
    \includegraphics[width=\linewidth]{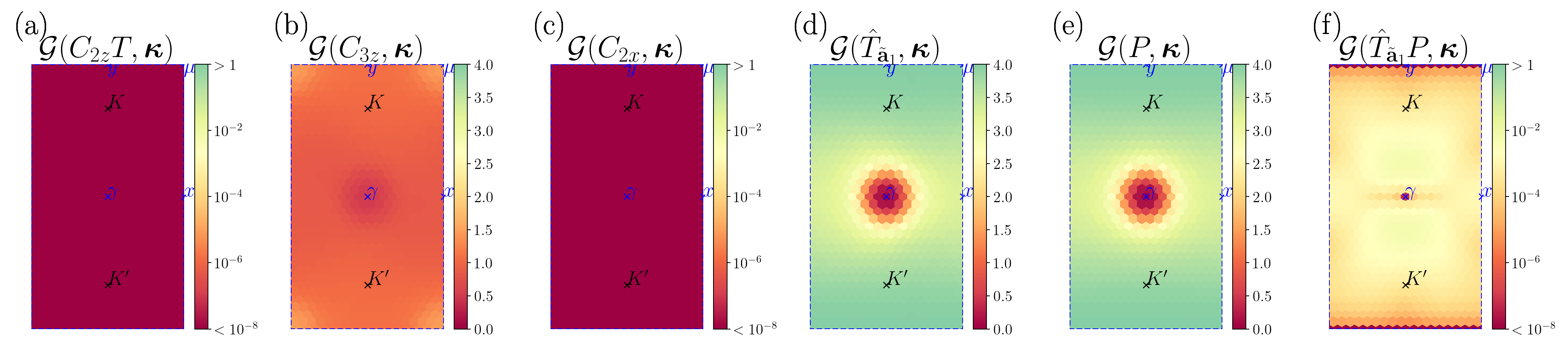}
    \caption{The symmetry breaking strength of six types of lattice symmetries $\mathcal{G}(g, \bm{\kappa})$ calculated from the Hartree-Fock solution at $w_0/w_1 = 0.8$ on $36\times 36$ momentum lattice. Unlike Fig.~\ref{fig:flatband_stripe_sym_break}, the kinetic Hamiltonian is considered here. The maximum value of $\mathcal{G}(\hat{T}_{\tilde{\mathbf{a}}_1}P, \bm{\kappa})$ is around $0.006$ in subfigure (f). Note that (a), (c) and (f) use log scale for $\mathcal{G}(g,\bm{\kappa})$.}
    \label{fig:add_non_flatband_stripe_sym_break}
\end{figure}

In Fig.~\ref{fig:add_non_flatband_stripe_sym_break}, we first present the symmetry breaking strength $\mathcal{G}(g, \bm{\kappa})$ for the six types of lattice symmetries given in Table \ref{tab:real_space_sym}. Similar to the flat band results, both $C_{2z}T$ and $C_{2x}$ symmetries are preserved. However, since the kinetic Hamiltonian satisfies $[H_0, P] \neq 0$ and $\{H_0, P\} = 0$, the total Hamiltonian does not commute with the particle-hole transformation, therefore there is no particle-hole symmetry. 
In Fig.~\ref{fig:add_non_flatband_stripe_sym_break} (f), we find that the presence of kinetic energy also breaks $\hat{T}_{\tilde{\mathbf{a}}_1}P$ symmetry, although the symmetry breaking is very weak (maximum value of $\mathcal{G}(\hat{T}_{\tilde{\mathbf{a}}_1}P, \bm{\kappa})$ is around $0.006$). We expect that some properties of the real space density distribution which requires $\hat{T}_{\tilde{\mathbf{a}}_1}P$ symmetry are no longer strictly correct, but will be approximately satisfied.

\begin{figure}[!htbp]
    \centering
    \includegraphics[width=0.3\linewidth]{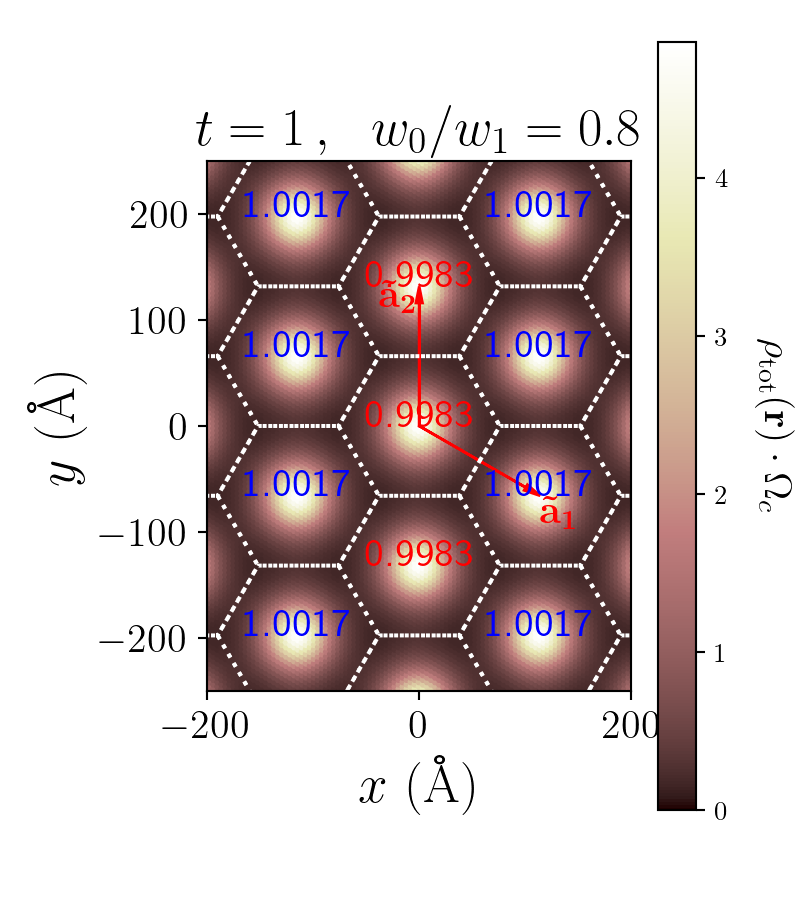}
    \caption{The total charge distribution in real space at $w_0/w_1 = 0.8$, with kinetic energy considered. The numbers are total electron numbers in each moir\'e unit cell, in which we observe a charge ``density wave''. The modulation of total charge between difference unit cells is less than $0.2\%$.}
    \label{fig:add_non_flatband_stripe_real_total}
\end{figure}

\begin{figure}[!htbp]
    \centering
    \includegraphics[width=\linewidth]{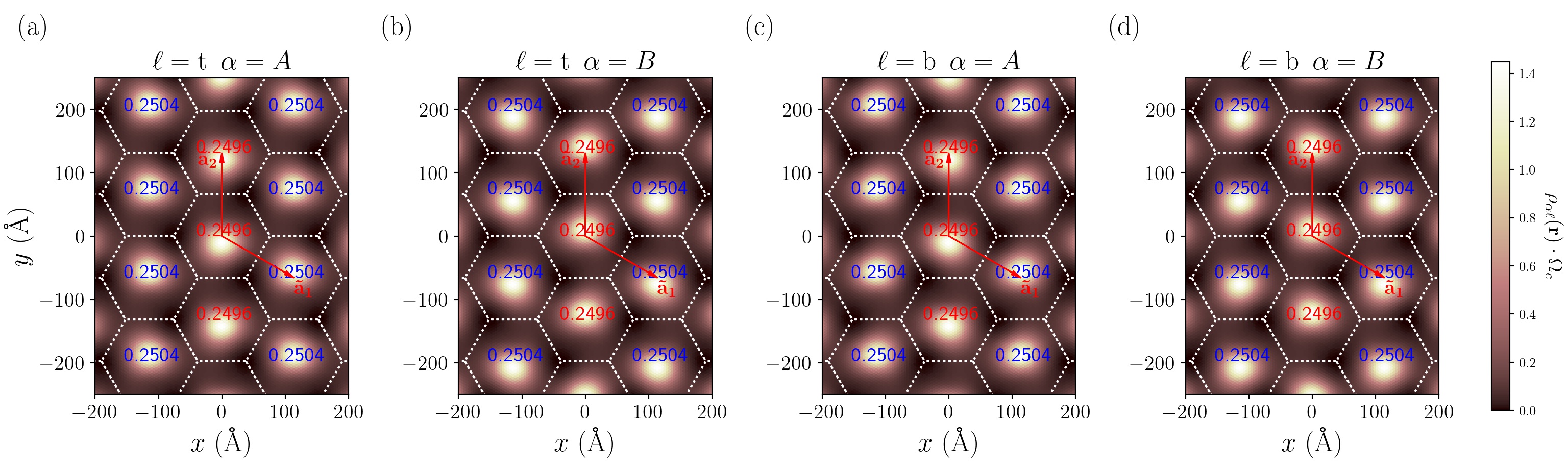}
    \caption{The electron density distribution in real space at $w_0/w_1 = 0.8$, and the kinetic energy is considered. Similar to Fig.~\ref{fig:flatband_stripe_real_sublattice_layer}, the numbers represent the total electron charge for each component $Q_{\alpha\ell}$ in each moir\'e unit cell.}
    \label{fig:add_non_flatband_stripe_real_sublattice_layer}
\end{figure}

Similar to Fig.~\ref{fig:flatband_stripe_real_total} in main text, Fig.~\ref{fig:add_non_flatband_stripe_real_total} presents the total electron density in real space. We can already notice that the total charge $Q$ in the unit cell around $\mathbf{r} = 0$ is different from the $Q$ in the unit cell around $\mathbf{r} = \tilde{\mathbf{a}}_1$. The charge on every $AA$ stacking site is slightly modulated, although the total charge difference between two moir\'e unit cells are differed by less than $0.2\%$. This could also be observed in the charge density for different sublattice and layer components in Fig.~\ref{fig:add_non_flatband_stripe_real_sublattice_layer}. Indeed, the electron density in sublattice $A$ top layer $\rho_{\alpha=A,\ell={\rm t}}(\mathbf{r})$ is not equal to the density distribution in sublattice $B$ bottom layer $\rho_{\alpha=B,\ell={\rm b}}(\mathbf{r} + \tilde{\mathbf{a}}_1)$ due to the weakly breaking $\hat{T}_{\tilde{\mathbf{a}}_1}P$, although these two values are very close to each other. 

\begin{figure}[!htbp]
    \centering
    \includegraphics[width=\linewidth]{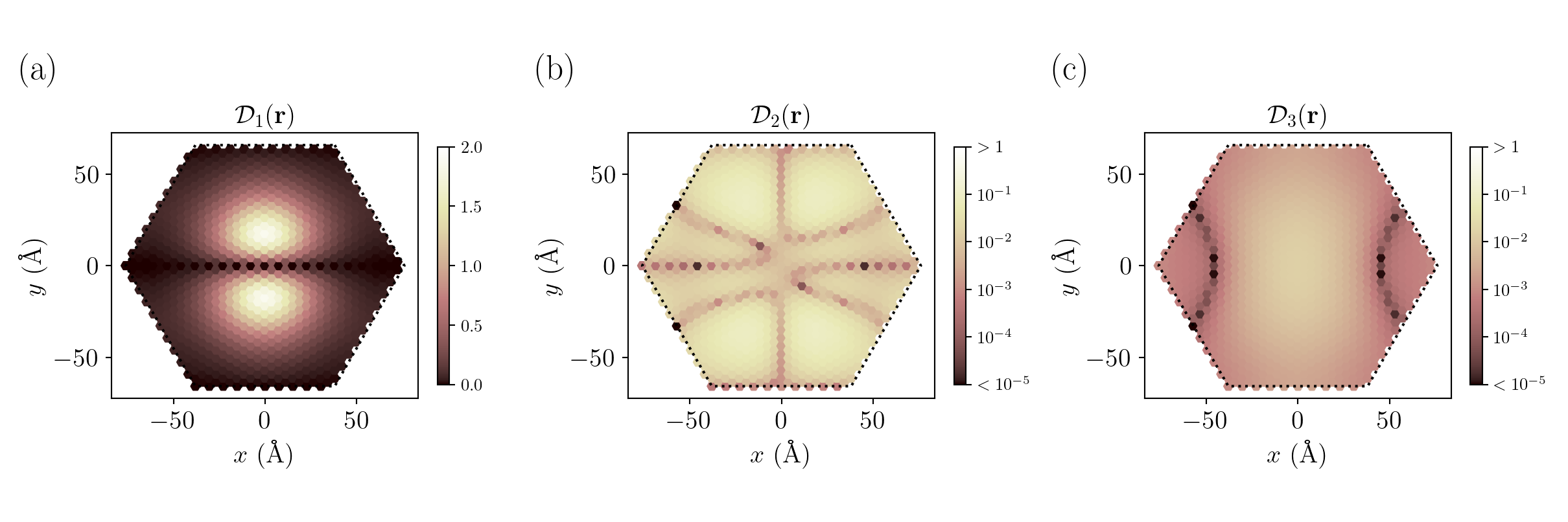}
    \caption{The translation symmetry breaking of the density distribution at $w_0/w_1 = 0.8$ when the kinetic Hamiltonian is included. The definition of each quantity shown in the three subfigures ($\mathcal{D}_1(\mathbf{r})$, $\mathcal{D}_2(\mathbf{r})$ and $\mathcal{D}_3(\mathbf{r})$ respectively) are given by Eqs.~(\ref{eqn:def_D1}), (\ref{eqn:def_D2}) and (\ref{eqn:add_def_D3}). Note that (b) and (c) use color log scale.}
    \label{fig:add_non_flatband_stripe_real_shift}
\end{figure}

To quantify the change of charge density distribution under translation transformation $\mathbf{r} \rightarrow \mathbf{r} + \tilde{\mathbf{a}}_1$, we evaluate the values of the functions $\mathcal{D}_{1}(\mathbf{r})$ and $\mathcal{D}_2(\mathbf{r})$ defined in Eqs.~(\ref{eqn:def_D1}) and (\ref{eqn:def_D2}) for the $C_{2z}T$ stripe phase solution with kinetic energy. Since the symmetry $\hat{T}_{\tilde{\mathbf{a}}_1}P$ symmetry is broken, this state is not invariant under the transformation $C_{2z}T\hat{T}_{\tilde{\mathbf{a}}_1}P$, and the total charge density will no longer be the same under translation $\mathbf{r}\rightarrow \mathbf{r} + \tilde{\mathbf{a}}_1$, as we have discussed in Sec.~\ref{sec:real_space}. To measure the change of the total charge density under such translation, we can also define the following quantity:
\begin{equation}\label{eqn:add_def_D3}
    \mathcal{D}_3(\mathbf{r}) = \Omega_c \Bigg{|}\sum_{\alpha\ell}\Big{[}\rho_{\alpha\ell}(\mathbf{r}) - \rho_{\alpha\ell}(\mathbf{r} + \tilde{\mathbf{a}}_1)\Big{]} \Bigg{|}\,.
\end{equation}
Clearly $\mathcal{D}_3 = 0$ when the total charge distribution is exactly the same in two moir\'e unit cells. We evaluate the values of $\mathcal{D}_{1}(\mathbf{r})$, $\mathcal{D}_2(\mathbf{r})$ and $\mathcal{D}_3(\mathbf{r})$ over a moir\'e unit cell, and the results can be found in Fig.~\ref{fig:add_non_flatband_stripe_real_shift}. From Figs.~\ref{fig:add_non_flatband_stripe_real_shift} (b) and (c), we find that both the total charge density and single layer charge density are changed notably after the real space translation. The maximum value of total charge density change between moir\'e unit cells as measured by $\mathcal{D}_3(\mathbf{r})$ is around $0.016$, as expected by the weak breaking of the $P$ symmetry due to the kinetic term.

\subsubsection{QAH phase}
We have also analyzed the symmetries and the real space charge distributions of the quantum anomalous Hall states at $t=0, 1$ and $w_0/w_1 = 0.4$ on a $36\times 36$ momentum lattice. Similar to the $C_{2z}T$ stripe phase, this state is also spin and valley polarized as shown in App.~\ref{sec:add_polarization}. 

\begin{figure}[!htbp]
    \centering
    \includegraphics[width=\linewidth]{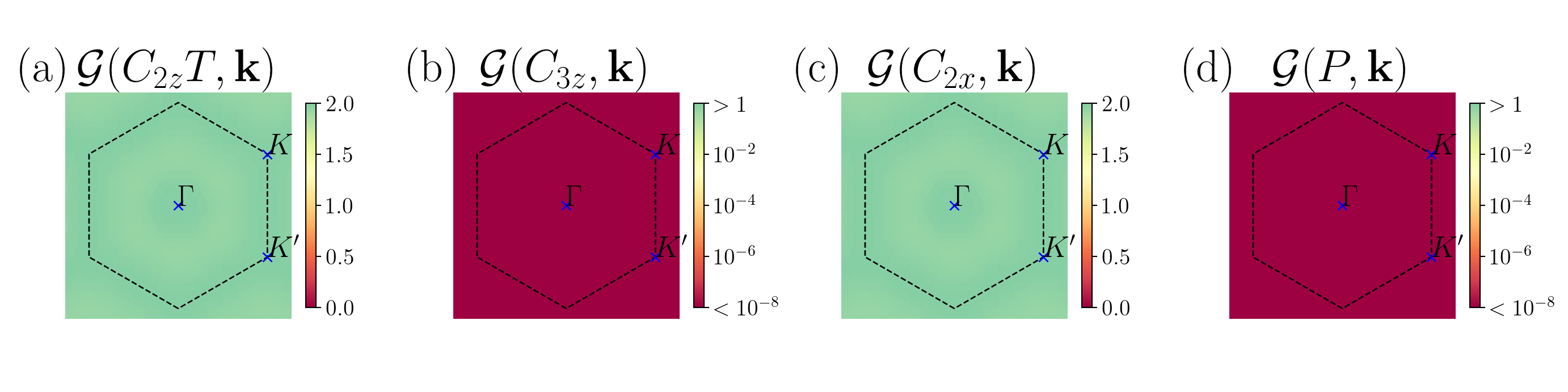}
    \caption{The symmetry breaking strength of four types of single valley symmetries ($C_{2z}T$, $C_{3z}$, $C_{2x}$ and $P$) calculated from the QAH state solution at flat band limit ($t=0$) and $w_0/w_1 = 0.4$ on $36\times 36$ momentum lattice. The black dashed line stands for the moir\'e Brillouin zone. Note that we use log scale for $\mathcal{G}(C_{3z}, \vk)$ (b) and $G(P, \vk)$ (d). We found that this QAH state has broken $C_{2z}T$ and $C_{2x}$ symmetries, while it is still invariant under $C_{3z}$ and $P$ transformations.}
    \label{fig:add_flatband_qah_sym_break}
\end{figure}

\begin{figure}[!htbp]
    \centering
    \includegraphics[width=\linewidth]{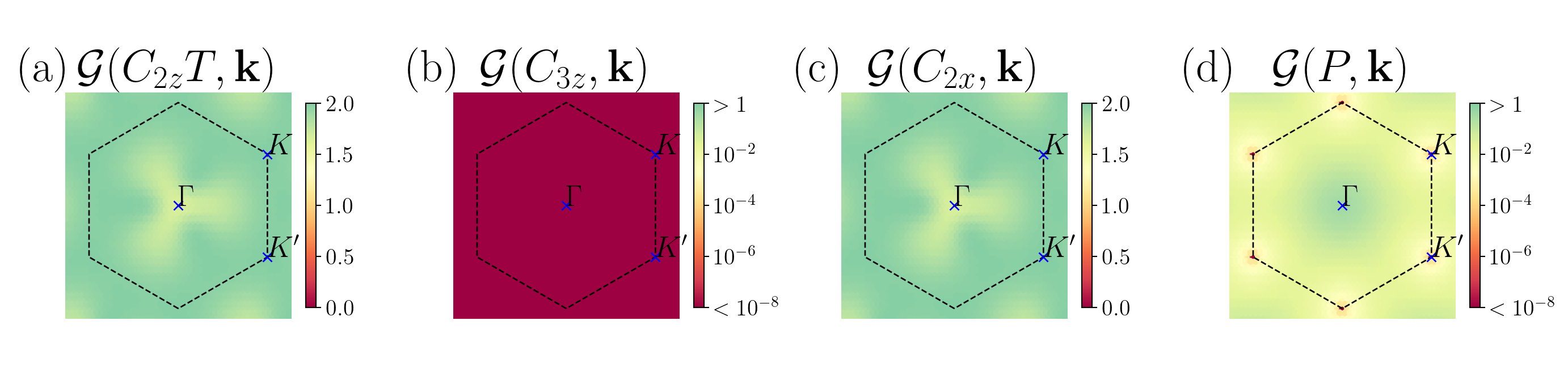}
    \caption{The symmetry breaking strength of four types of symmetries calculated from the QAH state solution with kinetic energy terms ($t=1$) at $w_0/w_1 = 0.4$ on $36\times 36$ momentum lattice. Similar to Fig.~\ref{fig:add_flatband_qah_sym_break}, we use log scale in subfigures (b) and (d). When the kinetic terms are considered, the QAH state breaks the $P$ symmetry, and the maximum value of $\mathcal{G}(P, \vk)$ is about 0.35.}
    \label{fig:add_non_flatband_qah_sym_break}
\end{figure}

The QAH state does not break the translation symmetry. Therefore, we use $\vk$ to represent the momentum in moir\'e Brillouin zone, instead of $\bm{\kappa}$. Eq.~(\ref{eqn:def_sym_break}) in the main text can also be defined for the MBZ. 
In Figs.~\ref{fig:add_flatband_qah_sym_break} and \ref{fig:add_non_flatband_qah_sym_break}, we provide the values of the symmetry breaking strength of the state in the flat band limit (Fig.~\ref{fig:add_flatband_qah_sym_break}) and at $t=1$ (Fig.~\ref{fig:add_non_flatband_qah_sym_break}). 
Here we consider four types of single valley symmetries: $C_{2z}T$, $C_{3z}$, $C_{2x}$ and $P$. 
As shown in Fig.~\ref{fig:add_flatband_qah_sym_break}, the QAH state in the flat band limit breaks $C_{2z}T$ and $C_{2x} $symmetries, while it is invariant under $C_{3z}$ and $P$ transformation. 
If the kinetic term is added into consideration (Fig.~\ref{fig:add_non_flatband_qah_sym_break}), the $C_{3z}$ symmetry is still fulfilled, but all other three symmetries are broken.
It is reasonable to observe strong $C_{2z}T$ symmetry breaking in both Figs.~\ref{fig:add_flatband_qah_sym_break} (a) and \ref{fig:add_non_flatband_qah_sym_break} (a), since the breaking of $C_{2z}T$ is a property of states with nonzero winding numbers. Besides, the $P$ transformation commutes with the projected interacting Hamiltonian $H_I$ and anti-commutes with the kinetic Hamiltonian $H_0$, and therefore the total Hamiltonian at $t = 1$ does not commute with $P$. Hence, the QAH state at $t=1$ is not symmetric under the $P$ transformation, as shown in Fig.~\ref{fig:add_non_flatband_qah_sym_break} (d).

\begin{figure}[!htbp]
    \centering
    \includegraphics[width=\linewidth]{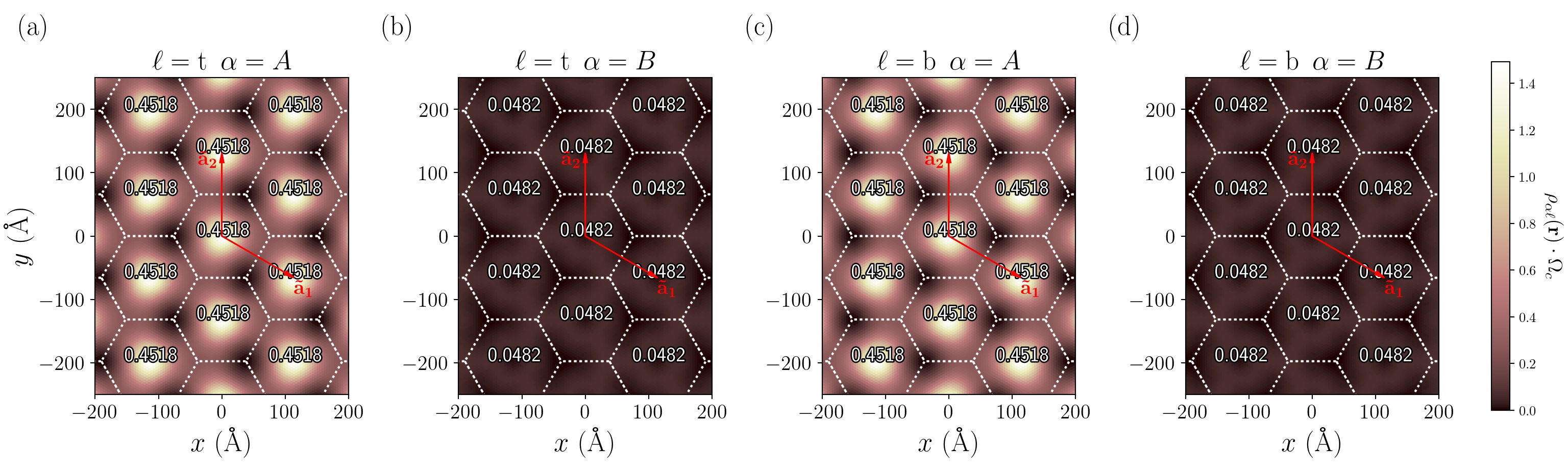}
    \caption{The electron density distribution of the QAH state in real space at flat band limit and $w_0/w_1 = 0.4$. The numbers represent the total electron charge of the corresponding component $Q_{\alpha\ell}$ in each moir\'e unit cell.}
    \label{fig:add_flatband_qah_real_sublattice_layer}
\end{figure}

\begin{figure}[!htbp]
    \centering
    \includegraphics[width=\linewidth]{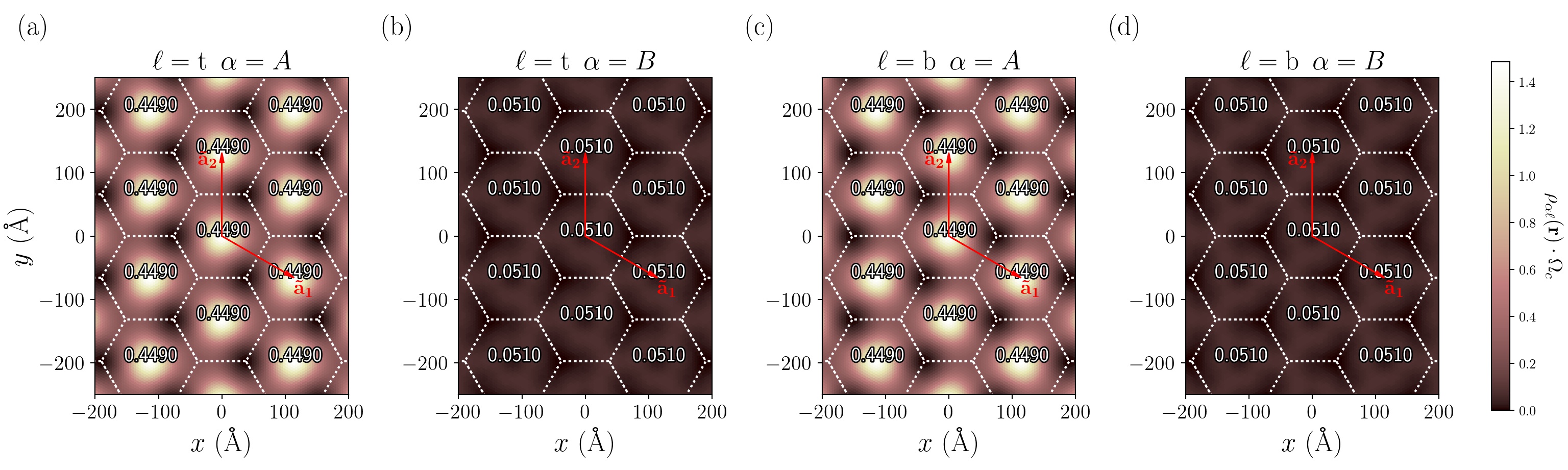}
    \caption{The electron density distribution of the QAH state at $t=1$ and $w_0/w_1 = 0.4$. The numbers represent the total electron charge of each component $Q_{\alpha\ell}$ in every moir\'e unit cell.}
    \label{fig:add_non_flatband_qah_real_sublattice_layer}
\end{figure}

We can also apply Eq.~(\ref{eqn:density_fourier}) to these QAH states to obtain the charge distributions in real space. In Figs.~\ref{fig:add_flatband_qah_real_sublattice_layer} and \ref{fig:add_non_flatband_qah_real_sublattice_layer}, we provide the numerical results of $\rho_{\alpha\ell}(\mathbf{r})$ of each sublattice and layer components for the QAH states at $t=0$ and $t=1$, respectively. The white numbers represent the total charge of each component in every moir\'e unit cell $Q_{\alpha\ell}$, which is defined in Eq.~(\ref{eqn:def_charge_sublattice_layer}) in the main text. In both of the cases, the electrons can be found on $A$ sublattices with a much higher probability than on $B$ sublattices, since the Chern band wavefunctions in TBG has a substantial sublattice polarization, as discussed in Ref.~\cite{bultinck_ground_2020}.

\end{document}